\documentclass[proof]{pasj01}
\Received{$\langle$reception date$\rangle$}
\Accepted{$\langle$acception date$\rangle$}
\Published{$\langle$publication date$\rangle$}
\usepackage{graphicx}
\usepackage{multirow}
\usepackage{comment}
\usepackage{xcolor}
\usepackage{ulem}

\begin{document}

\title{Nobeyama 45-m Mapping Observations toward Orion A.\  I. Molecular Outflows}

\author{Yoshihiro \textsc{Tanabe}\altaffilmark{1}}
\author{Fumitaka \textsc{Nakamura}\altaffilmark{2,3,4}}
\author{Takashi \textsc{Tsukagoshi}\altaffilmark{2}}
\author{Yoshito \textsc{Shimajiri}\altaffilmark{2,5,6}}
\author{Shun \textsc{Ishii}\altaffilmark{2,7}}
\author{Ryohei \textsc{Kawabe}\altaffilmark{2,3,4}}
\author{Jesse R. \textsc{Feddersen}\altaffilmark{8}}
\author{Shuo \textsc{Kong}\altaffilmark{8}}
\author{Hector G. \textsc{Arce}\altaffilmark{8}}
\author{John \textsc{Bally}\altaffilmark{9}}
\author{John M. \textsc{Carpenter}\altaffilmark{7}}
\author{Munetake \textsc{Momose}\altaffilmark{1}}

\altaffiltext{1}{College of Science, Ibaraki University, 2-1-1 Bunkyo, Mito, Ibaraki 310-8512, Japan}
\altaffiltext{2}{National Astronomical Observatory of Japan, 2-21-1 Osawa, Mitaka, Tokyo 181-8588, Japan}
\altaffiltext{3}{The Graduate University for Advanced Studies (SOKENDAI), 2-21-1 Osawa, Mitaka, Tokyo
181-0015, Japan}
\altaffiltext{4}{Department  of  Astronomy,  The  University  of  Tokyo,  Hongo,  Tokyo  113-0033,  Japan}
\altaffiltext{5}{Laboratoire AIM, CEA/DSM-CNRS-Universit\'e Paris Diderot, IRFU/Service d’Astrophysique, CEA Saclay, F-91191 Gif-sur-Yvette, France}
\altaffiltext{6}{Department of Physics and Astronomy, Graduate School of Science and Engineering, Kagoshima University, 1-21-35 Korimoto, Kagoshima,
Kagoshima 890-0065, Japan}
\altaffiltext{7}{Joint ALMA Observatory, Alonso de C\'ordova 3107 Vitacura, Santiago, Chile}
\altaffiltext{8}{Department of Astronomy, Yale University, New Haven, CT 06511, USA}
\altaffiltext{9}{Department of Astrophysical and Planetary Sciences, University of Colorado, Boulder, CO, USA}

\email{yoshihiro.tanabe.ap@vc.ibaraki.ac.jp}


\maketitle

\begin{abstract}
We conducted an exploration of  ${}^{12}$CO molecular outflows in the Orion A giant molecular cloud to investigate outflow feedback using ${}^{12}$CO ({\it J} = 1--0) and ${}^{13}$CO ({\it J} = 1--0) data obtained by the Nobeyama 45-m telescope. 
In the region excluding the center of OMC 1, we identified 44 ${}^{12}$CO (including 17 newly detected) outflows based on the unbiased and systematic procedure of 
automatically determining the velocity range of the outflows and separating the cloud and outflow components.
The optical depth of the ${}^{12}$CO emission in the detected outflows is estimated to be approximately 5.
The total momentum and energy of the outflows, corrected for optical depth, are estimated to be 1.6 $\times$ 10$^2$ $\ M_{\odot}$ km s$^{-1}$ and 1.5 $\times$ 10$^{46}$ erg, respectively.  
The momentum and energy ejection rate of the outflows are estimated to be 36\% and 235\% of the momentum and energy dissipation rates of the cloud turbulence, respectively.
Furthermore, the ejection rates of the outflows are comparable to those of the expanding molecular shells estimated by Feddersen et al. (2018, ApJ, 862, 121).
Cloud turbulence cannot be sustained by the outflows and shells unless the energy conversion  efficiency is as high as 20\%.
\end{abstract}

\section{Introduction}

During the early phase of star formation, molecular outflows are ubiquitous phenomena in both low- and high-mass star-forming regions  (e.g., see Lada \yearcite{1985ARA&A..23..267L} and Arce et al. \yearcite{2007prpl.conf..245A}). 
Outflows inject energy and momentum into the parent molecular cloud and may play an important role in the self-regulation of star formation (e.g., see Norman \& Silk \yearcite{1980ApJ...238..158N} and Bally \yearcite{2016ARA&A..54..491B}). 
This substantial injected energy can affect the cloud structure and evolution (Solomon et al. \yearcite{1981ApJ...245L..19S}).
Feedback from outflows may maintain the turbulence in molecular clouds (Shu et al. \yearcite{1987ARA&A..25...23S}; Nakamura \& Li \yearcite{2007ApJ...662..395N}), protecting the clouds from gravitational collapse (Shu et al. \yearcite{1987ARA&A..25...23S}).

In the past decade, many researchers have investigated molecular outflows and their feedback in parent molecular clouds by comparing the energy ejection rates of outflows and the energy dissipation rates of cloud turbulence.
For example, 
Arce et al. (\yearcite{2010ApJ...715.1170A}) conducted an outflow survey in Perseus and concluded that the outflows had sufficient energy to feed the observed turbulence in the entire Perseus cloud complex (see also Hatchell et al. \yearcite{2007A&A...472..187H}, \yearcite{2009A&A...502..139H}).
Nakamura et al. (\yearcite{2011ApJ...737...56N}, \yearcite{2011ApJ...726...46N}) investigated the outflows in the L1688 cloud of the $\rho$ Ophiuchi regions and the Serpens South cloud, and they found that the energy ejection rate of outflows in each cloud was comparable to the energy dissipation rate of the cloud turbulence (see also White at al. \yearcite{2015MNRAS.447.1996W}).
Li et al. (\yearcite{2015ApJS..219...20L}) conducted an unbiased outflow survey in Taurus and concluded that the outflow feedback was sufficient to maintain the observed turbulence in the current epoch. 
The abovementioned studies were all conducted in relatively low-mass ($\leq10^4\ M_{\odot}$ Enoch et al. \yearcite{2006ApJ...638..293E}) molecular clouds containing low-mass star forming regions.
In this study, we investigate the feedback of outflows in the Orion A giant molecular cloud (referred to simply as Orion A hereafter), which contains high-mass star-forming regions.

Orion A is the nearest high-mass star-forming region, with a distance estimated to be 414 $\pm$ 7 pc by Menten et~al. (\yearcite{2007A&A...474..515M}).
Orion A is often divided into several subregions (e.g., see Bally \yearcite{2008hsf1.book..459B} and Feddersen et al. \yearcite{2018ApJ...862..121F}).
OMC 2/3 is an intermediate-mass star-forming region located in the northernmost part of Orion A (Chini et al. \yearcite{1997ApJ...474L.135C}).
More than 500 young stellar objects (YSOs) have been found in this region in previous observations (e.g., see Chini et al. \yearcite{1997ApJ...474L.135C} and Megeath et al. \yearcite{2012AJ....144..192M}).
OMC 1 has the brightest intensity and the largest velocity width.
This region contains the H\,\emissiontype{II} region M42 created by the Trapezium cluster as well as the BN/KL Nebula.
OMC 4/5 is located south of OMC 1. 
Shimajiri et al. (\yearcite{2015ApJS..217....7S}) identified 225 cores in dust continuum at $\lambda=1.1$\ mm in this region.
L1641-N and NGC 1999 contain well-known young clusters such as L1641-N and V380 Ori, respectively.
Approximately 80 YSOs have been identified in the L1641-N cluster (Fang et al. \yearcite{2009A&A...504..461F}).
The V380 Ori cluster contains many Harbig-Haro (HH) objects (Allen \& Davis \yearcite{2008hsf1.book..621A}).

Previous outflow surveys of Orion A were limited to particular subregions, such as OMC 2/3 
(e.g., Chini et al. \yearcite{1997ApJ...474L.135C}, Aso et al. \yearcite{2000ApJS..131..465A}, Williams et al. \yearcite{2003ApJ...591.1025W}, Takahashi et al. \yearcite{2008ApJ...688..344T}, and Bern{\'e} et al. \yearcite{2014ApJ...795...13B})
, V380 Ori (Davis et al. \yearcite{2000MNRAS.318..952D}), and L1641-N (Stanke \& Williams \yearcite{2007AJ....133.1307S} and Nakamura et al. \yearcite{2012ApJ...746...25N}). 
In particular, the OMC 2/3 regions were observed relatively well.
 Aso et al. (\yearcite{2000ApJS..131..465A}) and Williams et al. (\yearcite{2003ApJ...591.1025W}) conducted outflow surveys of OMC 2/3 in ${}^{12}$CO\ ({\it J} = 1--0) with angular resolutions of $\sim$\timeform{20"} and $\sim$\timeform{10"}, and they identified eight and nine outflows, respectively. 
Williams et al. (\yearcite{2003ApJ...591.1025W}) found that the energy ejection rate of the outflows was comparable to the energy dissipation rate of the cloud turbulence in OMC 2/3. 
Takahashi et al. (\yearcite{2008ApJ...688..344T}) also conducted a survey of the same region with ${}^{12}$CO\ ({\it J} = 3--2) and detected 14 outflows.
They identified outflows from the analyses of channel maps and position-velocity diagrams.
Despite their efforts, the search procedures of outflows in these studies were not systematic.

In this paper, we present the results of our systematic outflow survey across the entirety of Orion A cloud and discuss the impact of the outflows on their parent molecular cloud.
This work is part of  the "Star Formation Legacy Project" using the Nobeyama 45-m telescope; an overview of the project is given in a separate paper (Nakamura et al. in prep.).
An outline of the remainder of this paper is as follows.
Section 2 describes the details of our Nobeyama 45-m observations and data.
In section 3, we present a systematic procedure for searching outflows. 
The results of our outflow search and the physical parameters of the identified outflows are given in section 4.
In section 5, we discuss the outflow feedback into Orion A.
Section 6 summarizes the main results of this study.

\section{Observations and data}\label{obs}

We conducted 2 deg$^2$ mapping observations of ${}^{12}$CO ({\it J} = 1--0, 115.271202 GHz) and ${}^{13}$CO ({\it J} = 1--0, 110.201354 GHz) of Orion A using the FOREST (Minamidani et al.\ \yearcite{2016SPIE.9914E..1ZM}) receiver mounted on the NRO 45-m telescope. 
The observations were made between December 2014 and March 2017.
The telescope beam size (HPBW) was $\sim$\timeform{14"} at 115 GHz and the typical pointing accuracy was $\timeform{3"}$. 
We employed the On-The-Fly (OTF) scan mode (Sawada et al.\ \yearcite{2008PASJ...60..445S}) for the mapping observations. 
We adopted a spheroidal function with a spatial grid size of $\timeform{7".5}$ as a convolution function.
To improve sensitivity and coverage, we combined the FOREST data and with previously published data (Shimajiri et al. \yearcite{2011PASJ...63..105S}, \yearcite{2014A&A...564A..68S}, \yearcite{2015ApJS..221...31S} and Nakamura et al. \yearcite{2012ApJ...746...25N}) collected by the BEARS receiver (Sunada et al.\ \yearcite{2000SPIE.4015..237S}). 
The final maps of the ${}^{12}$CO and ${}^{13}$CO have an effective resolutions (FWHM) of $\sim\timeform{22"}$, corresponding to $\sim$0.05 pc at a distance of 414 pc, and an effective velocity resolution of $\sim$0.2 km s$^{-1}$. 
The typical root mean square (rms) noise levels of the ${}^{12}$CO and ${}^{13}$CO cubes are 0.47 K and 0.18 K in units of $T_{\rm MB}$, respectively.  
More details on the observations and data reduction are described by Kong et al. (\yearcite{2018ApJS..236...25K}) and Nakamura et al. (in prep.) .

\begin{table}
\caption{Observed lines and data sensitivity}
  \begin{tabular}{cccccc} \hline
   Molecule&Transition&Rest Frequency&Effective Resolution&Velocity Resolution&Noise Level\\
   &&(GHz)&(arcsec)&($\rm km\ s^{-1}$)&(K)\\\hline
   {$\rm {}^{12}CO$}&{\it J} = 1--0&115.271202&21.6&0.20&0.47\\
   {$\rm {}^{13}CO$}&{\it J} = 1--0&110.201354&22.0&0.22&0.18\\\hline
\label{}
\end{tabular}
\end{table}

\begin{figure}[h]
 \includegraphics[keepaspectratio, width=16cm]{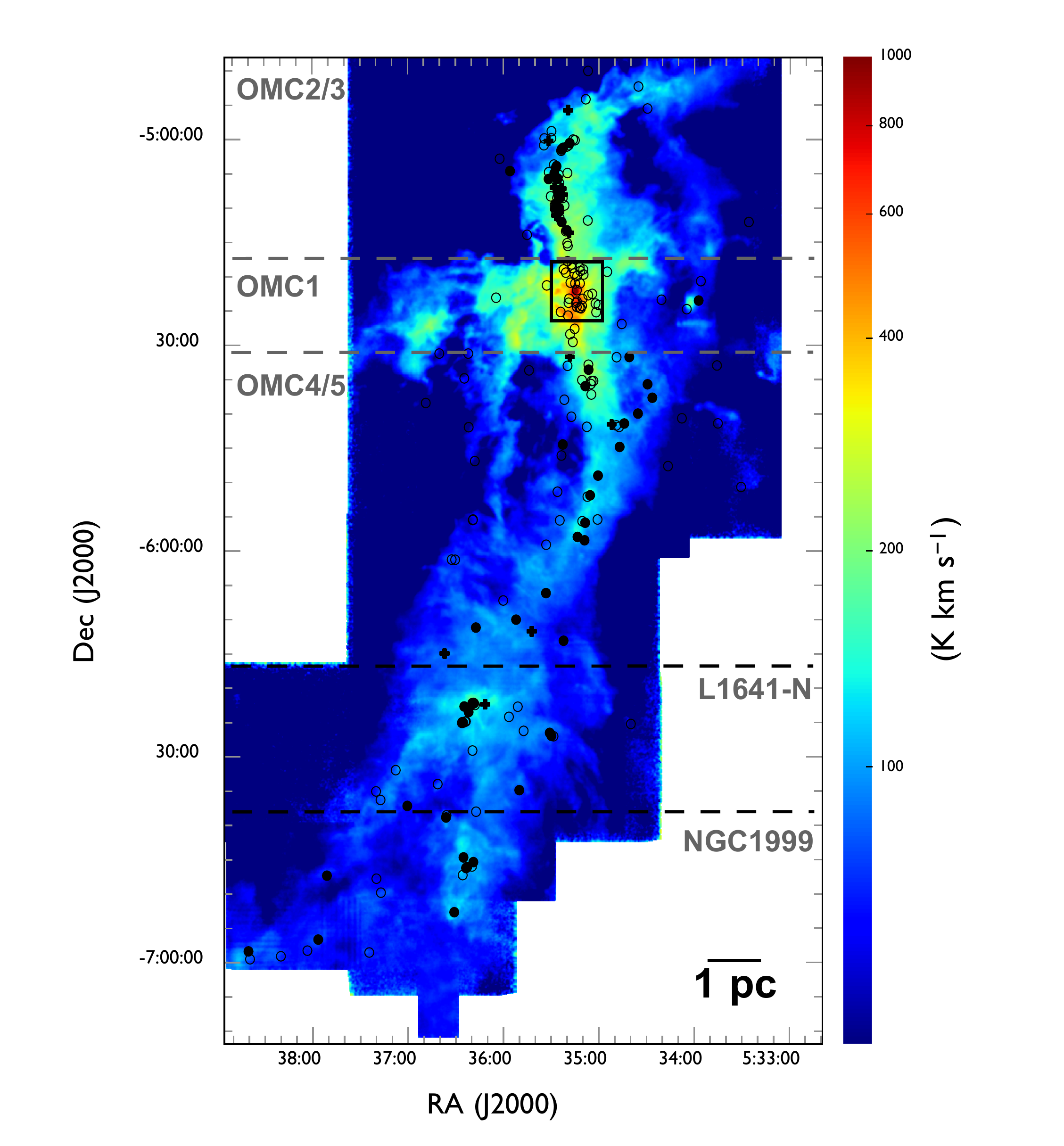}
\caption{Outflow driving source candidates superposed on the ${}^{12}$CO\ ({\it J} = 1--0) total integrated intensity map of Orion A.
The integrated velocity ranges from 2.0 km s$^{-1}$ to 20.0 km s$^{-1}$ in $v_{\rm LSR}$.
 The open circles, filled circles, and filled crosses indicate the positions of the protostars without H$_2$ jets, protostars with H$_2$ jets, and pre-main-sequence stars with H$_2$ jets, respectively (Megeath et al. \yearcite{2012AJ....144..192M} and Davis et al. \yearcite{2009A&A...496..153D}).
 The black square represents the center of OMC 1.}	
	\label{12}
\end{figure}

\begin{figure}[h]
 \includegraphics[keepaspectratio, width=16cm]{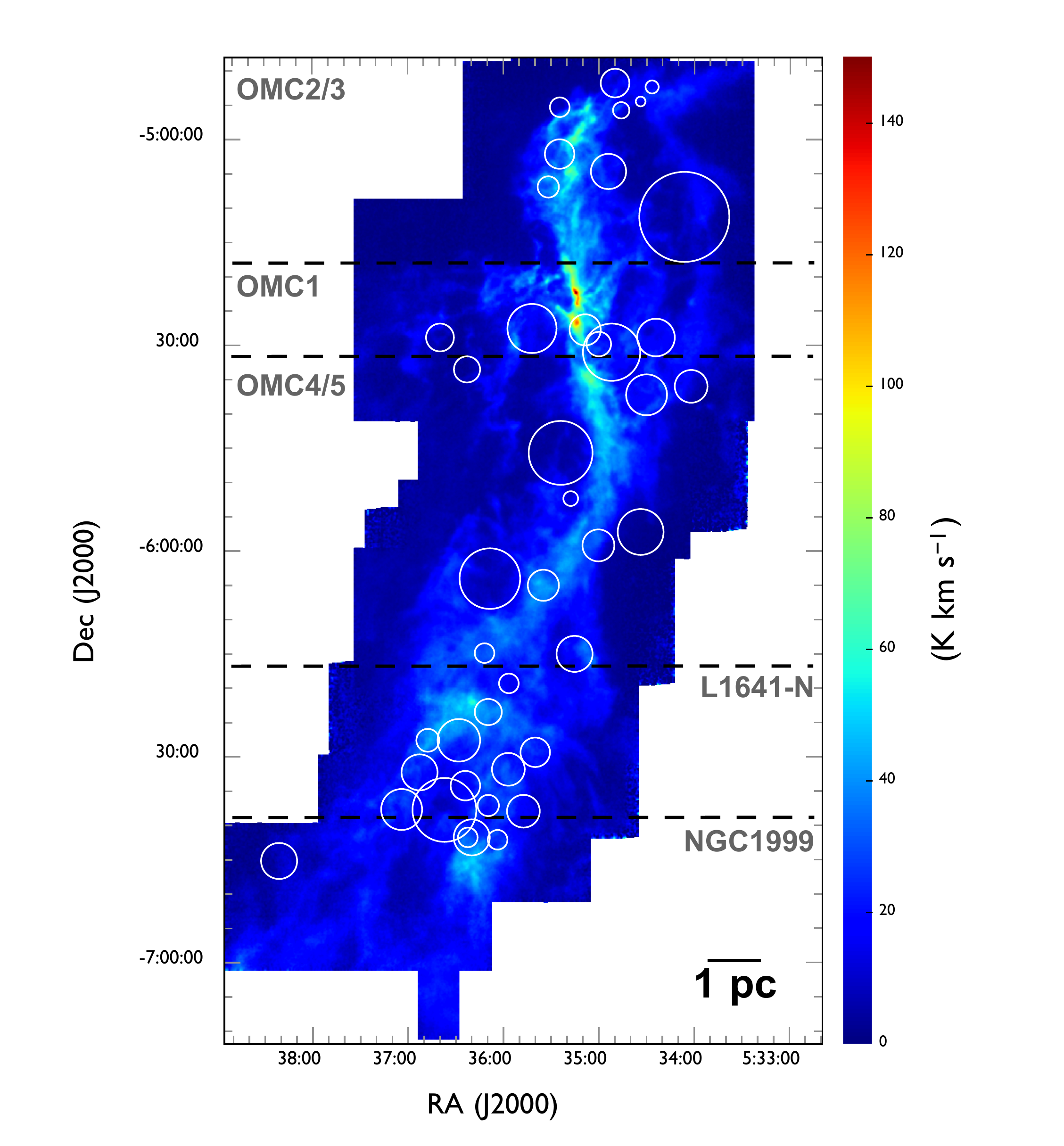}
 	\caption{
	${}^{13}$CO total integrated intensity map of Orion A.
	The integrated velocity ranges from 2.0 km s$^{-1}$ to 20.0 km s$^{-1}$ in $v_{\rm LSR}$.
	The white circles indicate molecular shells identified by Feddersen et al. (\yearcite{2018ApJ...862..121F}).
	The size of each circle reflects the shell size.
	}
	\label{13}
\end{figure}

\begin{figure}[h]
 \includegraphics[keepaspectratio, width=16cm]{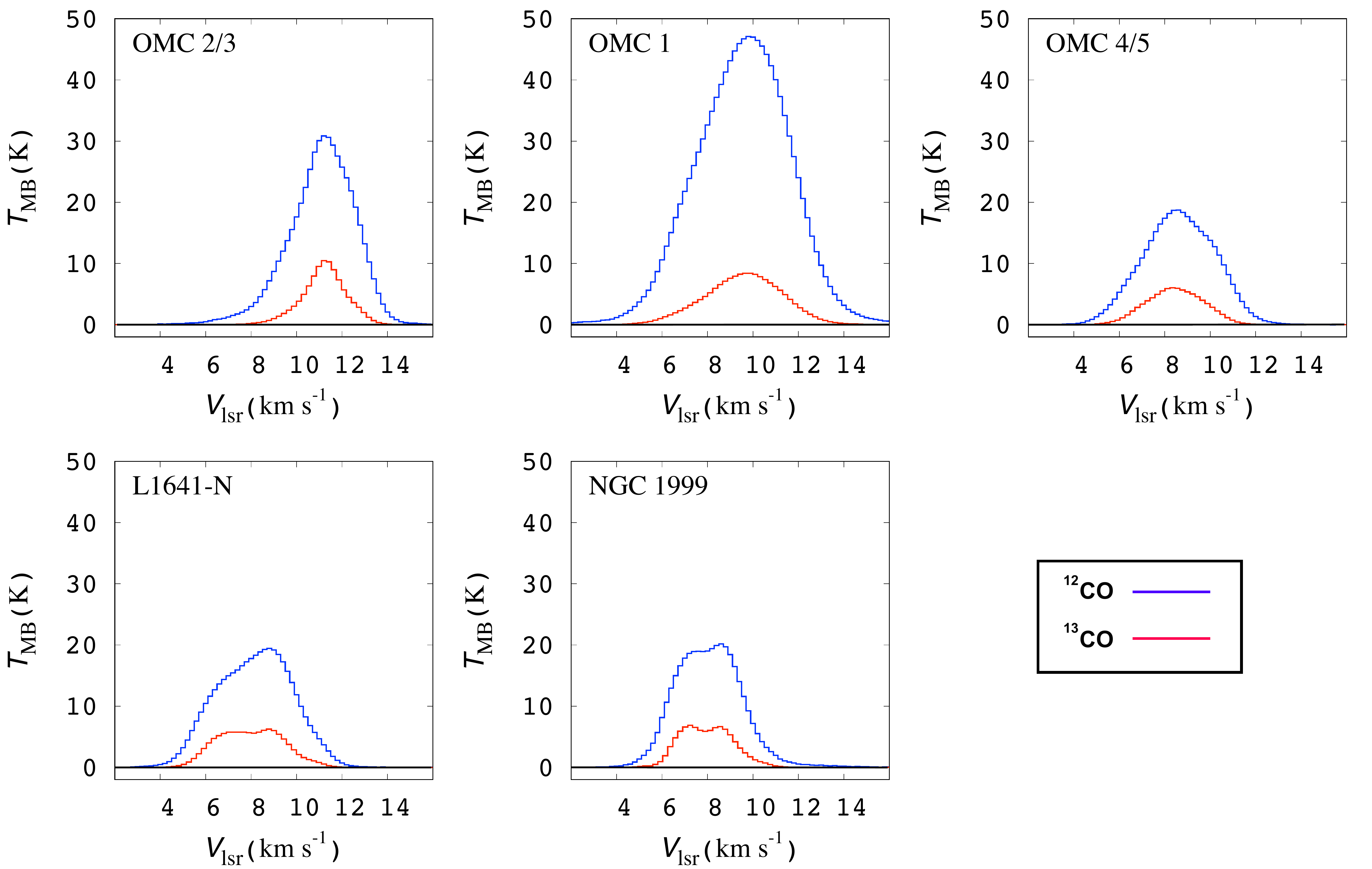}
 	\caption{Averaged spectra of ${}^{12}$CO\ ({\it J} = 1--0) (blue) and ${}^{13}$CO\ ({\it J} = 1--0) (red) in each subregion. All spectra are presented for emissions detected above 5$\sigma$.} 
	\label{sp}
\end{figure}

Figures \ref{12} and \ref{13} show the total integrated intensity maps of the ${}^{12}$CO\ ({\it J} = 1--0) and ${}^{13}$CO ({\it J} = 1--0) emissions of Orion A, respectively. 
An integral-shaped filament is clearly seen for both ${}^{12}$CO and ${}^{13}$CO.
The averaged line profiles of the ${}^{12}$CO and ${}^{13}$CO in each subregion are shown in figure \ref{sp}. 
The peak velocities of the ${}^{12}$CO and ${}^{13}$CO spectra in figure \ref{sp} shift from 11 km s$^{-1}$ to 8 km s$^{-1}$ in the north to south direction.
The averaged velocity widths at OMC 2/3, OMC 1, OMC 4/5, L1641-N, and NGC 1999 are 3.1, 4.7, 3.8, 4.0, and 3.3 km s$^{-1}$ for $^{12}$CO and 2.1, 4.0, 3.1, 3.5, and 2.8 km s$^{-1}$ for $^{13}$CO, respectively.
These values are obtained from Gaussian fitting to each spectrum, as shown in figure \ref{sp}.
More details on the data including the first and second moment maps and channel maps of the ${}^{12}$CO\ ({\it J} = 1--0) and ${}^{13}$CO\ ({\it J} = 1--0) are provided by Kong et al. (\yearcite{2018ApJS..236...25K}) and Nakamura et~al. in (prep.).

The black square shown in figure \ref{12} is the center of OMC 1 and indicates the area that is excluded from the following analysis because the identification of the outflows there is difficult with our method (see section \ref{prosec}).
There are two reasons why we excluded this area from our procedure.
1) YSOs in this region are so crowded that we cannot judge which YSO is responsible for any particular high-velocity component or is a candidate for outflows.
2) The local velocity width of the molecular clouds around each YSO is so broad that we cannot separate the outflow components from the main cloud components.

\section{Method for identifying molecular outflows}\label{prosec}
To investigate the outflow feedback into the parent cloud, 
we identify the CO outflows and estimate their physical parameters based on the data cube of the ${}^{12}$CO\ ({\it J} = 1--0) emission line. 
In this section, we describe our outflow search procedure.

\subsection{Outflow driving source candidates}
To identify the molecular outflows in the vicinity of  YSOs, we summarize a set of YSO candidates in our observed region.
We made a sample of candidate outflow driving sources from the {\it Spitzer} YSO catalog produced by Megeath et~al.\ (\yearcite{2012AJ....144..192M}). 
The YSOs were divided into two categories based on their infrared photometry: protostars and pre-main-sequence (PMS) stars with a circumstellar disk. 
We selected protostars as candidate driving sources of outflows because outflows tend to appear in an early phase of star formation. 
In our observed region, Megeath et~al.\ (\yearcite{2012AJ....144..192M}) identified 198 protostars,
 including 32 protostars in the center of OMC 1.

We also supplemented the candidates with sources that appear to drive H$_2$ jets. 
Davis et~al.\ (\yearcite{2009A&A...496..153D}) conducted an unbiased survey of the molecular hydrogen emission line of $v$ = 1--0 S(1) at  2.12 $\rm \mu$m originating from shocks in outflows (see also Stanke et~al.\ \yearcite{2002A&A...392..239S}).  
Davis et~al.\ (\yearcite{2009A&A...496..153D}) also identified driving sources of H$_2$ jets from the {\it Spitzer} YSO catalog based on the jets' morphologies and/or alignments, arguing that 65 YSOs were associated with H$_2$ jets in our observed region. 
Among the 65 YSOs, 53 were categorized as protostars and 12 were categorized as PMS star.

We eventually focused on 178 candidates outflow driving sources selected from Megeath's catalog, 65 of which Davis described as YSOs associated with jets.
Table \ref{count} summarizes the number of candidates in each category, and figure \ref{12} shows all candidates driving sources of the outflows superposed on the ${}^{12}$CO\ ({\it J} = 1--0) integrated intensity map.

\begin{table}[h!]
	\caption{Categorization of candidate outflow driving sources}\renewcommand{\arraystretch}{0.8}
		\begin{tabular}{cccccc}\hline
		&&\multicolumn{2}{c}{Categories}\\
		&& Protostar&PMS star &Total\\\hline
		\multirow{2}{*}{H$_2$ jet}&Yes&53&12&65\\
		&No&113&0&113\\
		&Total&166&12&178\\\hline
		\end{tabular}\label{count} 
\end{table}

\subsection{Search procedure\label{pro}} 
We searched for ${}^{12}$CO molecular outflows around candidates of driving sources using the following procedure.

First, we obtained a ${}^{12}$CO spectrum averaged over a circle with a radius of \timeform{30"} ($\sim$0.06 pc) centered on each candidate and applied least square fitting using a Gaussian function described by
\begin{equation}
f(v_{\rm lsr})=T_{\rm peak}\exp{[-(v_{\rm lsr}-v_{\rm sys})^2/2\sigma_v^2]}. 
\label{gauss}
\end{equation}
We obtained three parameters from the fit assuming equal weights: $T_{\rm peak}$ in K, $v_{\rm sys}$ in km s$^{-1}$, and standard deviation $\sigma_v$ in\ km s$^{-1}$.

Next, we produced the integrated intensity maps (10$\arcmin$ $\times$ 10$\arcmin$) of the blue-shifted emission ($v_{\rm sys} -v_{\rm lsr} \geq 2\ \sigma_v$) and red-shifted emission ($v_{\rm lsr} -v_{\rm sys} \geq 2\ \sigma_v$).
We defined emissions above 5$\sigma$ with its spatial extents larger than the beam size in these maps as high velocity blue- and red-shifted emissions associated with each YSO candidate. 
We measured the position angle (PA) of the $^{12}$CO peak of a high velocity emission 
from the candidate YSO.
Note that the difference between the blue and red axes is small ($\sim$5$^{\circ}$) for most outflows, thus we list only one value as the PA of each outflow.

 Lastly, we compared the position angle of the high-velocity emission to those of  the H$_2$ jets identified by Davis et~al.\ (\yearcite{2009A&A...496..153D}).
We  identified a blue- or red-shifted emission as an outflow associated with the candidate only when the position angles of the emission and H$_2$ jet exist within $\pm 20^{\circ}$ of each other.
For the high-velocity emissions not associated with H$_2$ jets, we simply regard them as outflows.

The idea behind the above procedure is that the velocity profile of the ambient clouds can be expressed by a single Gaussian while the outflows can be identified as the additional high velocity components.
Therefore, in the region where the clouds are composed of more than two velocity components with significantly different peak intensities, this procedure does not work well. 
These regions are mostly in the OMC 1 region, and they are beyond the scope of this study.

\begin{figure}[h!]
 \includegraphics[keepaspectratio,width=16cm]{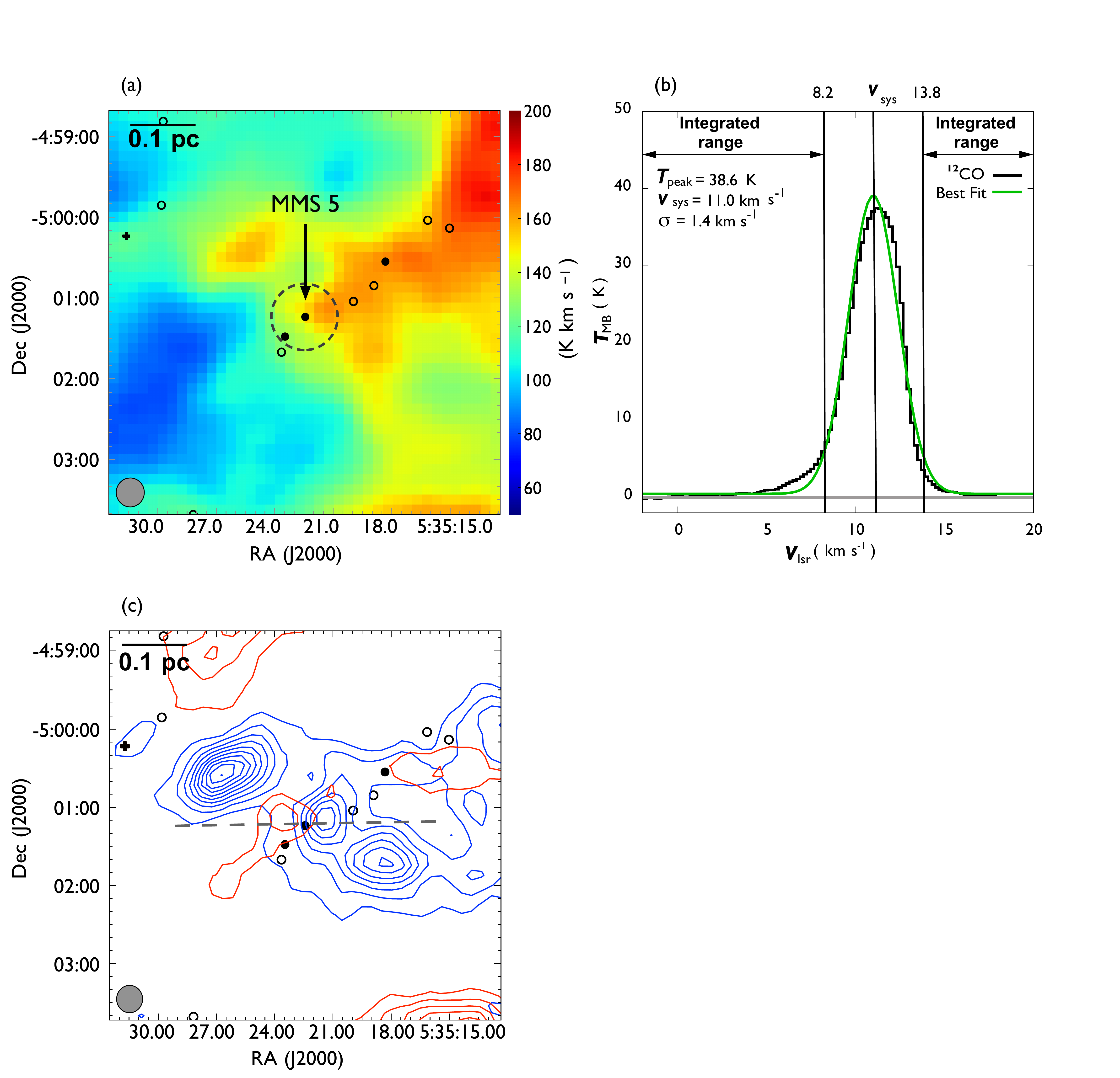}
	\caption{Results of our outflow search procedure for MMS 5.
 	(a) Outflow driving source candidates overlaid on the ${}^{12}$CO integrated intensity map. 
	(b) Average spectrum of the ${}^{12}$CO emission over the \timeform{30"} radius area around the center of the left panel and the result of Gaussian fitting. 
	 (c) ${}^{12}$CO integrated intensity maps for the distributions of blue- and red-shifted components. 
	In panels (a) and (c), the filled and open circles represent the same as in figure \ref{12}, and the effective angular resolution is indicated in the bottom-left. 
         The area where the ${}^{12}$CO spectrum is averaged in panel (b) is indicated by the dashed circle in panel (a).  
	In panel (b), three parameters for the fitting are indicated on the left, and the integrated velocity range for the blue- and red-shifted components is indicated by arrows.
	In panel (c), the blue and red contours show the distributions of the blue- and red-shifted components integrated from $v_{\rm LSR}$ = -1.2 km s$^{-1}$ to 8.2 km s$^{-1}$ and from 13.8 km s$^{-1}$ to 20.2 km s$^{-1}$, respectively. 
	The contour intervals are 3 K km s$^{-1}$ ($\sim$5$\sigma)$ starting at 3 K km s$^{-1}$. 
	The gray dashed line represents the axis of the outflow.
	In this example, we focus on the one candidate located at the center of panels (a) and (c), but we apply the same procedure independently for all  other candidates.
	For example, the peaks located to the north and south of MMS 5 in panel (c) are regarded as outflows associated with different sources (see figures \ref{3b} and \ref{5b}).}
	\label{ex}
\end{figure}

Figure \ref{ex} shows an example of the outflow search using the above procedure. 
Figure \ref{ex} (a) shows the ${}^{12}$CO integrated intensity map centered on the protostar MMS 5 ($\alpha_{\rm J2000}$,  $\delta_{\rm J2000}$) = ($\timeform{05h35m22s.43}, -$\timeform{5D01'14".1}), and figure \ref{ex} (b) shows the spectrum averaged over the \timeform{30"} radius area around MMS 5 and the results of the Gaussian fitting. 
Figure \ref{ex} (c) shows the integrated intensity maps of blue- and red-shifted emission.
The distributions of the blue- and red-shifted components are relatively similar to those of the MMS 5 outflow identified in previous studies (Aso et al. \yearcite{2000ApJS..131..465A}, Williams et al. \yearcite{2003ApJ...591.1025W}, and Takahashi et al. \yearcite{2008ApJ...688..344T}).

\section{Results}\label{res}

\subsection{Identified outflows}
\subsubsection{Overview of identified outflows}

\begin{figure}[h!]
 \includegraphics[keepaspectratio, width=16cm]{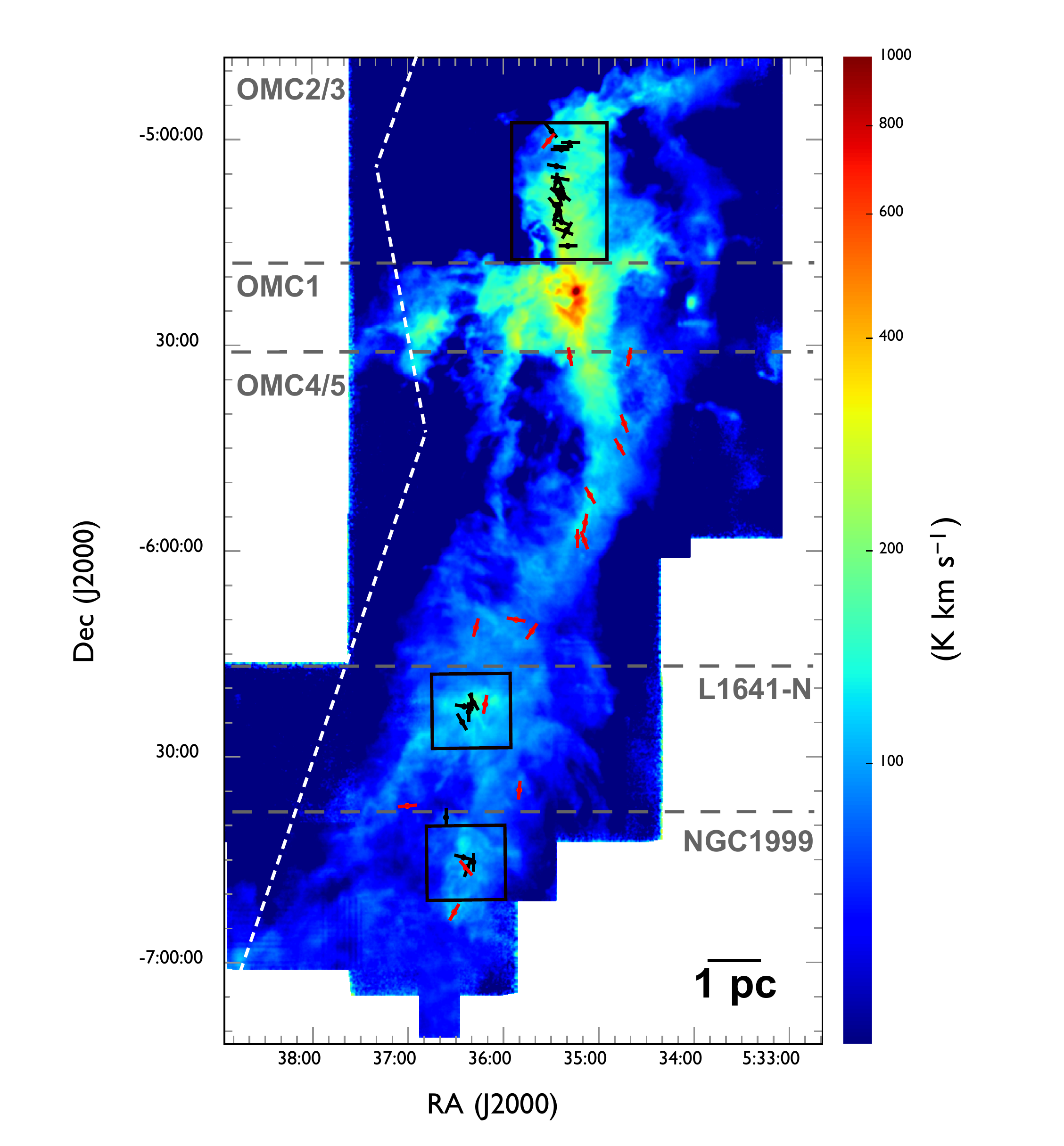}
	\caption{Locations of the detected outflows overlaid on the ${}^{12}$CO\ ({\it J} = 1--0) total integrated intensity map which is the same as in figure \ref{12}. 
	The red and black filled circles indicate the positions of newly detected and known outflow driving sources. 	
	Each solid lines indicates the position angles of an outflow.
	White dashed lines represent the directions of filaments in the clouds considered in this study (see section \ref{PA}).
	The black squares indicate OMC 2/3, L1641-N and NGC1999, for which close-up views are presented in figure \ref{up}.}
	\label{out}
\end{figure}

\begin{figure}[h!]
 \includegraphics[keepaspectratio, width=16cm]{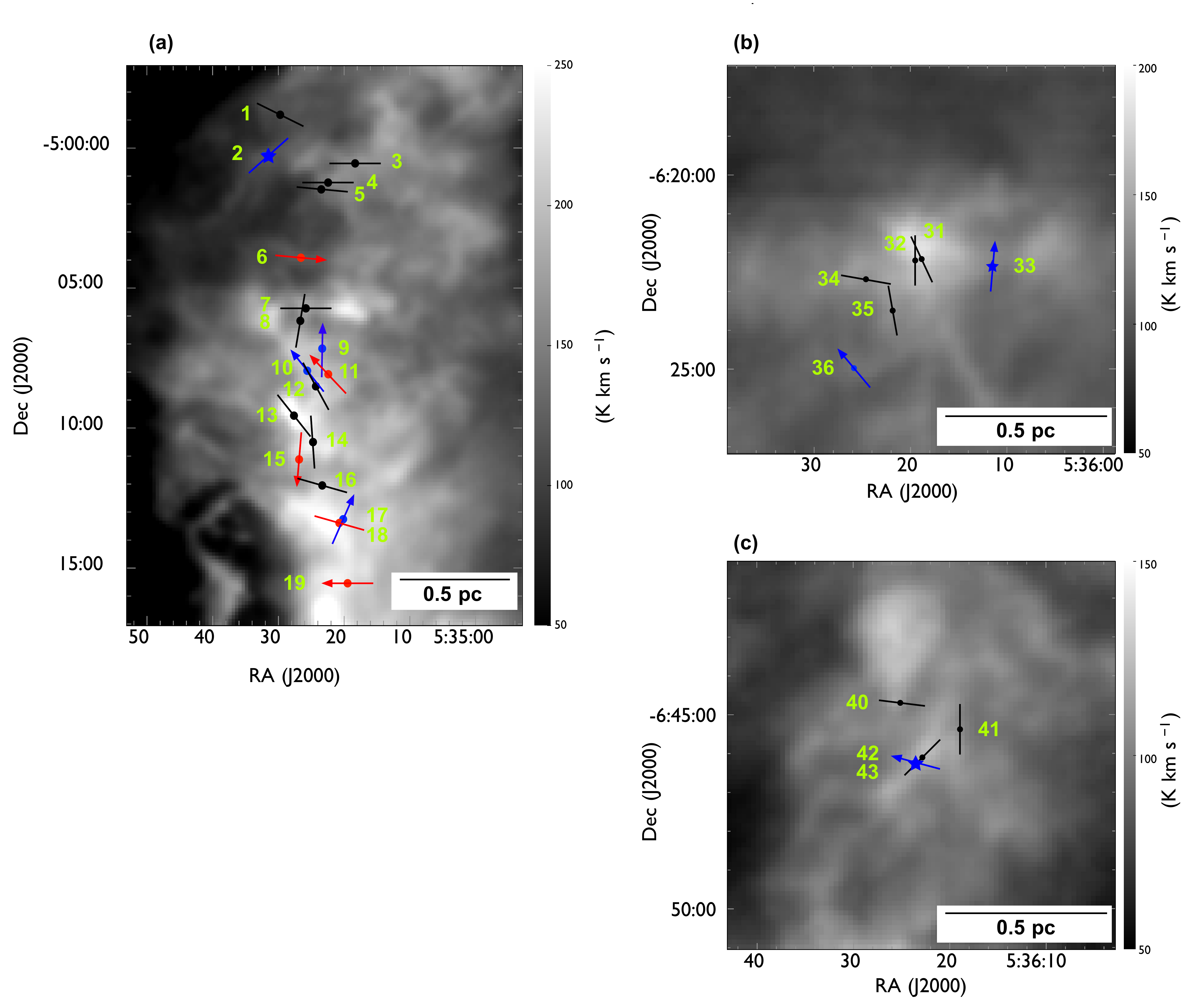}
	\caption{Close-up views of the detected outflows in (a) OMC 2/3, (b) L1641-N, and (c) NGC 1999 .
	Filled circles and star symbols indicate the positions of known and newly detected outflow driving sources, respectively.
	Each solid line or arrow indicates the position angles of an outflow.
	The black, blue, and red colors represent blue- and red-shifted pairs of, single blue-shifted, and single red-shifted outflows, respectively.
	The direction of teach arrow indicates the direction of a single outflow lobes.
	The background grayscale image is a total integrated intensity of ${}^{12}$CO\ ({\it J} = 1--0).}
		\label{up}
\end{figure}

Following the procedure described in section\ \ref{pro}, we identified 44 CO outflows in the ${}^{12}$CO\ ({\it J} = 1--0) line, of which 17 are newly detected. 
To distinguish them from those detected in the ${}^{13}$CO\ ({\it J} = 1--0) line, hereafter we refer to them as ${}^{12}$CO outflows.
Figures \ref{out} and \ref{up} show the locations of the detected outflows and their position angles, and table \ref{index} lists all the outflows.
25 outflows consist of a pair of blue- and red-shifted lobes, and 19 outflows have a single lobe (13 blue-shifted, and 6 red-shifted).
The integrated images and position-velocity (P-V) diagrams of the newly detected outflows are shown in figures \ref{1}-\ref{17}, and those of previously known outflows are shown in the Appendix. 
All newly detected outflows are associated with H$_2$ jets.

\subsubsection{Detection rate of outflows}
Table \ref{counta} lists the number and detection rate of outflows for the driving sources in each category.  
The detection rate of molecular outflows for protostars with H$_2$ jets is 62\%, which is $\sim$15 times higher than that for protostars without  H$_2$ jets (4\%).

In table \ref{counta}, of the 65 YSOs with H$_2$ jets, 40 (62\%) are also associated with outflows, and of the 44 YSOs associated with outflows, only 4 (9\%) are not accompanied by H$_2$ jets.
These correlations imply that the molecular outflows and H$_2$ jets occur in almost the same phase of star formation.

Table \ref{countb} lists the number and detection rate of outflows in each subregion. 
The detection rates of outflows among the subregions are consistent with each other within uncertainties, except for in OMC 1.
Note that we exclude the 32 YSOs distributed in the center of OMC 1 (see section \ref{obs} and figure \ref{12}), and the detection rate of the outflows in OMC 1 is 0 \% to represent an incomplete sample.

\begin{table}[h!]
	\caption{Outflow detection rate sorted by driving source category}
		\begin{tabular}{lccr}\hline
		Category&Number of outflows &Number of candidates&Detection rate\\\hline
		Protostar without jet&4&113&4\%\\
		Protostar with jet&33&53&62\%\\
		PMS star with jet&7&12&58\%\\\hline
		Total&44&178&25\%\\\hline
		\end{tabular}\label{counta} 
\end{table}

\begin{table}[h!]
	\caption{Outflow detection rate in each subregion}
		\begin{tabular}{lccr}\hline
    Subregion& Number of outflows &Number of candidates&Detection rate\footnotemark[*]\\\hline
	OMC 2/3&19&62&31 $\ \pm\ $7\%\\
    	OMC 4/5&  11&59&19 $\ \pm\ $6\%\\
	L1641-N&  8 &27&30 $\ \pm\ $10\%\\
	NGC 1999&6 &19&32 $\ \pm\ $13\%\\\hline
	OMC 1\footnotemark[\dag] &0&11&0\%\\\hline
		\end{tabular} \label{countb} 
			\begin{tabnote}
			\footnotemark[*]The uncertainties $\surd(N)$, where $N$ is the number of detected outflows.\\
			\footnotemark[\dag]Excluding the center of OMC 1.
\end{tabnote}			
\end{table}

\clearpage
\begin{table}[h!]
 \caption{List of outflows}
  \begingroup
\renewcommand{\arraystretch}{0.8}
   {\footnotesize\scalebox{0.95}{
     \begin{tabular}{lccccccccl} \hline
No.& $\alpha$\footnotemark[*]& $\delta$ \footnotemark[*]&Driving source&Common name&Velocity&PA\footnotemark[$\S$] &New&Reference\\
  &(J2000.0)&(J2000.0)&category\footnotemark[$\dag$]&of driving source&shift\footnotemark[$\ddag$]&&detection&\\\hline
\multicolumn{9}{c}{OMC 2/3}\\\hline
1&05 35 29.72&-04 58 48.8&P&&BR&65&-&1\\
2&05 35 31.62&-05 00 14.0&PMSJ&&B&130&Yes&-\\
3&05 35 18.32&-05 00 33.0&PJ&MMS 2&BR&90&-&1,2,3,4\\
4　&05 35 22.43&-05 01 14.1&PJ&MMS 5&BR&90&-&1,2,3,4\\
5　&05 35 23.47&-05 01 28.7&PJ&MMS 6&BR&85&-&1,3\\
6　&05 35 26.57&-05 03 55.1&PJ&MMS 7&BR&90&-&1,2,3,4\\
7　&05 35 25.82&-05 05 43.6&PJ&MMS 9&BR&90&-&1,2,3,4\\
8　&05 35 26.66&-05 06 10.3&P&HOPS 75&BR&170&-&1,2,3\\
9　&05 35 23.33&-05 07 09.8&PMSJ&MMS 11&B&160&-&1,4\\
10　&05 35 25.61&-05 07 57.3&PJ&MMS 14&B&40&-&1,2,4\\
11　&05 35 22.41&-05 08 04.8&PMSJ&MMS 15&R&45&-&1,2,4\\
12　&05 35 24.30&-05 08 30.6&PJ&MMS 103&BR&30&-&1,2,4\\
13　&05 35 27.63&-05 09 33.5&PJ&FIR\ 3&BR&40&-&1,2,4,5\\
14　&05 35 24.73&-05 10 30.2&PJ&VLA13&BR&5&-&1,5\\
15　&05 35 26.85&-05 11 07.3&PMSJ&&R&175&-&1\\
16　&05 35 23.33&-05 12 03.1&PJ&FIR\ 6\ b&BR&75&-&1,6\\
17　&05 35 20.14&-05 13 15.5&PJ&FIR\ 6\ d&B&155&-&6\\
18　&05 35 20.73&-05 13 23.6&P&FIR\ 6\ c&R&75&-&1,6\\
19　&05 35 19.47&-05 15 32.7&P&CSO\ 33&R&90&-&1\\\hline
\multicolumn{9}{c}{OMC 4/5}\\\hline
20　&05 35 18.28&-05 31 42.1&PMSJ&&B&15&Yes&-\\
21　&05 34 40.91&-05 31 44.4&PJ&&BR&145&Yes&-\\
22　&05 34 44.06&-05 41 25.9&PJ&&BR&40&Yes&-\\
23　&05 34 46.94&-05 44 50.9&PJ&&B&30&Yes&-\\
24　&05 35 05.49&-05 51 54.4&PJ&&B&30&Yes&-\\
25　&05 35 08.60&-05 55 54.3&PJ&&B&160&Yes&-\\
26　&05 35 13.41&-05 57 58.1&PJ&Ori\ 2-6&BR&0&Yes&-\\
27　&05 35 09.00&-05 58 27.6&PJ&&BR&25&Yes&-\\
28　&05 35 52.00&-06 10 01.8&PJ&&B&60&Yes&-\\
29　&05 36 17.26&-06 11 11.0&PJ&&BR&165&Yes&-\\
30　&05 35 42.18&-06 11 43.8&PMSJ&&R&135&Yes&-\\\hline
\multicolumn{9}{c}{L1641-N}\\\hline
31　&05 36 18.83&-06 22 10.2&PJ&&BR&25&-&7,8\\
32　&05 36 19.50&-06 22 12.3&PJ&&BR&0&-&7,8\\
33　&05 36 11.45&-06 22 22.1&PMSJ&&B&175&Yes&7,8\\
34　&05 36 24.61&-06 22 41.3&PJ&&BR&80&-&7,8\\
35　&05 36 21.84&-06 23 29.8&PJ&&BR&10&-&7,8\\
36　&05 36 25.86&-06 24 58.7&PJ&&B&40&-&7,8\\
37　&05 35 50.02&-06 34 53.4&PJ&&R&175&Yes&-\\
38　&05 37 00.45&-06 37 10.5&P&&B&95&Yes&-\\\hline
\multicolumn{9}{c}{NGC 1999}\\\hline
39　&05 36 36.10&-06 38 52.0&PJ&V380 Ori-NE&BR&0&-&9,10\\
40　&05 36 25.13&-06 44 41.8&PJ&IRS 63 (HH 147)&BR&80&-&11\\
41　&05 36 18.93&-06 45 22.7&PJ&VLA 3, HOPS 168&BR&0&-&11\\
42　&05 36 22.84&-06 46 06.2&PJ&VLA 1/2 (HH 1/2)&BR&140&-&11\\
43　&05 36 23.54&-06 46 14.5&PJ&IRS 121 (HH36)&B&75&Yes&11\\
44　&05 36 30.97&-06 52 40.9&PJ&&BR&150&Yes&-\\\hline
     \end{tabular}\label{index}}}
\begin{tabnote}
\footnotemark[*] Coordinates of the outflow driving source.\\
\footnotemark[$\dag$] Category of the outflow driving sources. "P," "PJ," and "PMSJ" represent protostar without H$_2$ jet, protostar with H$_2$ jet and PMS star with H$_2$ jet,  respectively.\\
\footnotemark[$\ddag$] The velocity shifts of outflows. "BR," "B," and "R" represent a blue and red-shifted pair of, single blue-shifted, and single red-shifted outflow, respectively.   \\
\footnotemark[$\S$] Outflow position angle on the plane of the sky (5$^{\circ}$ bin). \\
References-- 1 Takahashi et al. (\yearcite{2008ApJ...688..344T}); 2 Williams et al. (\yearcite{2003ApJ...591.1025W}); 3 Aso et al. (\yearcite{2000ApJS..131..465A});  4 Chini et al. \yearcite{1997ApJ...474L.135C}; 5 Shimaziri et al. (\yearcite{2008ApJ...683..255S}); 6 Shimajiri et al. (\yearcite{2009PASJ...61.1055S}); 7 Nakamura et al. (\yearcite{2012ApJ...746...25N}); 8 Stanke \& Williams (\yearcite{2007AJ....133.1307S}); 9 Davis et al. (\yearcite{2000MNRAS.318..952D}); 10 Choi et al. (\yearcite{2017ApJS..232...24C}); 11 Moro-Mart{\'{\i}}n et al. (\yearcite{1999ApJ...520L.111M}). \\
\end{tabnote}
\endgroup
     \end{table}
\clearpage

\begin{figure}[h]
 \includegraphics[keepaspectratio, width=16cm]{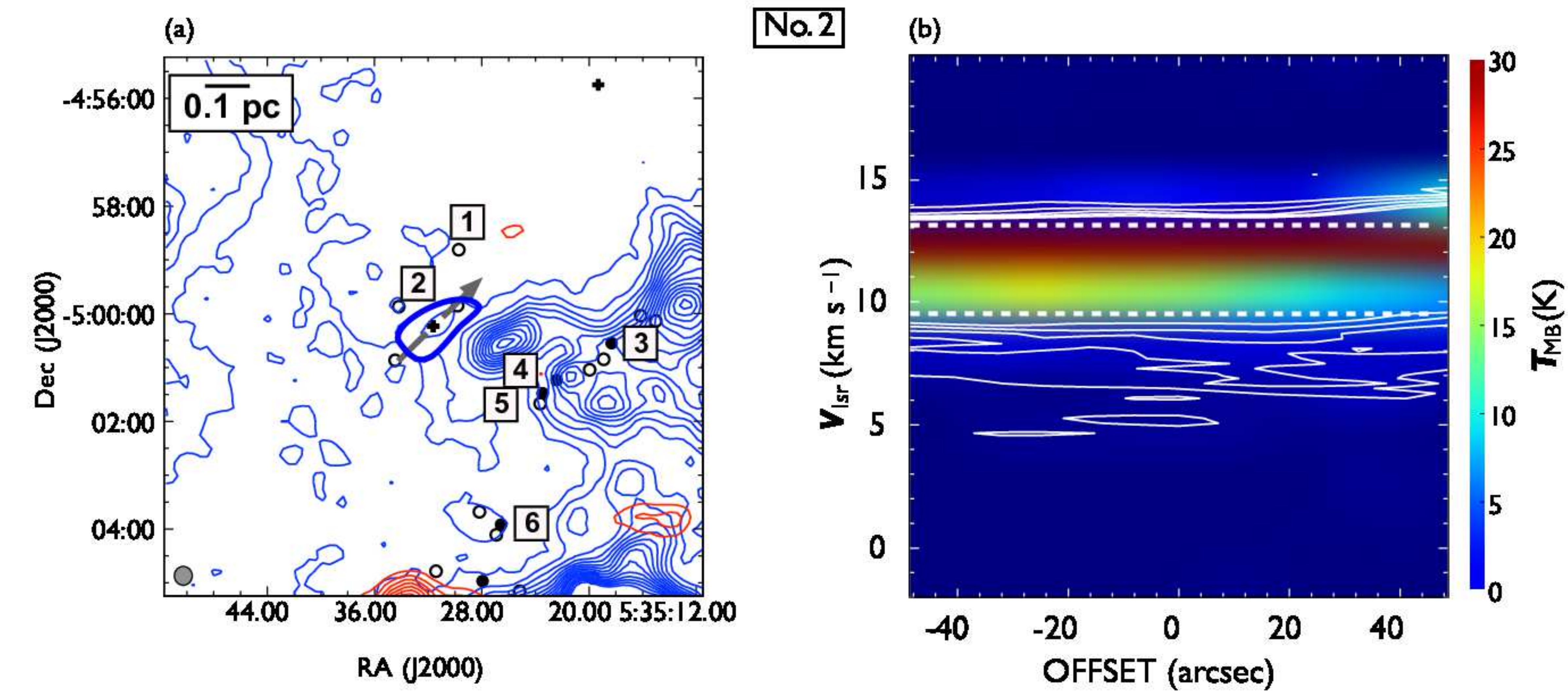}
 	\caption
	{
	(a) The distribution of outflow No. 2, driven by a PMS star with a jet located in the northern part of OMC 3, and (b) the P--V diagram along the outflow axis.
	Contour levels and symbols shown in panel (a) are the same as those in figure \ref{ex} (d), and the thick contour represents the projected area where we estimated the mass of the outflow. 
	In panel (a), the integrated velocity ranges of blue- and red-shifted components are 2.0 km s$^{-1}$ to 9.5 km s$^{-1}$ and 13.8 km s$^{-1}$ to 20.2 km s$^{-1}$, respectively.
	The upper (lower) limit for the blue and red ranges are indicated by the dashed lines in panel (b).
	The gray dashed arrow in panel (a) represents the cut along which the P--V diagram in panel (b) was obtained and also represents the position angle of the outflow.
	The direction of the arrow indicates the positive offset direction in the P--V diagram.
	The effective angular resolution of the ${}^{12}$CO data is shown in the lower left corner.
 	In panel (b), the contour intervals are 1, 2, 3, 4, and 5 K in $T_{\rm MB}$.
	While the blue-shifted lobe can be seen with a north-west to south-east elongation, the red-shifted component is not detected.
	 }
	\label{1}
\end{figure}
\begin{figure}[h!]
 \includegraphics[keepaspectratio, width=16cm]{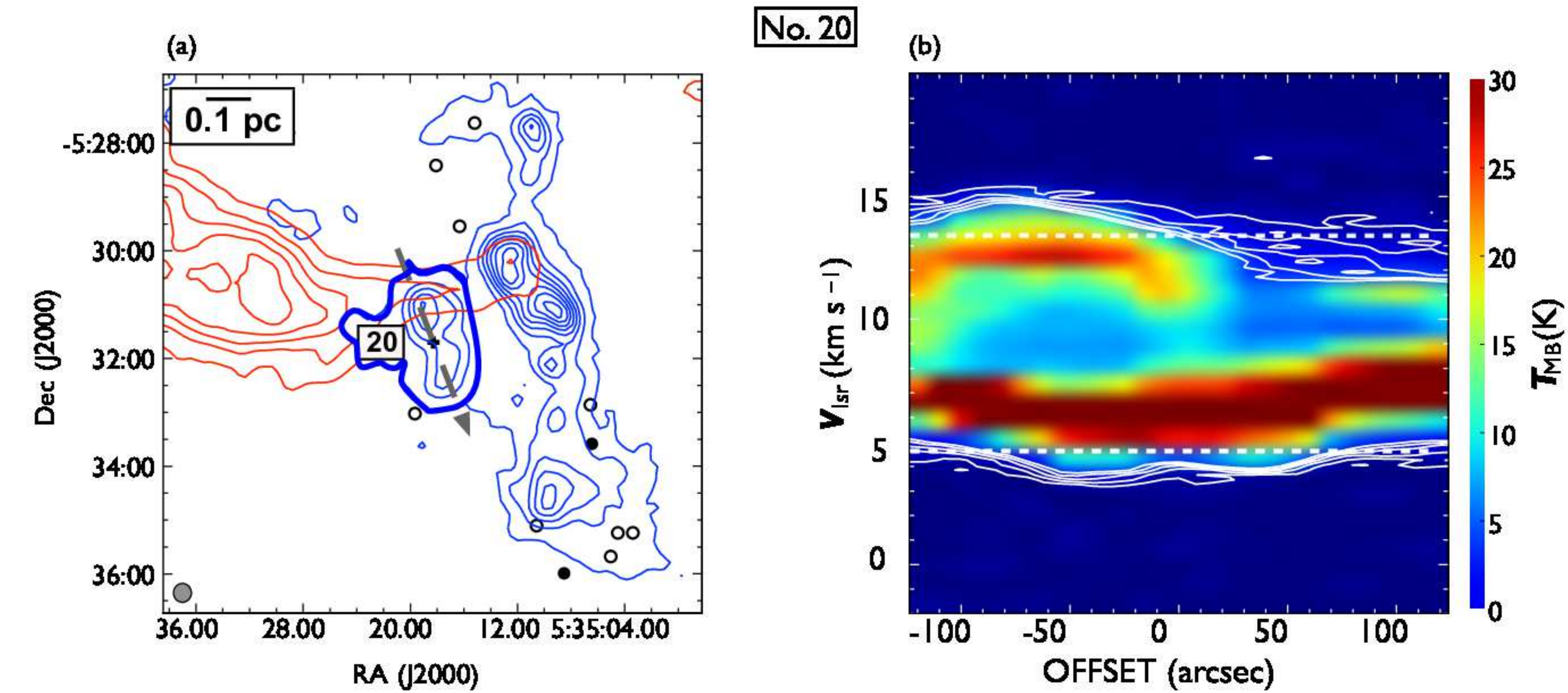}
 	\caption
	{
	The same as in figure \ref{1} but for outflow No. 20. In panel (a), the blue- and red-shifted integrated intensity velocity ranges are -1.9 km s$^{-1}$ to 4.8 km s$^{-1}$ and 13.1 km s$^{-1}$ to 20.2 km s$^{-1}$, respectively.
	This outflow, located at $\sim$\timeform{10'} south of the center of OMC 1, has only a blue-shifted lobe and is driven by a PMS star with a jet.
	}
	\label{2}
\end{figure}

\begin{figure}[h]
 \includegraphics[keepaspectratio, width=16cm]{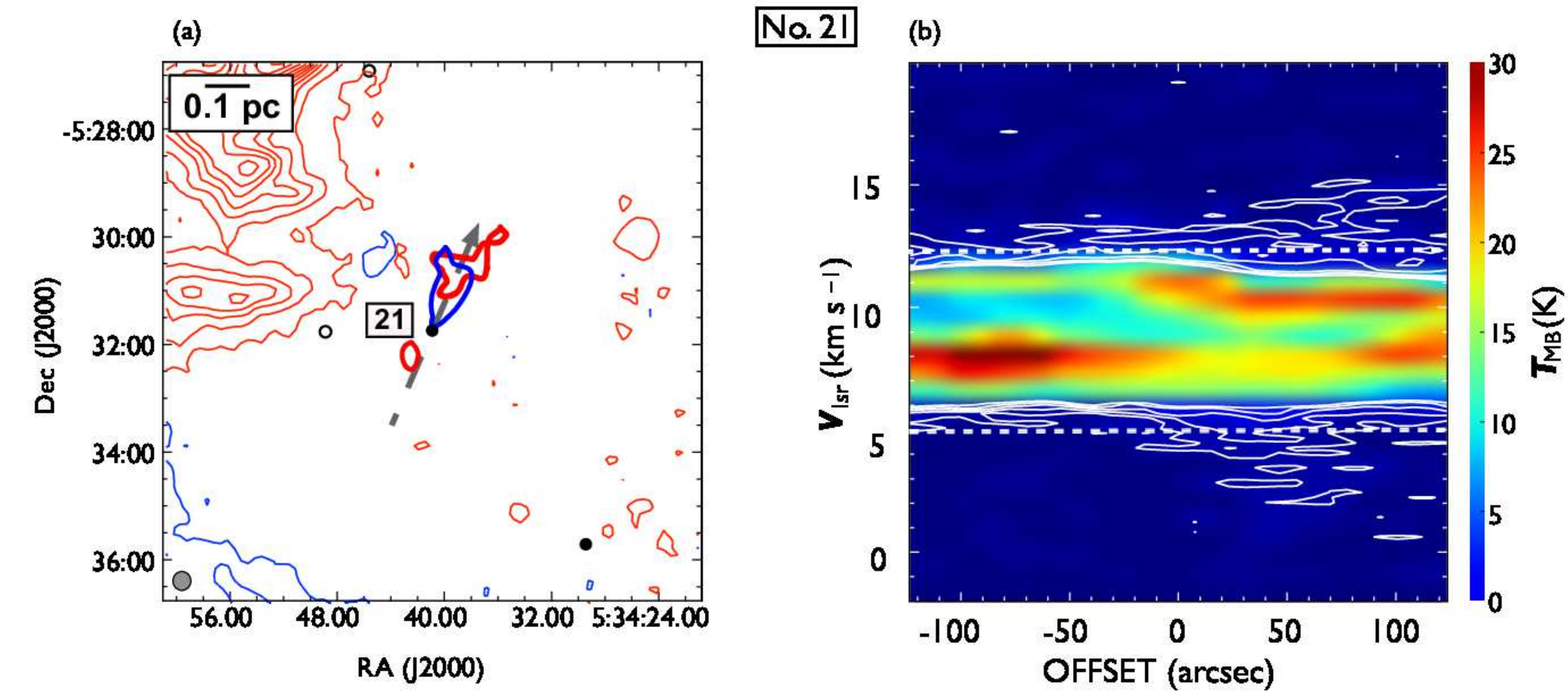}
 	\caption
	{
	The same as in figure \ref{1} but for outflow No. 21. In panel (a), the blue- and red-shifted integrated intensity velocity ranges are -1.9 km s$^{-1}$ to 4.8 km s$^{-1}$ and 12.1 km s$^{-1}$ to 20.2 km s$^{-1}$, respectively.
	It can be seen that the the blue-shifted lobe and red-shifted lobe have a north-west elongation; the red-shifted lobe also has a south-east elongation. 
	}
	\label{3}
\end{figure}

\begin{figure}[h]
 \includegraphics[keepaspectratio, width=16cm]{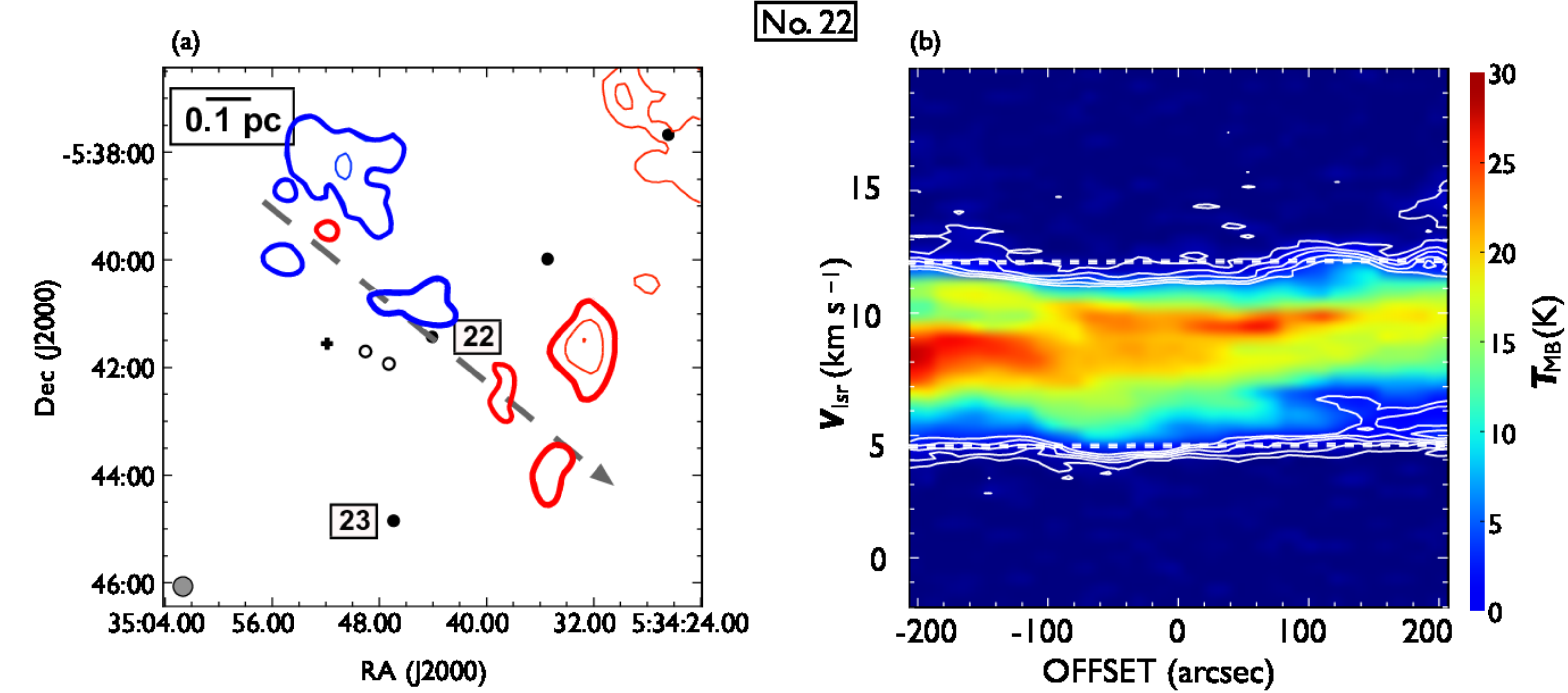}
 	\caption
	{
	The same as in figure \ref{1} but for outflow No. 22. In panel (a), the blue- and red-shifted integrated intensity velocity ranges are -1.9 km s$^{-1}$ to 4.4 km s$^{-1}$ and 12.2 km s$^{-1}$ to 20.2 km s$^{-1}$, respectively.
	This bipolar outflow has multiple blue- and red-shifted lobes with a north-east to south-west elongation.
	}
	\label{4}
\end{figure}

\begin{figure}[h]
 \includegraphics[keepaspectratio, width=16cm]{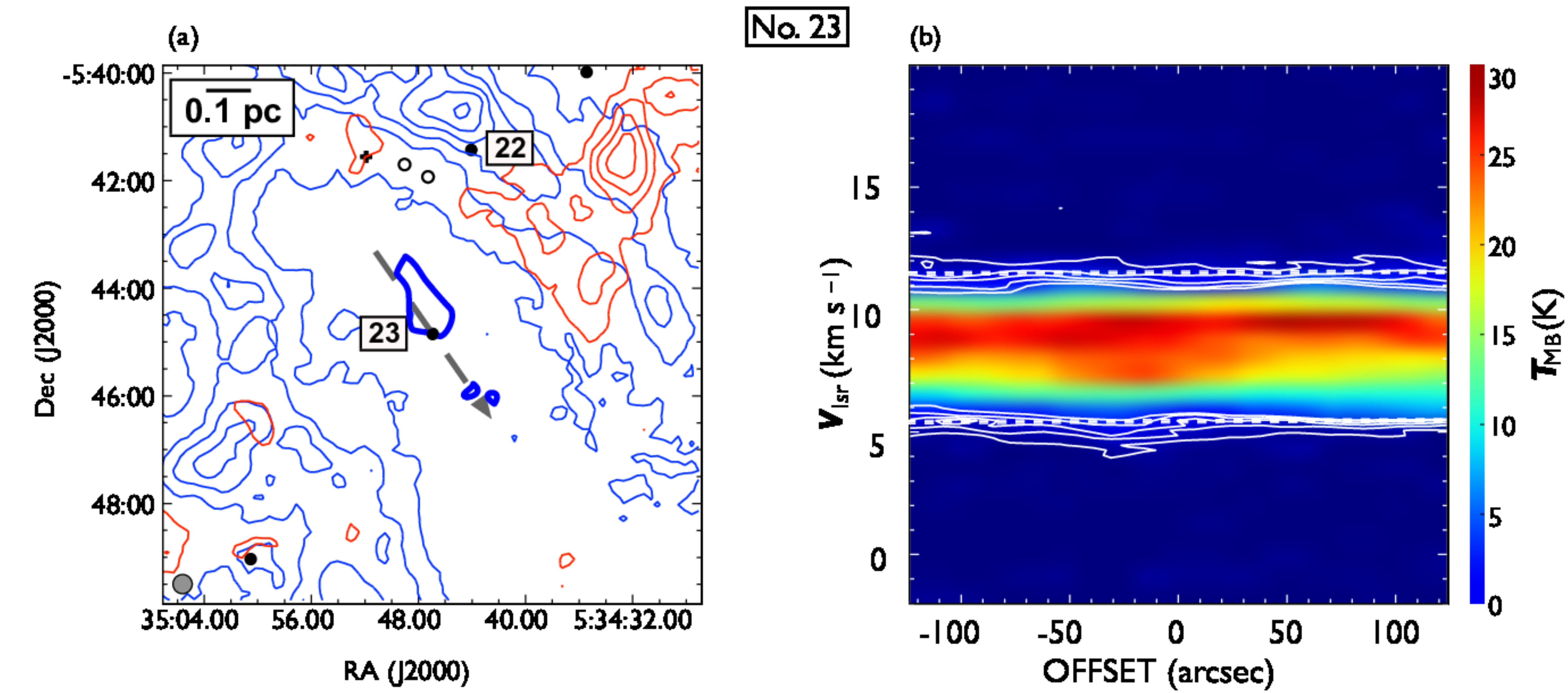}
 	\caption
	{
	The same as in figure \ref{1} but for outflow No. 23. In panel (a), the blue- and red-shifted integrated intensity velocity ranges are -1.9 km s$^{-1}$ to 5.4 km s$^{-1}$ and 11.5 km s$^{-1}$ to 20.2 km s$^{-1}$, respectively.
	}
	\label{5}
\end{figure}

\begin{figure}[h]
 \includegraphics[keepaspectratio, width=16cm]{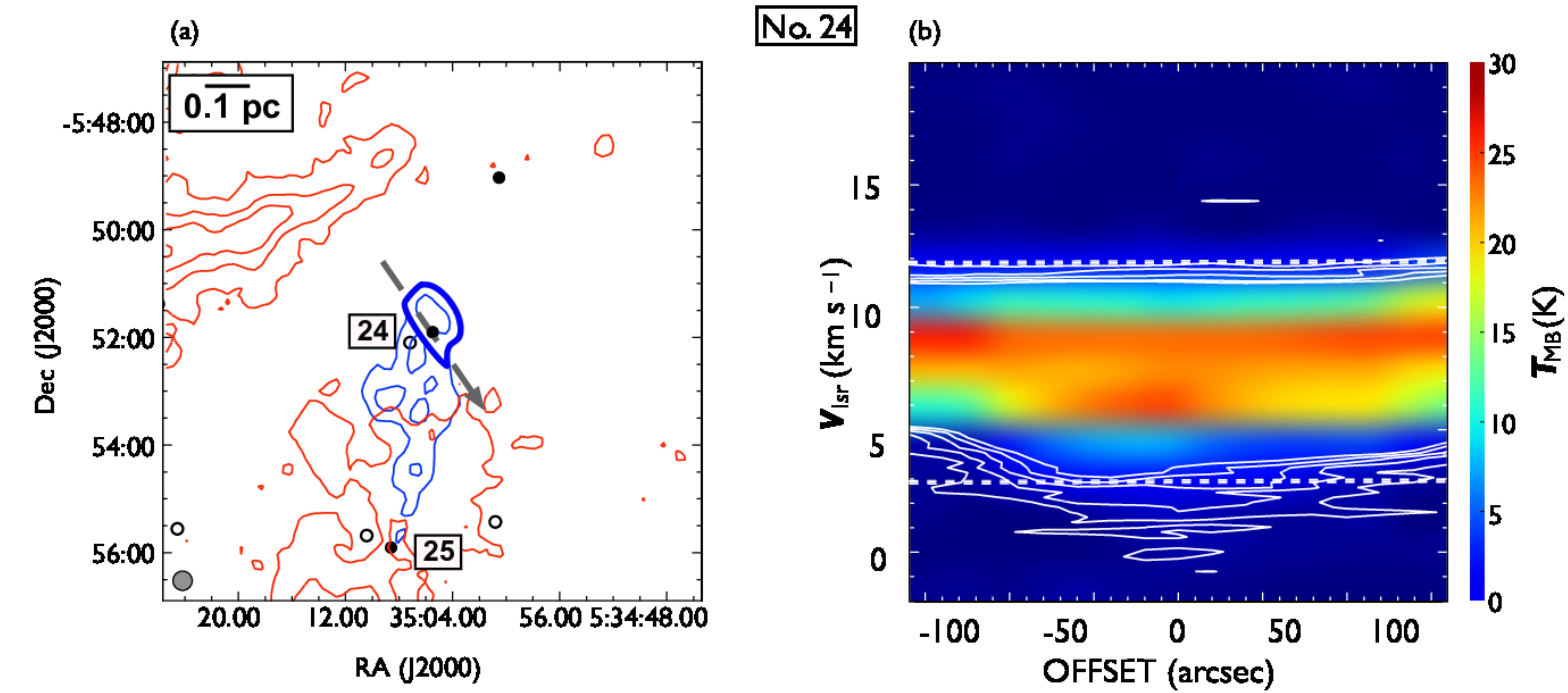}
 	\caption
	{
	The same as in figure \ref{1} but for outflow No. 24. In panel (a), the blue- and red-shifted integrated intensity velocity ranges are -1.9 km s$^{-1}$ to 3.8 km s$^{-1}$ and 11.8 km s$^{-1}$ to 18.0 km s$^{-1}$, respectively.
	This blue-single outflow has a strong blue-shifted lobe with a north-east to south-west elongation.
	The blue-shifted emission south of the central source is considered to be part of outflow No. 25.
	}\label{6}
\end{figure}

\begin{figure}[h]
 \includegraphics[keepaspectratio, width=16cm]{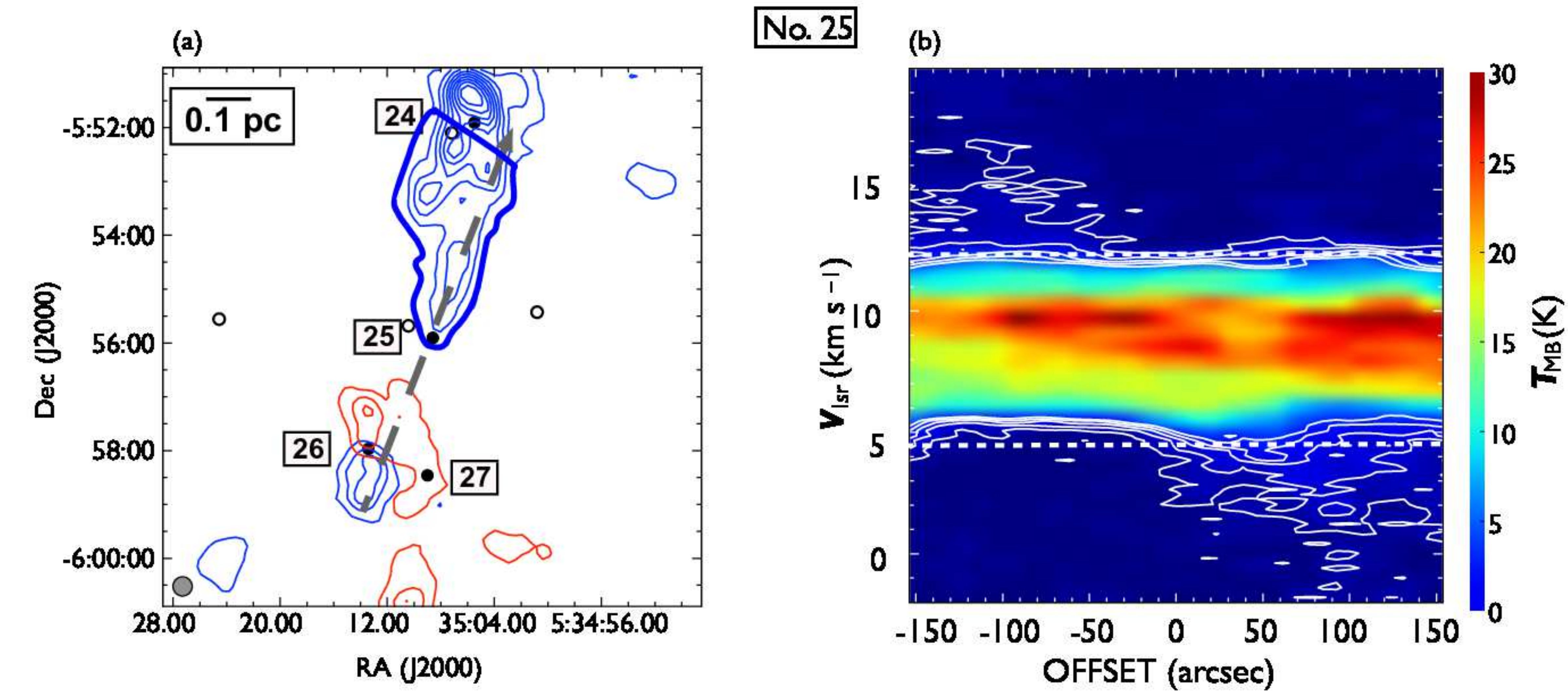}
 	\caption
	{
	The same as in figure \ref{1} but for outflow No. 25. In panel (a), the blue- and red-shifted integrated intensity velocity ranges are -1.9 km s$^{-1}$ to 4.6 km s$^{-1}$ and 12.2 km s$^{-1}$ to 20.2 km s$^{-1}$, respectively.
	This single outflow has a strong blue-shifted lobe in the northern direction.
	The red-shifted emission south of the central source is considered to be part of outflow No. 27. 
	}\label{7}
\end{figure}

\begin{figure}[h]
 \includegraphics[keepaspectratio, width=16cm]{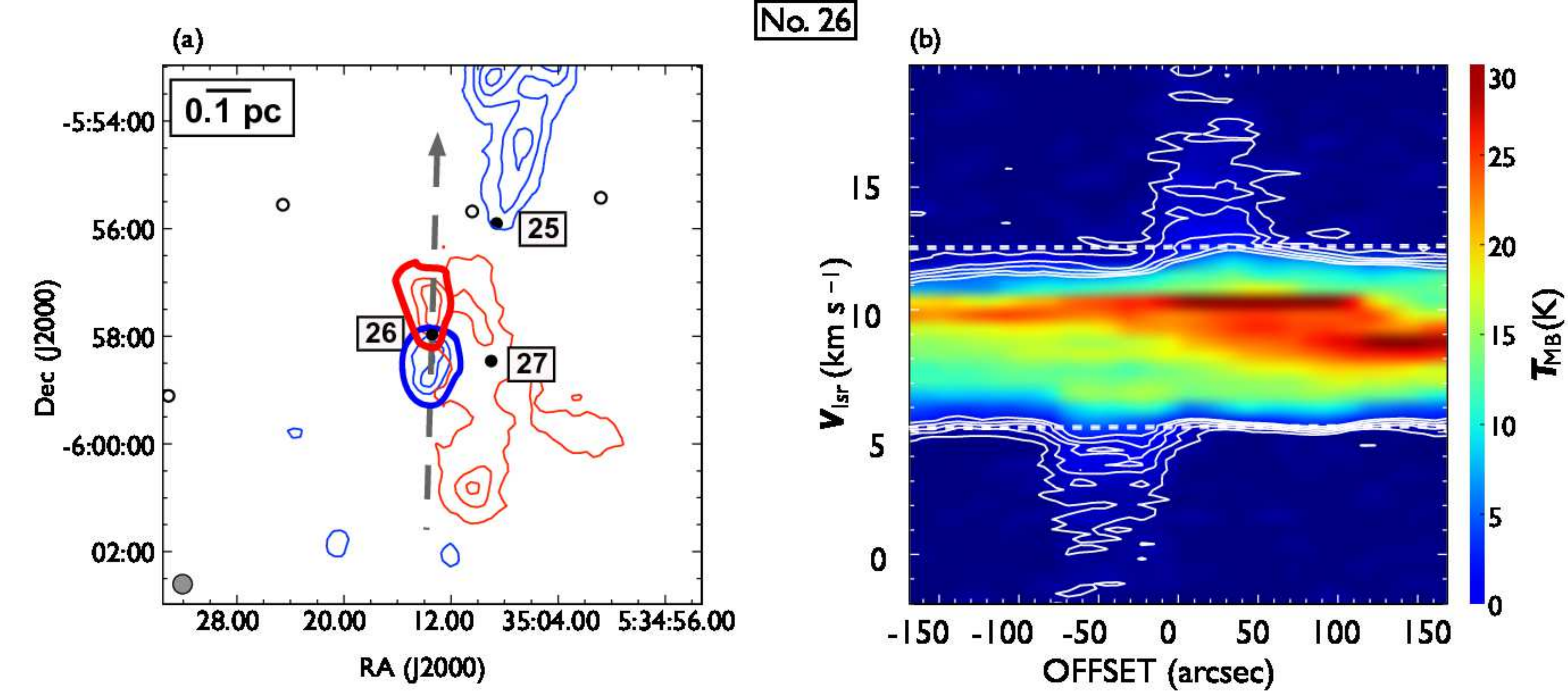}
 	\caption
	{
	The same as in figure \ref{1} but for outflow No. 26. In panel (a), the blue- and red-shifted integrated intensity velocity ranges are -1.9 km s$^{-1}$ to 4.9 km s$^{-1}$ and 12.8 km s$^{-1}$ to 20.2 km s$^{-1}$, respectively.
	The blue-shifted lobe in the south and red-shifted lobe in the north are clearly visible.
	}\label{8}
\end{figure}

\begin{figure}[h]
 \includegraphics[keepaspectratio, width=16cm]{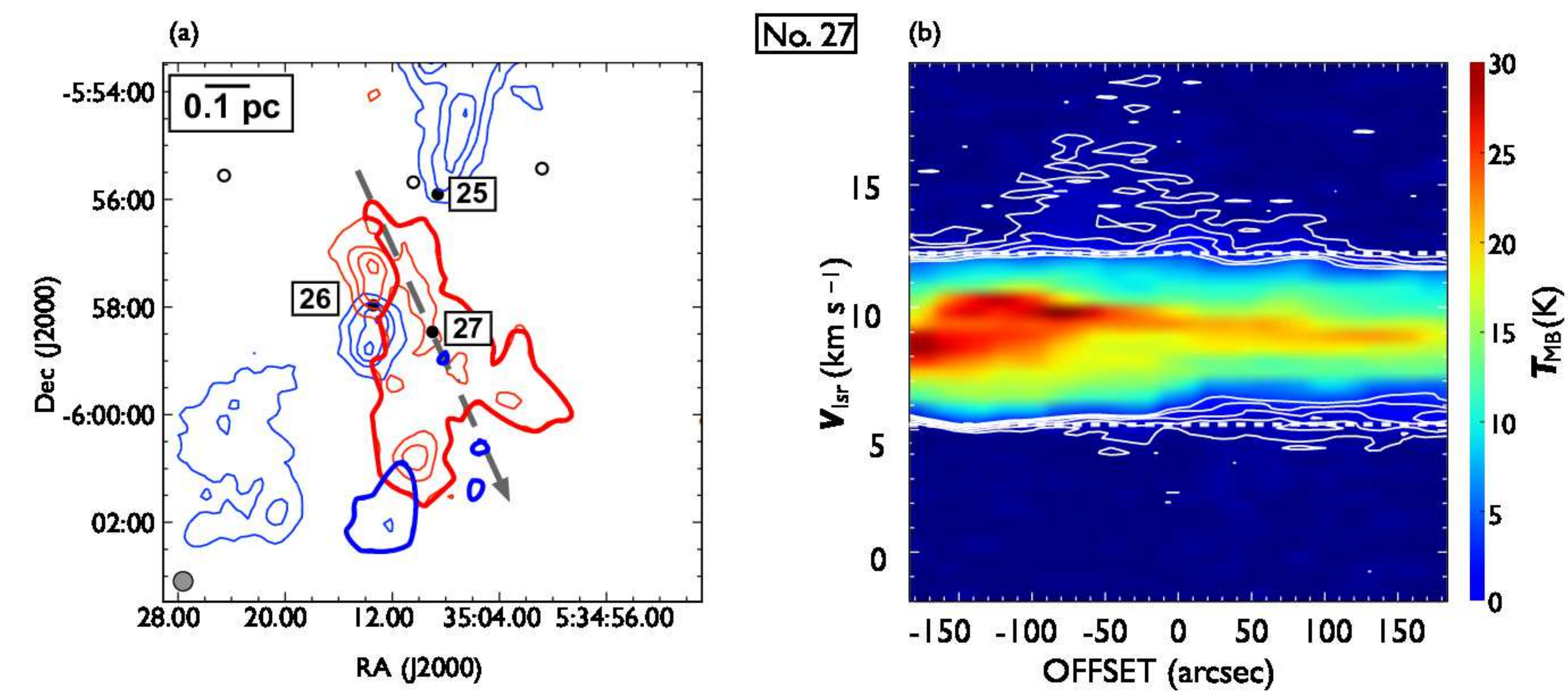}
 	\caption
	{
	The same as in figure \ref{1} but for outflow No. 27. In panel (a), the blue- and red-shifted integrated intensity velocity ranges are -1.9 km s$^{-1}$ to 5.2 km s$^{-1}$ and 12.0 km s$^{-1}$ to 20.2 km s$^{-1}$, respectively.
	This outflow has strong red-shifted lobes with a north-east to south-west elongation and faint blue-shifted lobes with a south-west elongation.
	}
	\label{9}
\end{figure}

\begin{figure}[h]
 \includegraphics[keepaspectratio, width=16cm]{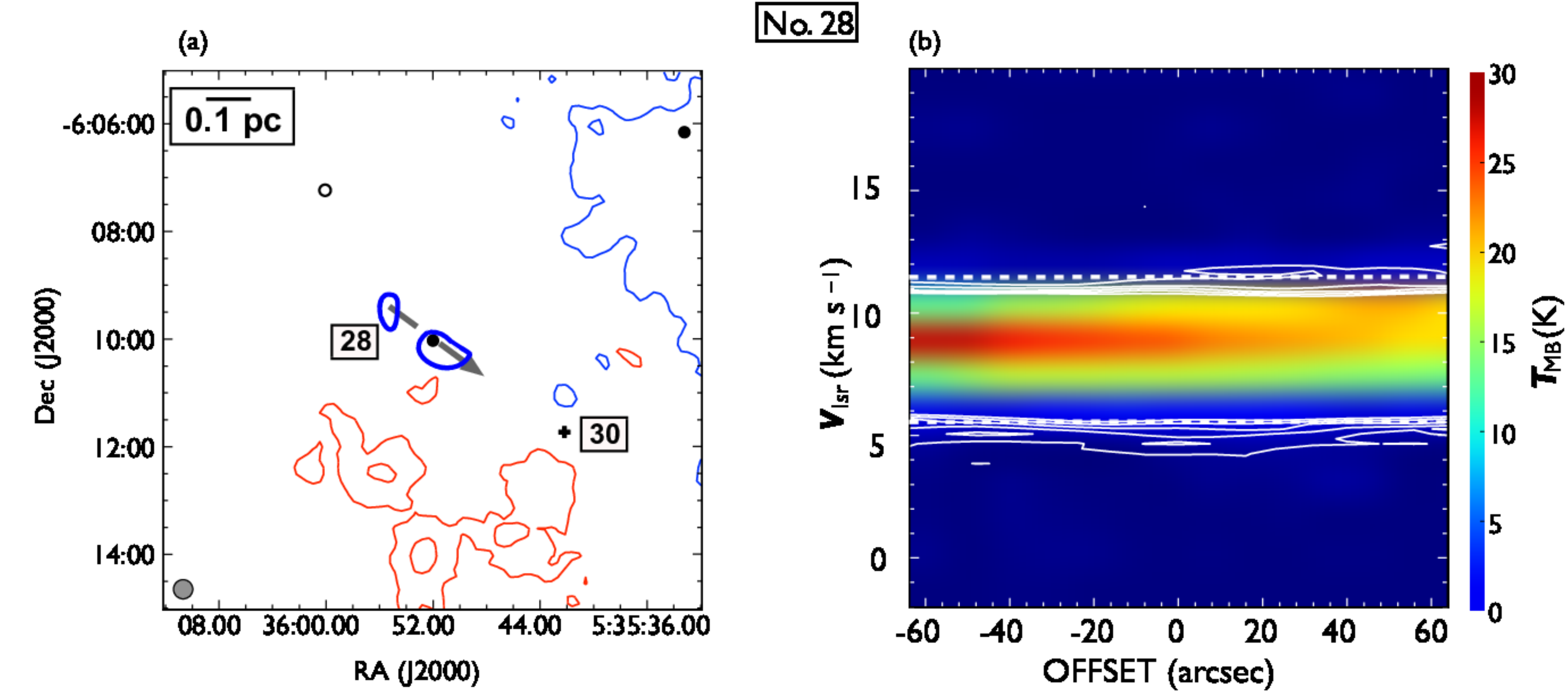}
 	\caption
	{
	The same as in figure \ref{1} but for outflow No. 28. In panel (a), the blue- and red-shifted integrated intensity velocity ranges are -1.9 km s$^{-1}$ to 5.5 km s$^{-1}$ and 11.4 km s$^{-1}$ to 20.2 km s$^{-1}$, respectively.
	}\label{10}
\end{figure}

\begin{figure}[h]
 \includegraphics[keepaspectratio, width=16cm]{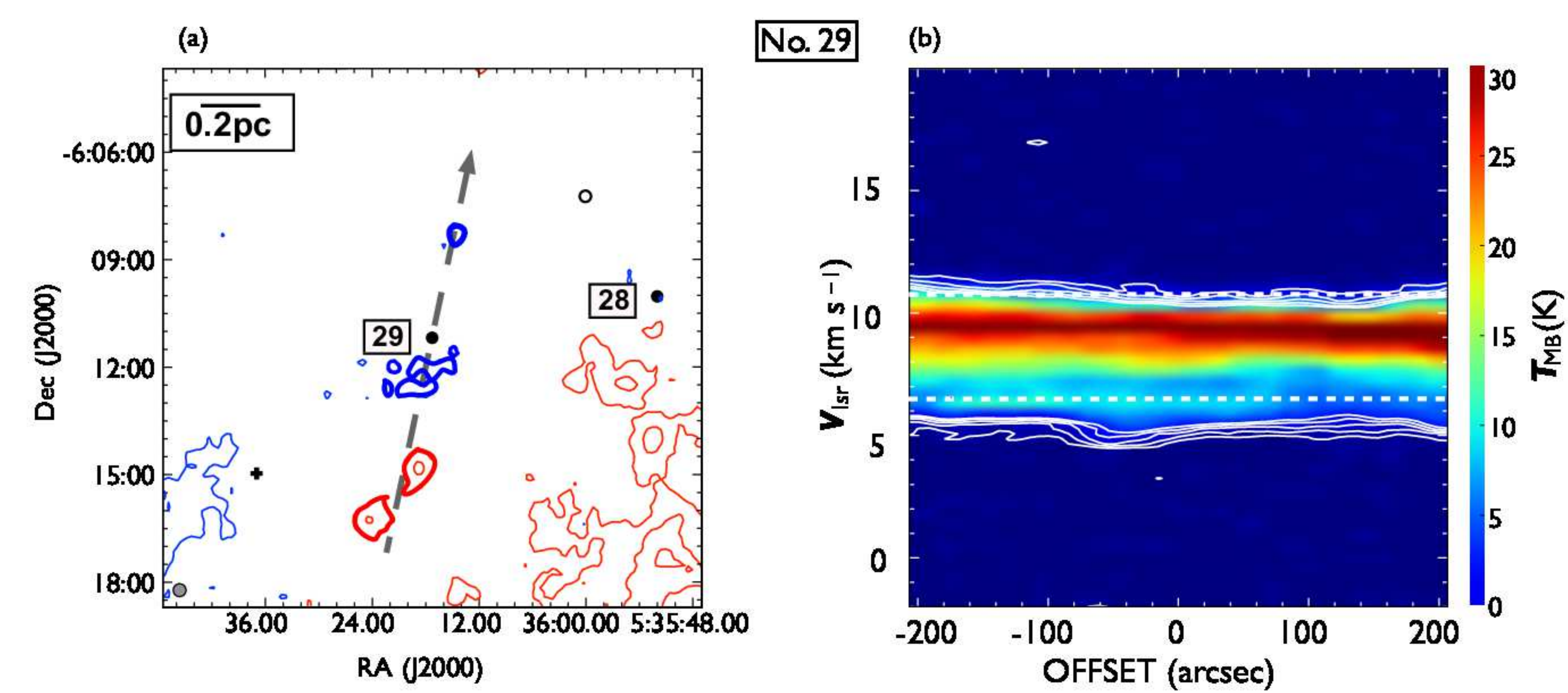}
 	\caption
	{
	The same as figure \ref{1}, but for outflow No. 29. In panel (a), the blue- and red-shifted integrated intensity velocity ranges are -1.9 km s$^{-1}$ to 6.7 km s$^{-1}$ and 10.9 km s$^{-1}$ to 20.2 km s$^{-1}$, respectively.
	This outflow is a faint bipolar outflow with multiple lobes in the north and south.
	}\label{11}
\end{figure}

\begin{figure}[h]
 \includegraphics[keepaspectratio, width=16cm]{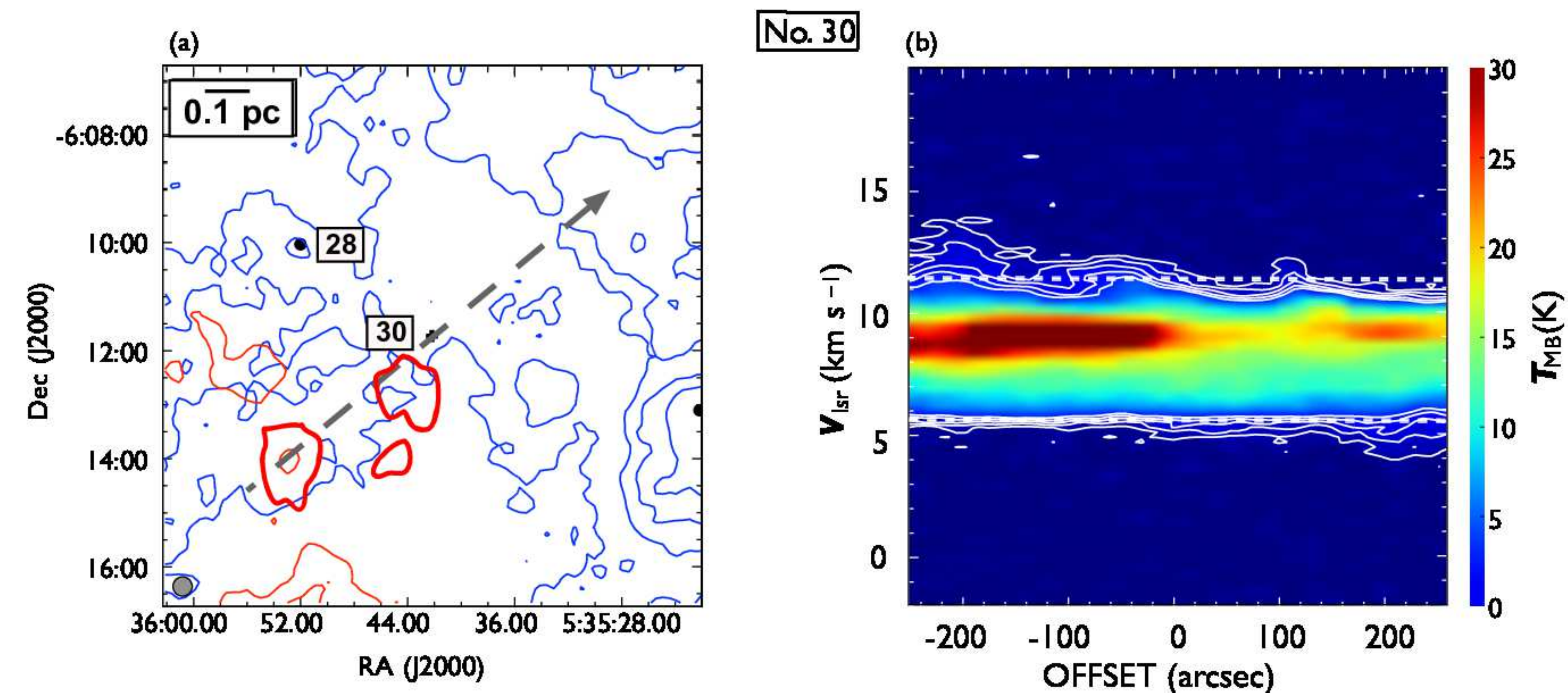}
 	\caption
	{
	The same as figure \ref{1}, but for outflow No. 30. In panel (a), the blue- and red-shifted integrated intensity velocity ranges are -1.9 km s$^{-1}$ to 5.7 km s$^{-1}$ and 11.5 km s$^{-1}$ to 20.2 km s$^{-1}$, respectively.
	This red-single outflow has the plural red-shifted lobes with a north-west elongation.
	}\label{122}
\end{figure}

\begin{figure}[h]
 \includegraphics[keepaspectratio, width=16cm]{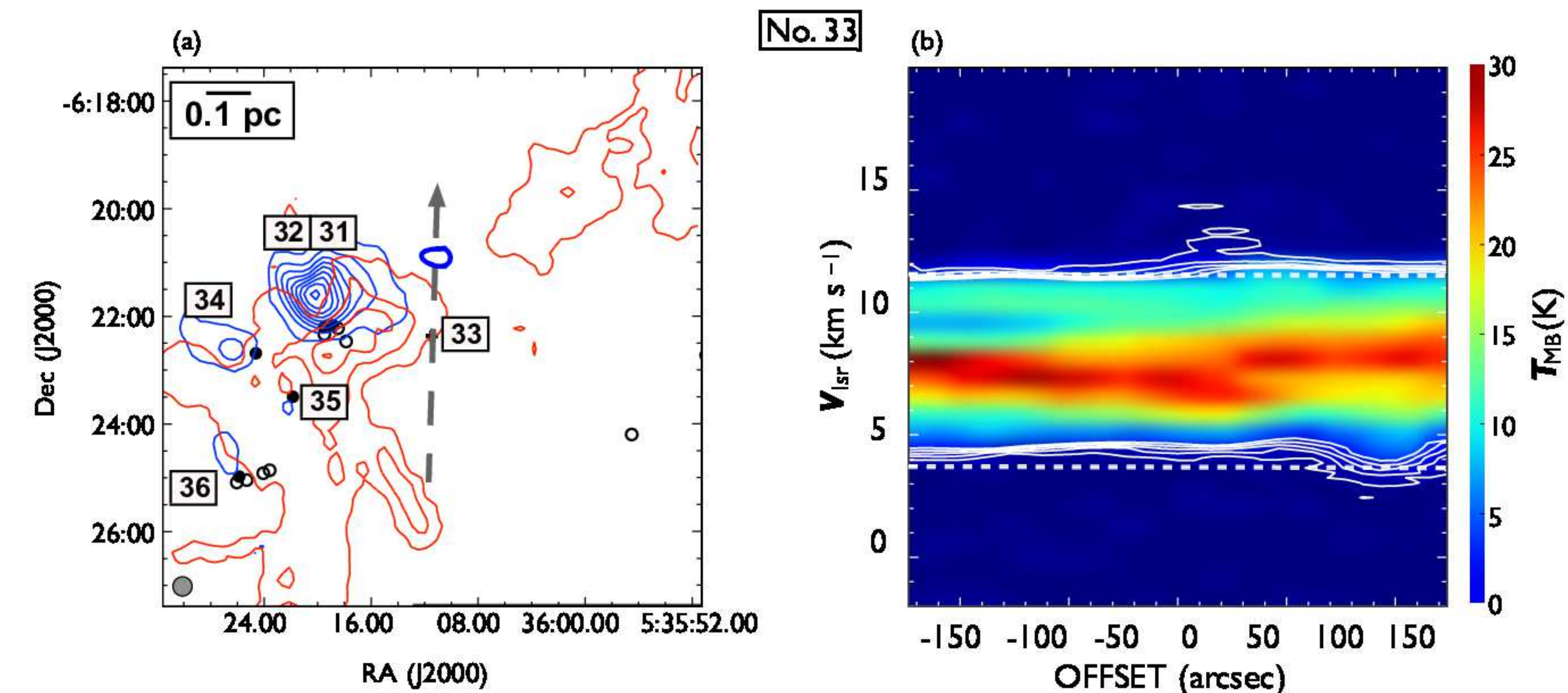}
 	\caption
	{
	The same as in figure \ref{1} but for outflow No. 33. In panel (a), the blue- and red-shifted integrated intensity velocity ranges are -1.9 km s$^{-1}$ to 3.7 km s$^{-1}$ and 11.5 km s$^{-1}$ to 18.0 km s$^{-1}$, respectively.
	This faint single blue outflow is driven by the central L1641 region PMS star with a jet. 
	}\label{132}
\end{figure}

\begin{figure}[h]
 \includegraphics[keepaspectratio, width=16cm]{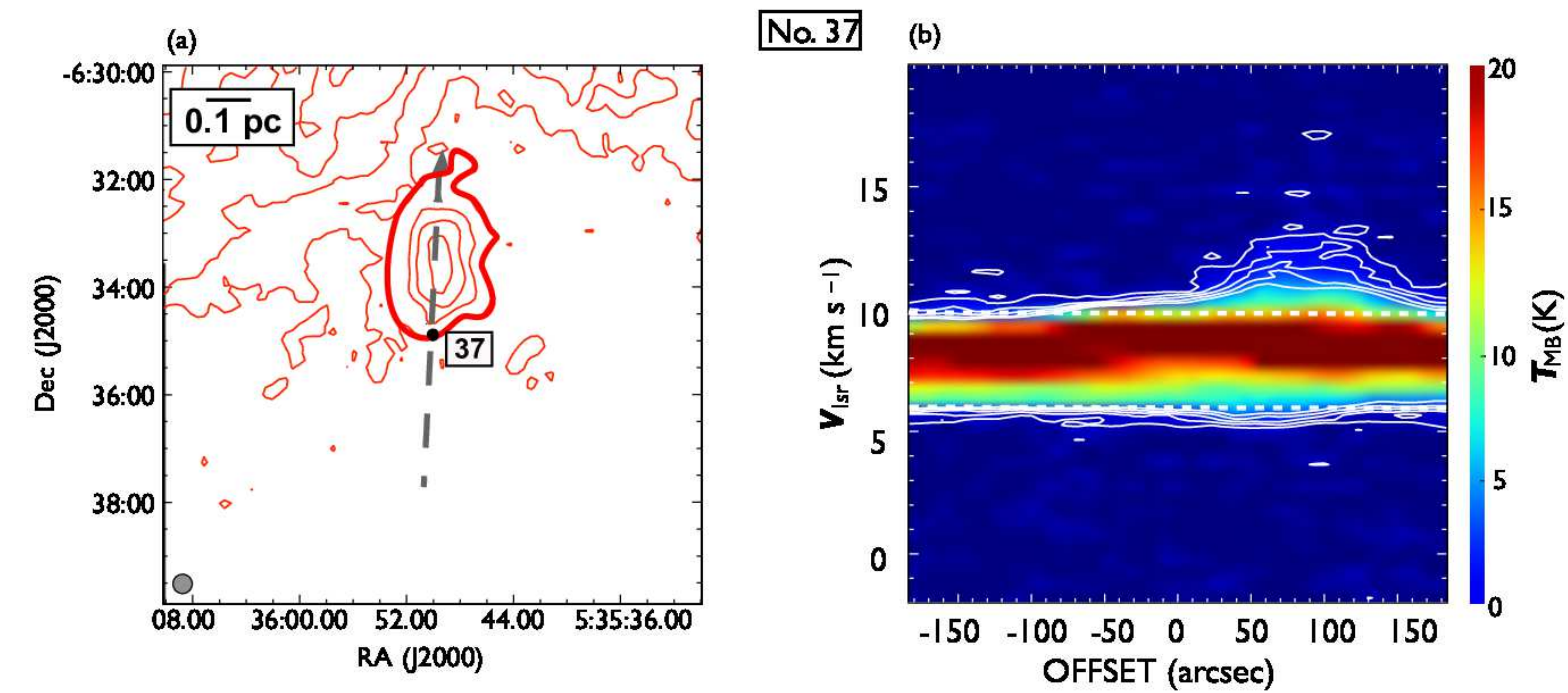}
 	\caption
	{
	The same as in figure \ref{1} but for outflow No. 37. In panel (a), the blue- and red-shifted integrated intensity velocity ranges are -1.9 km s$^{-1}$ to 6.0 km s$^{-1}$ and 10.9 km s$^{-1}$ to 20.2 km s$^{-1}$, respectively.
	 The red-shifted lobe clearly can be seen to the north.
	}\label{14}
\end{figure}

\begin{figure}[h]
 \includegraphics[keepaspectratio, width=16cm]{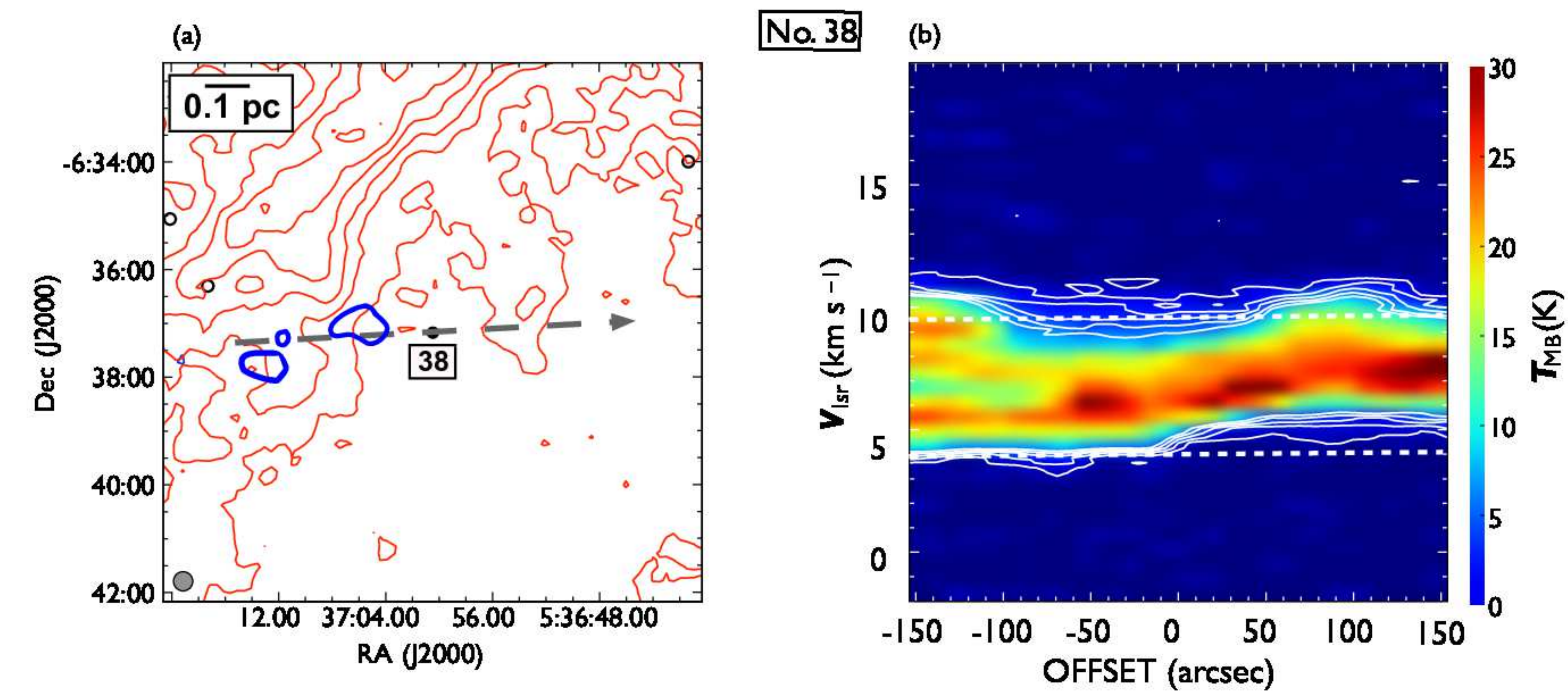}
 	\caption
	{
	The same as in figure \ref{1} but for outflow No. 38. In panel (a), the blue- and red-shifted integrated intensity velocity ranges are -1.9 km s$^{-1}$ to 3.9 km s$^{-1}$ and 9.5 km s$^{-1}$ to 18.0 km s$^{-1}$, respectively.
	This faint single blue outflow has the three blue-shifted components to the east.
	}\label{15}
\end{figure}

\begin{figure}[h]
 \includegraphics[keepaspectratio, width=16cm]{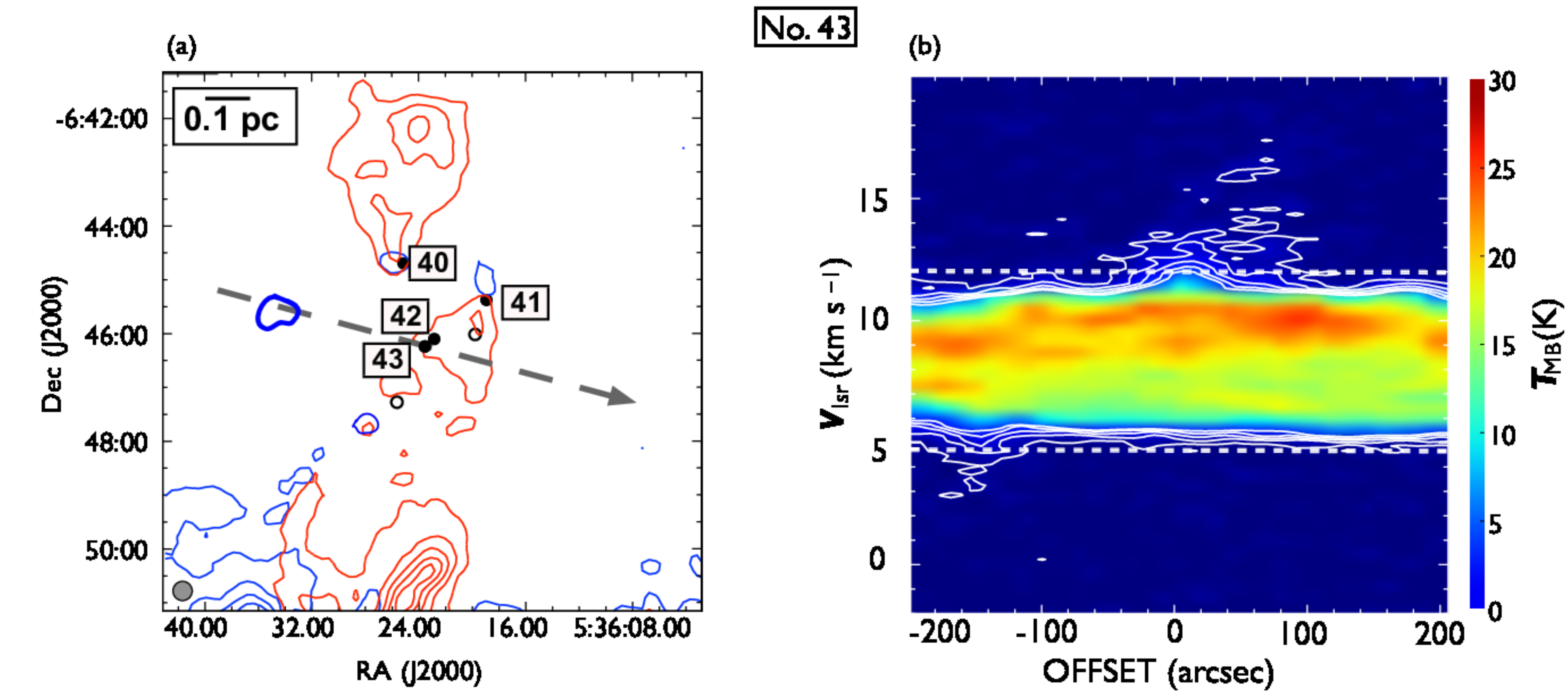}
 	\caption
	{
	The same as in figure \ref{1} but for outflow No. 43. In panel (a), the blue- and red-shifted integrated intensity velocity ranges are -1.9 km s$^{-1}$ to 4.8 km s$^{-1}$ and 12.1 km s$^{-1}$ to 20.2 km s$^{-1}$, respectively.
	This single blue outflow is driven by the protostar just to the south of the HH 1/2 VLA source.
	}\label{16}
\end{figure}

\begin{figure}[h]
 \includegraphics[keepaspectratio, width=16cm]{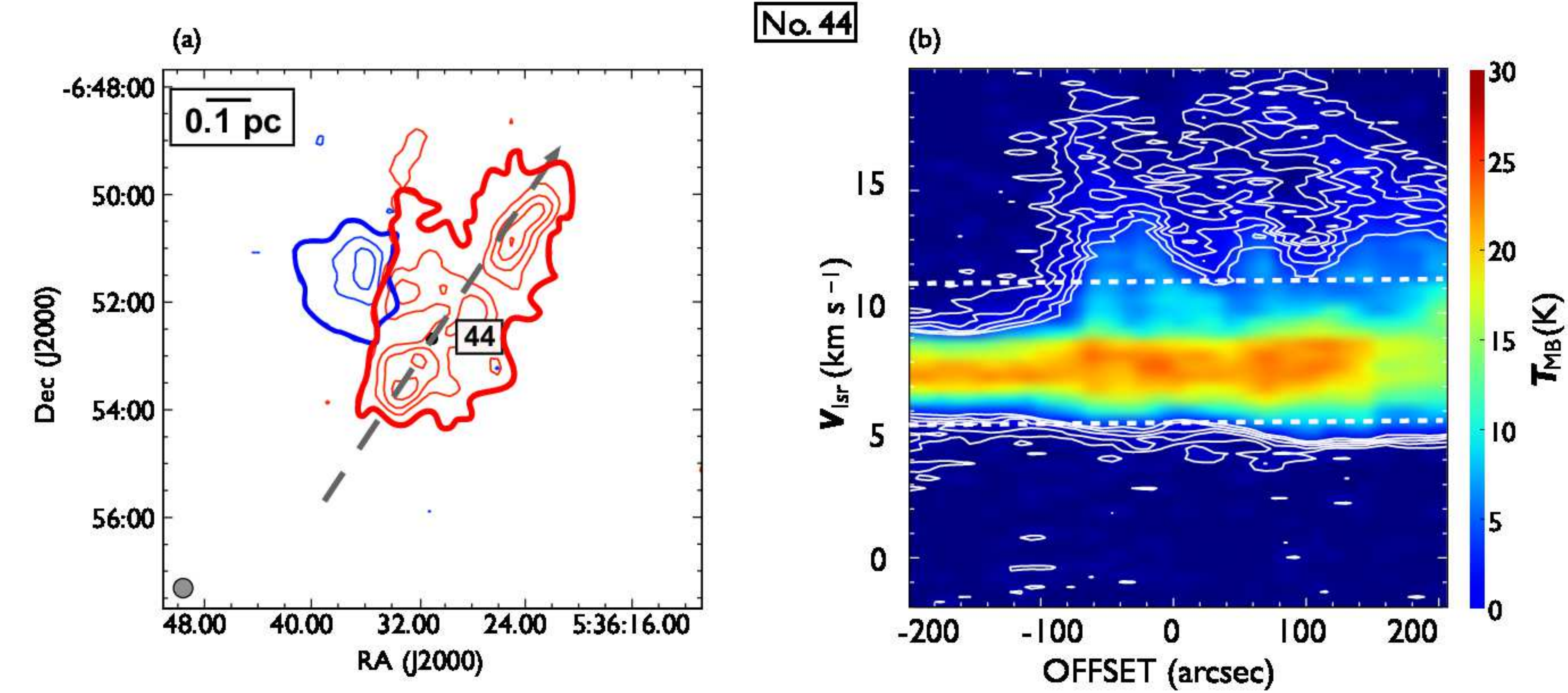}
 	\caption
	{
	The same as in figure \ref{1} but for outflow No. 44. In panel (a), the blue- and red-shifted integrated intensity velocity ranges are -1.9 km s$^{-1}$ to 5.4 km s$^{-1}$ and 11.2 km s$^{-1}$ to 20.2 km s$^{-1}$, respectively.
	This outflow has a strong red-shifted component with a north-west to south-east elongation and a faint blue-shifted component with a north-east elongation. 
	We define the position angle of No. 44 as the direction of the red-shifted component.
	}
	\label{17}
\end{figure}

\subsection{Optical depth of ${}^{12}$CO outflows}\label{tausec}
\begin{figure}[h]
 \includegraphics[keepaspectratio,width=16cm]{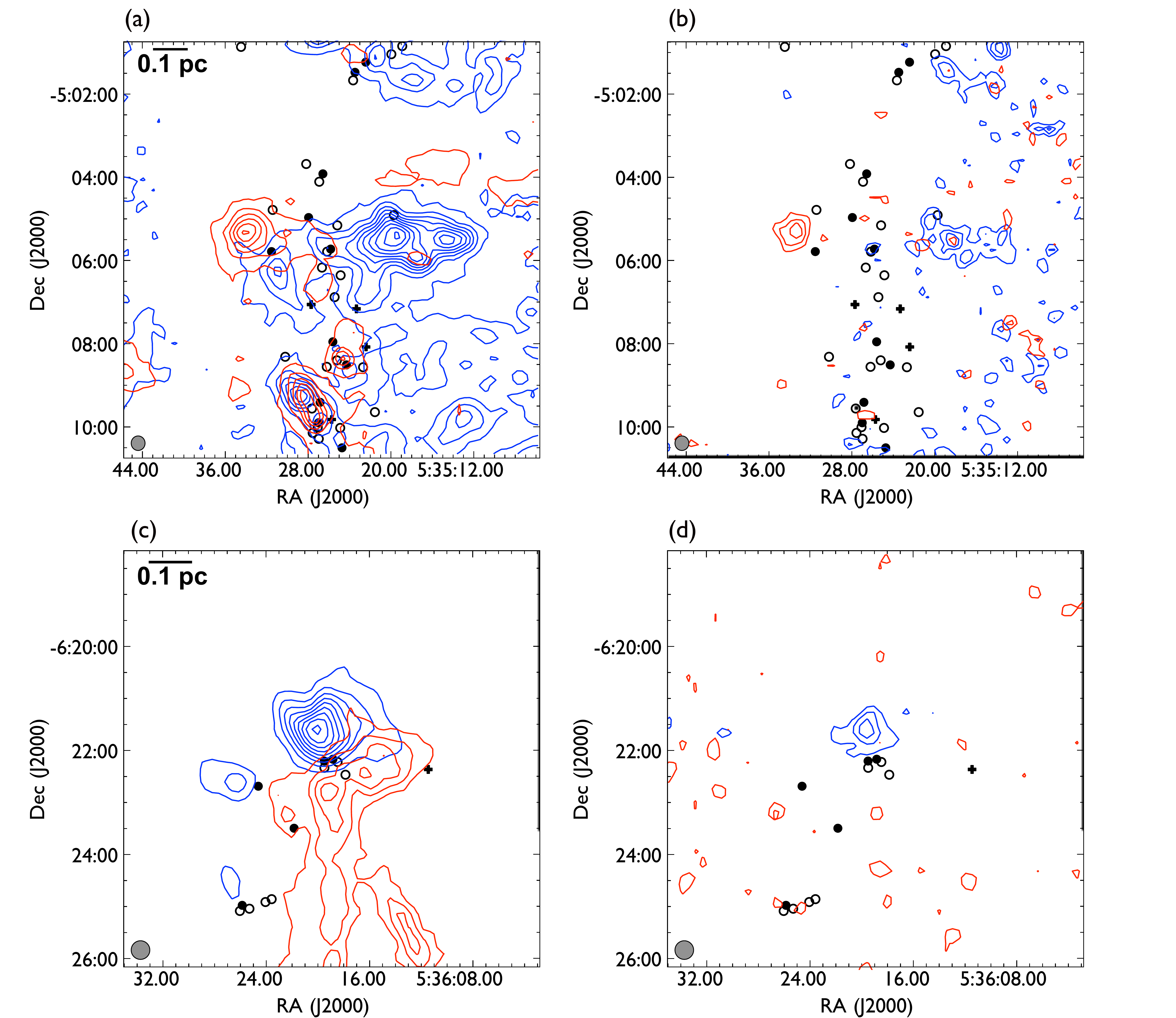}
	\caption{The pair of  ${}^{12}$CO and ${}^{13}$CO outflows.
	Panels (a) and (c) show ${}^{12}$CO outflows following the procedure described in section \ref{pro}. 
	Contours and markers are the same as those in figure \ref{ex} (d).  
	Panels (b) and (d) show ${}^{13}$CO outflows with the same velocity ranges as (a) and (c), respectively. 
	The contour increments are 0.3 K km s$^{-1}$($\sim$1$\sigma$) starting at 0.9 K km s$^{-1}$($\sim$3$\sigma$).
	}
	\label{F13}
\end{figure}

To estimate the optical depth of ${}^{12}$CO\ ({\it J} = 1--0) outflows, we also search for outflows  in the ${}^{13}$CO\ ({\it J} = 1--0) map at the locations where we identified ${}^{12}$CO\ ({\it J} = 1--0) outflows.
We make the ${}^{13}$CO integrated intensity map with the same velocity range as ${}^{12}$CO, (i.e., $|v_{\rm lsr}-v_{\rm sys}|\geq\ 2\sigma_v$). 
The ${}^{13}$CO outflows are identified as the regions where the ${}^{13}$CO integrated intensities are above 3$\sigma$ and the peak of a ${}^{13}$CO integrated intensity is located within the ${}^{12}$CO outflow lobe. 
The results are shown in figure \ref{F13}. 
We identify three outflows in the ${}^{13}$CO\ ({\it J} = 1--0) map (hereafter we refer to them as ${}^{13}$CO outflows): the blue and red lobes of outflow No. 7 and the blue lobes of outflow Nos. 31 and 32.
In these outflows, we estimate the optical depth using the following equation:
\begin{equation}
\frac{I_{\rm {}^{12}CO}}{I_{\rm {}^{13}CO}}=\frac{1-\exp(-\tau_{\rm {}^{12}\rm CO})}{1-\exp(-\tau_{\rm {}^{13}CO})},
\label{ratio}
\end{equation}
where  $I_{{\rm {}^{12}CO}}$ is the averaged integrated intensity of ${}^{12}$CO inside the 5$\sigma$ level, and $I_{{\rm {}^{13}CO}}$ is the averaged integrated intensity of ${}^{13}$CO within the contour of the 5$\sigma$ level of ${}^{12}$CO. 
$\tau_{\rm {}^{12}CO}$ and $\tau_{\rm {}^{13}CO}$ are the averaged optical depths of the ${}^{12}$CO and ${}^{13}$CO lines, respectively.
The above equation can be used when ${}^{12}$CO and ${}^{13}$CO are in local thermal equilibrium (LTE) with the same excitation temperature.
Assuming $\tau_{\rm {}^{12}CO}/\tau_{\rm {}^{13}CO} = 62$ based on the value of $[{}^{12}\rm C]/[{}^{13}C] $ (Langer \&  Penzias \yearcite{1993ApJ...408..539L}), equation \ref{ratio} provides us with the $\tau_{\rm {}^{12}CO}$ values of the outflows, as listed in table \ref{tau}. 
The average of the $\tau_{\rm {}^{12}CO}$ over the four lobes of the outflows is $\sim$5, suggesting that the optical depth of ${}^{12}$CO is not small in the detected outflows.
Hereafter, we adopt $\tau_{{}^{12}\rm CO}=5$ in other outflows.
This assumption is possible for outflows without ${}^{13}$CO detection; ${}^{12}$CO\ ({\it J} = 1--0) brightness temperature in outflows without ${}^{13}$CO detection is lower than those with ${}^{13}$CO detection, and we cannot give any stringent constraint on $\tau_{{}^{12}\rm CO}$.

\begin{table}[h!]
\vspace{5mm}
 \caption{Optical depth of outflows}
	\begin{tabular}{cccc}\hline
	No.&Lobe&$\tau_{{}^{13}\rm CO}$&$\tau_{{}^{12}\rm CO}$\\\hline
	7 & Blue & 0.075& 4.65\\
       7& Red &0.097 & 6.01\\
     31-32 &  Blue&0.076& 4.71\\\hline
   \end{tabular}
 \label{tau}
\end{table}

\subsection{Outflow and cloud properties}\label{properties}
\subsubsection{Physical parameters of outflows}\label{properties1}
To estimate the physical parameters of each detected outflow, we assume that the {$\rm {}^{12}CO$} molecules are in LTE.
The results are summarized in table \ref{property}, but we do not correct for the inclination angle of outflows.
When the outflows are distributed randomly, the average of inclinations is 57.3$^{\circ}$, 
and the velocity of outflows will be increased by a factor of 1.85, while the timescale of outflows will be decreased by a factor of 0.64. 
The uncertainty arising from unknown inclination angles of outflows is large.
For example, in the rather extreme case of 80$^{\circ}$, the velocity will be higher by a factor of 5.76.

 \begin{table}[h!]
  \caption{{$\rm{}^{12}CO(1-0)$} Outflow properties}
  {\footnotesize
  \scalebox{0.8}{
   \begin{tabular}{lcccccccccccc}\hline
  No.	&	Lobe	&	$M_{\rm flow}$\footnotemark[*]	&	$P_{\rm flow}$	&	$E_{\rm flow}$	&	$\Delta v_{\rm max}$	&	$R_{\rm max}$&$t_{\rm d}	$&$\dot{M}_{\rm flow}$&$\dot{P}_{\rm flow}$&$\dot{E}_{\rm flow}$&\\
  &&$(M_{\odot})$&$(M_{\odot} \rm\ km\ s^{-1})$&$(10^{43} \rm erg)$&$\rm (km\ s^{-1})$&$\rm(pc)$&($10^4$\ yr)&$(10^{-6}\ M_{\odot}\ {\rm yr}^{-1})$&$(10^{-6}M_{\odot}\ \rm km\ s^{-1}\ yr^{-1})$&$(10^{30}\rm erg\ s^{-1})$\\\hline
\multicolumn{11}{c}{OMC 2/3}\\\hline
1	&	B	&	0.07 	$\	\pm\ $	0.03 	&	0.26 	$\	\pm\ $	0.11 	&	1.00 	$\	\pm\ $	0.45 	&	5.3	&	0.13	&	2.4	&	3.0 	&	10.6 	&	13.1 	\\
	&	R	&	0.11 	$\	\pm\ $	0.04 	&	0.24 	$\	\pm\ $	0.10 	&	0.63 	$\	\pm\ $	0.27 	&	3.9	&	0.23	&	5.9	&	1.5 	&	4.0 	&	3.5 	\\\hline
2	&	B	&	0.16 	$\	\pm\ $	0.07 	&	0.61 	$\	\pm\ $	0.27 	&	2.40 	$\	\pm\ $	1.15 	&	6.6	&	0.08	&	1.2	&	14.1 	&	50.3 	&	65.9 	\\\hline
3	&	B	&	0.03 	$\	\pm\ $	0.02 	&	0.13 	$\	\pm\ $	0.09 	&	0.64 	$\	\pm\ $	0.45 	&	6.4	&	0.06	&	0.9	&	2.5 	&	14.1 	&	21.1 	\\
	&	R	&	0.16 	$\	\pm\ $	0.06 	&	0.55 	$\	\pm\ $	0.21 	&	1.96 	$\	\pm\ $	0.79 	&	6.2	&	0.19	&	3.0	&	5.0 	&	18.6 	&	20.6 	\\\hline
4	&	B	&	0.20 	$\	\pm\ $	0.07 	&	0.96 	$\	\pm\ $	0.37 	&	4.87 	$\	\pm\ $	1.72 	&	6.7	&	0.08	&	1.1	&	17.6 	&	87.0 	&	136.8 	\\
	&	R	&	0.07 	$\	\pm\ $	0.03 	&	0.23 	$\	\pm\ $	0.11 	&	0.79 	$\	\pm\ $	0.29 	&	5.9	&	0.06	&	1.0	&	6.0 	&	22.6 	&	24.6 	\\\hline
5	&	B	&	0.37 	$\	\pm\ $	0.14 	&	1.92 	$\	\pm\ $	0.79 	&	10.54 $\	\pm\ $	4.04 	&	10.3	&	0.32	&	3.1	&	12.1 	&	61.9 	&	108.6 	\\
	&	R	&	0.01 	$\	\pm\ $	0.01 	&	0.02 	$\	\pm\ $	0.01 	&	0.06 	$\	\pm\ $	0.05 	&	3.7	&	0.05	&	1.2	&	0.5 	&	1.5 	&	1.5 	\\\hline
6	&	B	&	0.01 	$\	\pm\ $	0.01 	&	0.02 	$\	\pm\ $	0.01 	&	0.04 	$\	\pm\ $	0.02 	&	3.5	&	0.08	&	2.2	&	0.4 	&	0.9 	&	0.6 	\\
	&	R	&	1.04 	$\	\pm\ $	0.23 	&	4.34 	$\	\pm\ $	1.04 	&	18.23 $\	\pm\ $	4.78 	&	7.4	&	0.55	&	7.3	&	14.6 	&	59.4 	&	79.5 	\\\hline
7	&	B	&	2.29 	$\	\pm\ $	0.36 	&	10.60 $\	\pm\ $	2.09 	&	53.39 $\	\pm\ $	11.54 	&	13.0	&	0.42	&	3.1	&	73.2 	&	341.8 	&	539.4 	\\
	&	R(west)	&	1.55 $\	\pm\ $	0.30 	&	5.80 	$\	\pm\ $	1.63 	&	22.70 $\	\pm\ $	5.85 	&	8.4	&	0.34	&	4.0	&	39.2 	&	145.2 	&	181.4 	\\
	&	R(east)	&	0.13 $\	\pm\ $	0.04 	&	0.46 	$\	\pm\ $	0.15 	&	1.61 	$\	\pm\ $	0.52 	&	4.0	&	0.27	&	6.1	&	1.5 	&	7.5 	&	5.0 	\\\hline
8	&	B	&	0.14 	$\	\pm\ $	0.05 	&	0.72 	$\	\pm\ $	0.28 	&	3.76 	$\	\pm\ $	1.60 	&	7.9	&	0.13	&	1.6	&	9.1 	&	45.3 	&	73.9 	\\
	&	R	&	0.20 	$\	\pm\ $	0.05 	&	0.74 	$\	\pm\ $	0.18 	&	2.83 	$\	\pm\ $	0.73 	&	4.7	&	0.13	&	2.7	&	7.0 	&	27.7 	&	32.7 	\\\hline
9	&	B	&	0.14 	$\	\pm\ $	0.04 	&	0.69 	$\	\pm\ $	0.22 	&	3.51 	$\	\pm\ $	1.17 	&	6.8	&	0.1	&	1.4	&	9.6 	&	49.3 	&	77.5 	\\\hline
10	&	B	&	0.63 	$\	\pm\ $	0.16 	&	2.68 	$\	\pm\ $	0.71 	&	11.36 $\	\pm\ $	3.30 	&	6.9	&	0.34	&	4.8	&	13.1 	&	55.8 	&	73.9 	\\\hline
11	&	R	&	0.18 	$\	\pm\ $	0.04 	&	0.66 	$\	\pm\ $	0.17 	&	2.51 	$\	\pm\ $	0.55 	&	5.2	&	0.09	&	1.7	&	10.1 	&	38.7 	&	46.3 	\\\hline
12	&	B	&	0.10 	$\	\pm\ $	0.05 	&	0.46 	$\	\pm\ $	0.24 	&	2.13 	$\	\pm\ $	1.16 	&	6.3	&	0.06	&	0.9	&	10.6 	&	51.3 	&	72.9 	\\
	&	R	&	0.31 	$\	\pm\ $	0.10 	&	1.30 	$\	\pm\ $	0.44 	&	5.76 	$\	\pm\ $	1.99 	&	7.7	&	0.08	&	1.0	&	30.2 	&	130.3 	&	178.6 	\\\hline
13	&	B(north)	&	0.40 	$\	\pm\ $	0.15 	&	2.29 	$\	\pm\ $	0.97 	&	14.32 	$\	\pm\ $	6.81 	&	12.7	&	0.14	&	1.1	&	36.7 	&	208.2 	&	419.5 	\\
	&	B(south)	&	0.30 	$\	\pm\ $	0.10 	&	1.54 	$\	\pm\ $	0.54 	&	8.20 	$\	\pm\ $	3.19 	&	8.5	&	0.12	&	1.4	&	21.6 	&	110.2 	&	187.6 	\\
	&	R(north)	&	0.36 	$\	\pm\ $	0.11 	&	1.69 	$\	\pm\ $	0.57 	&	8.30 	$\	\pm\ $	3.17 	&	9.1	&	0.13	&	1.4	&	26.2 	&	120.7 	&	188.6 	\\
	&	R(south)	&	0.32 	$\	\pm\ $	0.10 	&	1.48 	$\	\pm\ $	0.50 	&	7.07 	$\	\pm\ $	2.70 	&	9.1	&	0.10	&	1.1	&	30.2 	&	134.3 	&	208.7 	\\\hline
14	&	B(south)	&	0.09 	$\	\pm\ $	0.03 	&	0.39 	$\	\pm\ $	0.14 	&	1.76 	$\	\pm\ $	0.69 	&	5.8	&	0.08	&	1.4	&	6.5 	&	28.2 	&	40.7 	\\
	&	R(south)	&	0.13 	$\	\pm\ $	0.05 	&	0.59 	$\	\pm\ $	0.24 	&	2.73 	$\	\pm\ $	1.19 	&	6.3	&	0.10	&	1.6	&	8.6 	&	36.7 	&	54.8 	\\\hline
15	&	R	&	0.16 	$\	\pm\ $	0.06 	&	0.64 	$\	\pm\ $	0.26 	&	2.57 	$\	\pm\ $	1.10 	&	5.0	&	0.26	&	5.1	&	3.0 	&	12.6 	&	16.1 	\\\hline
16	&	B	&	0.38 	$\	\pm\ $	0.12 	&	1.70 	$\	\pm\ $	0.56 	&	7.60 	$\	\pm\ $	2.80 	&	6.9	&	0.22	&	3.1	&	12.1 	&	54.8 	&	76.5 	\\
	&	R	&	0.22 	$\	\pm\ $	0.08 	&	0.87 	$\	\pm\ $	0.35 	&	3.70 	$\	\pm\ $	1.48 	&	7.0	&	0.24	&	3.4	&	6.0 	&	25.7 	&	34.7 	\\\hline
17	&	B	&	0.49 	$\	\pm\ $	0.19 	&	2.36 	$\	\pm\ $	0.98 	&	11.78 $\	\pm\ $	5.01 	&	8.3	&	0.54	&	6.3	&	7.5 	&	37.2 	&	58.9 	\\\hline
18	&	R	&	0.27 	$\	\pm\ $	0.12 	&	1.25 	$\	\pm\ $	0.55 	&	5.83 	$\	\pm\ $	2.72 	&	6.7	&	0.12	&	1.8	&	15.1 	&	69.4 	&	105.6 	\\\hline
19	&	R	&	0.36 	$\	\pm\ $	0.13 	&	1.80 	$\	\pm\ $	0.68 	&	8.93 	$\	\pm\ $	3.55 	&	6.8	&	0.16	&	2.3	&	15.6 	&	78.0 	&	122.2 	\\\hline
\multicolumn{10}{c}{OMC 4/5}\\\hline
20	&	B	&	0.06 	$\	\pm\ $	0.02 	&	0.32 	$\	\pm\ $	0.09 	&	1.64 	$\	\pm\ $	0.47 	&	5.7	&	0.12	&	2.1	&	3.0 	&	15.1 	&	24.6 	\\\hline
21	&	B	&	0.04 	$\	\pm\ $	0.02 	&	0.12 	$\	\pm\ $	0.08 	&	0.44 	$\	\pm\ $	0.30 	&	5.6	&	0.15	&	2.7	&	1.3 	&	4.5 	&	5.1 	\\
	&	R	&	0.05 	$\	\pm\ $	0.03 	&	0.29 	$\	\pm\ $	0.17 	&	1.93 	$\	\pm\ $	1.16 	&	7.6	&	0.17	&	2.2	&	2.0 	&	13.3 	&	28.1 	\\\hline
22	&	B	&	0.07 	$\	\pm\ $	0.03 	&	0.30 	$\	\pm\ $	0.12 	&	1.30 	$\	\pm\ $	0.50 	&	5.0	&	0.28	&	5.5	&	1.2 	&	5.4 	&	7.5 	\\
	&	R	&	0.04 	$\	\pm\ $	0.03 	&	0.22 	$\	\pm\ $	0.15 	&	1.22 	$\	\pm\ $	0.34 	&	7.0	&	0.51	&	7.1	&	0.5 	&	3.0 	&	5.4 	\\\hline
23	&	B	&	0.09 	$\	\pm\ $	0.03 	&	0.29 	$\	\pm\ $	0.11 	&	0.99 	$\	\pm\ $	0.38 	&	4.0	&	0.18	&	4.5	&	1.9 	&	6.5 	&	7.0 	\\\hline
24	&	B	&	0.18 	$\	\pm\ $	0.09 	&	1.29 	$\	\pm\ $	0.67 	&	9.47 	$\	\pm\ $	5.17 	&	11.7	&	0.25	&	2.1	&	8.6 	&	61.9 	&	144.9 	\\\hline
25	&	B	&	0.24 	$\	\pm\ $	0.11 	&	1.27 	$\	\pm\ $	0.61 	&	7.02 	$\	\pm\ $	2.23 	&	10.1	&	0.28	&	2.7	&	9.0 	&	47.3 	&	83.0 	\\\hline
26	&	B	&	0.13 	$\	\pm\ $	0.07 	&	0.70 	$\	\pm\ $	0.38 	&	4.06 	$\	\pm\ $	1.68 	&	10.7	&	0.14	&	1.3	&	9.8 	&	53.2 	&	97.4 	\\
	&	R	&	0.22 	$\	\pm\ $	0.11 	&	1.25 	$\	\pm\ $	0.67 	&	8.60 	$\	\pm\ $	5.05 	&	12.3	&	0.15	&	1.2	&	18.3 	&	102.4 &	223.0 	\\\hline
27	&	B	&	0.07 	$\	\pm\ $	0.04 	&	0.26 	$\	\pm\ $	0.16 	&	1.07 	$\	\pm\ $	0.63 	&	4.6	&	0.25	&	5.3	&	1.2 	&	5.0 	&	6.4 	\\
	&	R	&	0.45 	$\	\pm\ $	0.26 	&	2.33 	$\	\pm\ $	1.41 	&	12.83 $\	\pm\ $	4.77 	&	12.4	&	0.35	&	2.7	&	16.6 	&	85.2 	&	148.9 	\\\hline
28	&	B	&	0.05 	$\	\pm\ $	0.03 	&	0.18 	$\	\pm\ $	0.11 	&	0.62 	$\	\pm\ $	0.38 	&	4.1	&	0.16	&	3.8	&	1.3 	&	4.6 	&	5.2 	\\\hline
29	&	B	&	0.08 	$\	\pm\ $	0.04 	&	0.32 	$\	\pm\ $	0.18 	&	1.33 	$\	\pm\ $	0.73 	&	4.7	&	0.38	&	7.9	&	1.0 	&	4.1 	&	5.3 	\\
	&	R	&	0.05 	$\	\pm\ $	0.02 	&	0.16 	$\	\pm\ $	0.07 	&	0.49 	$\	\pm\ $	0.22 	&	3.7	&	0.40	&	10.7	&	0.5 	&	1.5 	&	1.4 	\\\hline
30	&	R	&	0.22 	$\	\pm\ $	0.10 	&	0.80 	$\	\pm\ $	0.37 	&	2.99 	$\	\pm\ $	1.47 	&	5.6	&	0.30	&	5.3	&	4.1 	&	15.1 	&	18.0 	\\\hline

   \end{tabular}
   }}
   \label{property}
\end{table}

\addtocounter{table}{-1}
 \begin{table}[h!]
  \caption{continued}
    {\footnotesize
      \scalebox{0.8}{
   \begin{tabular}{cccccccccccc} \hline
  No.	&	Lobe	&	$M_{\rm flow}$\footnotemark[*]&	$P_{\rm flow}$	&	$E_{\rm flow}$	&	$\Delta v_{\rm max}$	&	$R_{\rm max}$&$t_{\rm d}	$&$\dot{M}_{\rm flow}$&$\dot{P}_{\rm flow}$&$\dot{E}_{\rm flow}$&\\
  &&$(M_{\odot})$&$(M_{\odot} \rm\ km\ s^{-1})$&$(10^{43} \rm erg)$&$\rm (km\ s^{-1})$&$\rm(pc)$&($10^4$\ yr)&$(10^{-6}\ M_{\odot}\ {\rm yr}^{-1})$&$(10^{-6}M_{\odot}\ \rm km\ s^{-1}\ yr^{-1})$&$(10^{30}\rm erg\ s^{-1})$\\\hline
\multicolumn{10}{c}{L1641-N}\\\hline
31	&	B(north-west)	&	0.36 	$\	\pm\ $	0.09 	&	2.09 	$\	\pm\ $	0.65 	&	13.66 $\	\pm\ $	5.06 	&	19.8	&	0.17	&	0.8	&	43.3 	&	261.4 	&	528.5 	\\
	&	R(south-east)	&	0.14 	$\	\pm\ $	0.08 	&	0.77 	$\	\pm\ $	0.46 	&	4.48 	$\	\pm\ $	2.72 	&	8.2	&	0.65	&	7.8	&	2.0 	&	10.1  	&	18.1 	\\
	&	B(east)		&	0.05 	$\	\pm\ $	0.02 	&	0.24 	$\	\pm\ $	0.07 	&	1.19 	$\	\pm\ $	0.35 	&	5.4	&	0.16	&	2.8	&	1.5 	&	8.6 	        &	13.1 	\\\hline
32	&	B(north)		&	0.36 	$\	\pm\ $	0.08 	&	1.94 	$\	\pm\ $	0.49 	&	10.92 $\	\pm\ $	3.19 	&	11.0	&	0.19	&	1.7	&	21.4 	&	114.5 	&	203.4 	\\
	&	R(south)		&	0.48 	$\	\pm\ $	0.28 	&	2.73 	$\	\pm\ $	1.66 	&	16.07 $\	\pm\ $	10.19 &	15.4	&	1.01	&	6.3	&	7.5 	&	42.8 		&	79.5 	\\\hline
33	&	B			&	0.02 	$\	\pm\ $	0.01 	&	0.07 	$\	\pm\ $	0.04 	&	0.29 	$\	\pm\ $	0.15 	&	4.7	&	0.15	&	3.1	&	0.5 	&	2.5 		&	3.0 	\\\hline
34	&	B			&	0.04 	$\	\pm\ $	0.02 	&	0.16 	$\	\pm\ $	0.08 	&	0.74 	$\	\pm\ $	0.30 	&	5.6	&	0.15	&	2.6	&	1.3 	&	6.0 		&	9.0 	\\
	&	R			&	0.03 	$\	\pm\ $	0.01 	&	0.12 	$\	\pm\ $	0.04 	&	0.55 	$\	\pm\ $	0.07 	&	5.8	&	0.08	&	1.3	&	1.8 	&	9.0 		&	14.0 	\\\hline
35	&	B			&	0.01 	$\	\pm\ $	0.01 	&	0.03 	$\	\pm\ $	0.02 	&	0.13 	$\	\pm\ $	0.08 	&	4.7	&	0.04	&	0.8	&	0.9 	&	3.6 		&	4.9 	\\
	&	R			&	0.02 	$\	\pm\ $	0.01 	&	0.12 	$\	\pm\ $	0.07 	&	0.72 	$\	\pm\ $	0.34 	&	9.9	&	0.06	&	0.5	&	4.0 	&	22.0 		&	41.7 	\\\hline
36	&	B			&	0.01 	$\	\pm\ $	0.01 	&	0.04 	$\	\pm\ $	0.03 	&	0.19 	$\	\pm\ $	0.12 	&	5.0	&	0.08	&	1.6	&	0.6 	&	2.6 		&	3.7 	\\\hline
37	&	R			&	0.42 	$\	\pm\ $	0.08 	&	1.00 	$\	\pm\ $	0.23 	&	2.60 	$\	\pm\ $	0.73 	&	4.9	&	0.25	&	5.1	&	8.2 	&	19.6 		&	16.1 	\\\hline
38	&	B			&	0.04 	$\	\pm\ $	0.02 	&	0.11 	$\	\pm\ $	0.05 	&	0.33 	$\	\pm\ $	0.14 	&	3.6	&	0.22	&	5.8	&	0.6 	&	1.9 		&	1.8 	\\\hline
\multicolumn{10}{c}{NGC 1999}\\\hline
39	&       B			&      0.32    $\      \pm\ $  0.13    &       1.63    $\      \pm\ $  0.81    &       9.62    $\      \pm\ $  4.13  	&       11.6    &       0.23    &       1.9     &       16.0    &       85.5    &       157.5   \\
	&       R			&      0.16    $\      \pm\ $  0.09    &       0.85    $\      \pm\ $  0.49    &       4.94    $\      \pm\ $  2.24  	&       9.3     &       0.22    &       2.3     &       7.0     &       36.5    &       68.0    \\\hline
40	&       B			&      0.02    $\      \pm\ $  0.01    &       0.07    $\      \pm\ $  0.05    &       6.38    $\      \pm\ $  0.17  	&       4.9     &       0.06    &       1.2     &       1.5     &       5.5     &       6.5     \\
	&       R			&      0.27    $\      \pm\ $  0.12    &       1.34    $\      \pm\ $  0.61    &       6.83    $\      \pm\ $  1.71  	&       8.3     &       0.14    &       1.6     &       11.5    &       83.0    &       132.0   \\\hline
41	&       B			&      0.01    $\      \pm\ $  0.01    &       0.04    $\      \pm\ $  0.03    &       0.11    $\      \pm\ $  0.08  	&       3.0     &       0.06    &       2.0     &       0.5     &       2.0     &       2.0     \\
	&       R			&      0.08    $\      \pm\ $  0.05    &       0.44    $\      \pm\ $  0.27    &       2.53    $\      \pm\ $  1.65  	&       8.8     &       0.22    &       2.5     &       3.5     &       17.5    &       32.0    \\\hline
42	&       B			&      0.07    $\      \pm\ $  0.03    &       0.27    $\      \pm\ $  0.13    &       1.15    $\      \pm\ $  0.55  	&       4.2     &       0.46    &       10.7    &       0.5     &       2.5     &       3.5     \\
	&       R			&      0.10    $\      \pm\ $  0.06    &       0.44    $\      \pm\ $  0.29    &       2.09    $\      \pm\ $  0.39	&       6.6     &       0.46    &       6.9     &       1.5     &       6.5     &       9.5     \\\hline
43	&       B			&      0.04    $\      \pm\ $  0.02    &       0.16    $\      \pm\ $  0.10    &       0.73    $\      \pm\ $  0.47	&       6.0     &       0.37    &       6.0     &       0.5     &       2.5     &       4.0     \\\hline
44	&       B			&      0.07    $\      \pm\ $  0.04    &       0.30    $\      \pm\ $  0.14    &       1.26    $\      \pm\ $  0.58	&       4.7     &       0.34    &       7.1     &       1.0     &       4.0     &       5.5     \\
	&       R			&      1.42    $\      \pm\ $  0.53    &       9.13    $\      \pm\ $  3.77    &       63.03   $\      \pm\ $  20.54	&       13.3	   &       0.46    &       3.4     &       41.5    &       267.0   &       585.0   \\\hline
Total	&				&	21.5					 &	  84.2				   &	439.7  					&		&			& 		&      728.9   &     3585.0   &      6024.8\\\hline
   \end{tabular}
   }}\begin{tabnote}
  \footnotemark[*]The error is estimated by applying noise of $\pm$ 3$\sigma$ to pixels with emission above 3$\sigma$.
   	\end{tabnote}	
\end{table}

We adopt $T_{\rm peak}$, which is derived from the Gaussian fitting of equation (\ref{gauss}), as the excitation temperature ($T_{\rm ex}$) of each outflow. 
The average value of  $T_{\rm peak}$ of all outflows in our sample is 36 K. 
Note that  in the high-velocity wings, $T_{\rm ex}$ may be higher than $T_{\rm peak}$. 
For example, if $T_{\rm ex}$ is 2.0$\times T_{\rm peak}$ for all outflows, the mass of the outflows and other quantities proportional to mass will be higher by a factor of 1.9.
The column density of ${}^{12}$CO at each position and at each velocity component is expressed by 


\begin{eqnarray}
 N(v)&=&\rm \frac{3{\it k}^2}{8{\pi}^3{\it hB(J}+1)\nu\mu^2}\ \it f_{\tau}\ {\rm{\it T}_{ex}}\exp{\rm \left(\frac{{\it hB(J}+1)({\it J}+2)}{{\it kT}_{ex}}\right)} {\rm{\it T}_B}(\it v)\rm \Delta\it v,
\end{eqnarray}
where {\it k},  {\it h}, {\it B}, $\nu$, $\mu$, $\Delta v$ are the Boltsmann constant, Planck constant, rotational constant of ${}^{12}$CO, rest frequency of the ({\it J} = 1--0) transition, dipole moment of the ${}^{12}$CO molecule, and velocity channel width of the cube, respectively.
$f_{\tau}$ is the correction factor for the optical depth given by
\begin{equation}
f_{\tau}=\rm\frac{\tau_{\rm {}^{12}CO}}{1-\exp(-\tau_{\rm {}^{12}CO})},
\end{equation}
where $\tau_{\rm {}^{12}CO}$ is the optical depth of ${}^{12}$CO derived in section \ref{tausec}.
The mass of each outflow can be estimated from the channel maps integrated in the velocity range $|v_{\rm lsr}-v_{\rm sys}|\geq\ 2\sigma_v$:
\begin{eqnarray}
M_{\rm flow}=\sum_{|v_{\rm lsr}-v_{\rm sys}|\geq\ 2\sigma_v} m(v),
\end{eqnarray}  
where $m(v)$ is the mass of each velocity component, given by
\begin{eqnarray}
 m(v)&=&4.33\times10^{13}\frac{\bar{\mu} m_{\rm H}}{\chi_{\rm {}^{12}CO}}\it  \left(\frac{s(v)}{\rm cm^{2}}\right)   \it  f_{\tau}  \left(\frac{T_{\rm ex}}{\rm K}\right)    \rm exp{\left(\frac{5.53}{\it T_{\rm ex}}\right)} \left(\frac{\overline{\it T_{\rm B}}(\it v)}{\rm K}\right)  \left(\frac{\Delta\it v}{\rm km\ s^{-1}}\right).
 \label{mass}
\end{eqnarray} 
In equation (\ref{mass}), $\bar{\mu}$ = 2.4 is the mean molecular weight, $m_{\rm H}$ is the atomic mass of hydrogen, $\chi_{\rm {}^{12}CO}$  = 10$^{-4}$ (Frerking et al. \yearcite{1982ApJ...262..590F}) is the abundance of {$\rm {}^{12}CO$} relative to $\rm H_2$, $\overline{T_{\rm B}}(v)$ is the averaged brightness temperature of all pixels in $s(v)$, and $s(v)$\ is the projected area above the 3$\sigma$ level, given by
\begin{eqnarray}
s(v)=n_{\rm pix}{\left\{1.5\times10^{13} {\left(\frac{D}{\rm pc}\right)}{\left(\frac{\Delta\theta}{{\rm arcsec}}\right)}\right\}}^2\ [\rm cm^2],
\end{eqnarray}
where $n_{\rm pix}$ is the number of pixels in $s(v)$, $D$ is the distance to Orion A, and $\Delta\theta\ =\ \timeform{7".5}$ is the pixel size. 
After deriving $m(v)$, we calculate the outflow momentum $P_{\rm flow}$ and the outflow energy $E_{\rm flow}$ as follows:
\begin{eqnarray}
 P_{\rm flow}&=&\sum_{|v_{\rm lsr}-v_{\rm sys}|\geq\ 2\sigma_v} m(v){|v_{\rm lsr}-v_{\rm sys}|}\\
\lefteqn{\hspace{-57mm}\rm and}\nonumber\\
 E_{\rm flow}&=&\frac{1}{2}\sum_{|v_{\rm lsr}-v_{\rm sys}|\geq\ 2\sigma_v} m(v){|v_{\rm lsr}-v_{\rm sys}|}^2\ .
\end{eqnarray}\\
The maximum velocity of each outflow ($\Delta v_{\rm max}$) is taken from $|v_{\rm max}-v_{\rm sys}|$, where $v_{\rm max}$ is the highest velocity of each outflow whose emission is above 3$\sigma$.
The maximum size of each outflow $(R_{\rm max})$ is measured in the integrated map. 
The dynamical timescale of each outflow ($t_{\rm d}$) is defined as $R_{\rm max}/\Delta v_{\rm max}$. 
The mass outflow rate is calculated by $\dot{M}_{\rm flow}=M_{\rm flow}/t_{\rm d}$, momentum ejection rate by $\dot{P}_{\rm flow}=P_{\rm flow}/t_{\rm d}$, and energy ejection rate by $\dot{E}_{\rm flow}=E_{\rm flow}/t_{\rm d}$.

\begin{figure}[h!]
 \includegraphics[keepaspectratio, width=15cm]{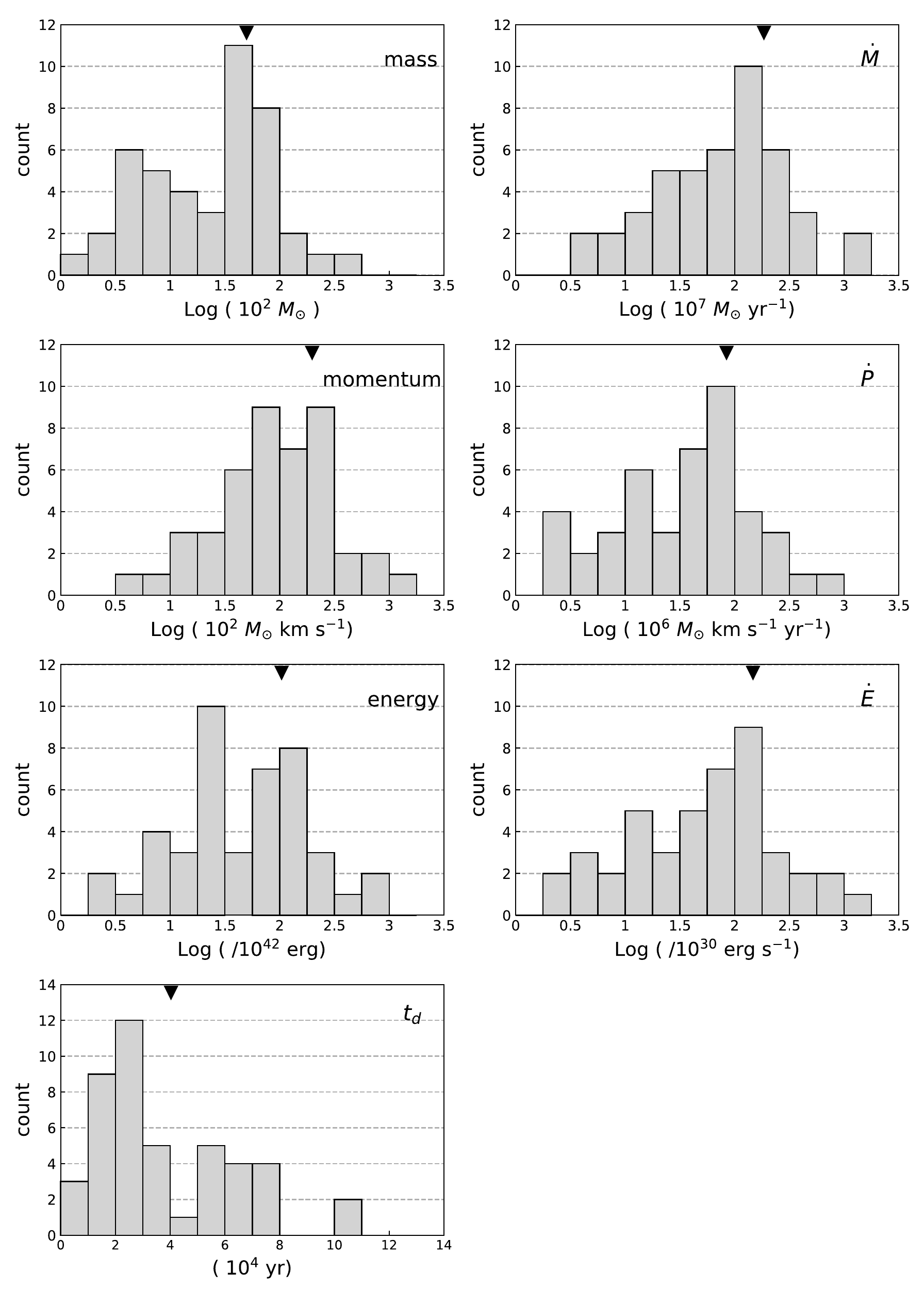}
 	\caption
	{
	Histograms of various outflow properties. 
	Each triangle represents the average of the corresponding property.
	} 
	\label{hist}
\end{figure}

Figure \ref{hist} shows histograms of mass, momentum, energy, mass outflow rate, momentum ejection rate, and energy ejection rate for all outflows and a histogram of dynamical timescale for all outflow lobes.
The average values of mass, momentum, energy, mass outflow rate, momentum ejection rate, and energy ejection rate of all outflows in our sample are 4.8$\times 10^{-1}\ M_{\odot}$, 1.9 $M_{\odot}\ {\rm km\ s^{-1}}$, 1.0$\times 10^{44}\ \rm erg$, 1.7 $\times$ 10$^{-5}\ M_{\odot}$  yr$^{-1}$, 8.1 $\times$ 10$^{-5}\ M_{\odot}$  km s$^{-1}$ yr$^{-1}$, and 1.4 $\times$ 10$^{32}$ erg s$^{-1}$, respectively.  
The average dynamical timescale of all outflow lobes is 3.8$\times 10^4$ yr.

\subsubsection{Properties of clouds}

Using the {$\rm {}^{13}CO$}\ ({\it J} = 1--0) line emission, we estimate the cloud kinetic properties in each subregion by the following method, and the results are summarized in table \ref{cloudpara}. 
An excitation temperature is calculated for each pixel by assuming that the {$\rm {}^{12}CO$} line is optically thick and using the equation from Rohlfs \& Wilson (\yearcite{1996tra..book.....R}):
\begin{equation}
T_{\rm ex}=\frac{5.53}{{\rm ln}(1+[5.53/(T_{\rm peak}({}^{12}\rm CO)+0.82)])}\ [\rm K],
\end{equation}
where $T_{\rm peak}$ is the peak intensity of ${}^{12}$CO at each pixel.
In the region where {$\rm {}^{13}CO$} is detected above 5$\sigma$, we calculate the optical depth of the ${}^{13}$CO\ ({\it J} = 1--0) line with the following equation
\begin{equation}
\tau_{\rm{}^{13}CO}(\nu)=-{\rm ln}\left[1-\frac{T_{\rm {}^{13}CO}(\nu)}{T_{\rm {}^{12}CO}(\nu)}\right].
\end{equation} 
The total column density of {$\rm {}^{13}CO$} can be expressed as
\begin{eqnarray}
N_{{\rm {}^{13}CO}}=4.74\times10^{13}{\rm {\it T}_{ex}}\exp{\rm \left(\frac{5.29}{{\it T}_{ex}}\right)}\int  f_{\tau}(\nu)  \left(\frac{ T_B(v)}{\rm K}\right)  \left(\frac{\rm d \it v}{\rm km\ s^{-1}}\right)\ [\rm cm^{-2}] .
\label{N13}
\end{eqnarray}
In equation \ref{N13}, $f_{\tau}(\nu)$ is the correction factor for opacity, which is defined as
\begin{equation}
f_{\tau}(\nu)=\rm\frac{\tau_{\rm {}^{13}CO}(\nu)}{1-\exp(-\tau_{\rm {}^{13}CO}(\nu))}\ .
\end{equation} 
The mass in each pixel is given by
\begin{equation}
m_{\rm cl}=N_{\rm {}^{13}CO}\frac{\bar{\mu} m_{\rm H}}{\chi_{\rm {}^{13}CO}}{\left\{1.5\times10^{13} {\left(\frac{D}{\rm pc}\right)}{\left(\frac{\Delta\theta}{{\rm arcsec}}\right)}\right\}}^2\ [\rm g],
\end{equation}
where $\rm \chi_{{}^{13}CO}$ is the abundance of {$\rm {}^{13}CO$} relative to $\rm H_2$: $\rm \chi_{{}^{13}CO}=(1/67)\times 10^{-4}$. 
We estimate the cloud mass for all pixels with $ {}^{13}$CO emission above 5 $\sigma$, and we find that the total mass of each subregion, $M_{\rm cl}$ = 5.8, 7.1, 9.6, 5.0, and 4.5 $\times 10^{3}M_{\odot}$ for the OMC 2/3, OMC 1, OMC 4/5, L1641-N, and NGC 1999 subregions (see figure\ref{12}), respectively. 

We also estimate the turbulent energy of Orion A following the method presented by Li et al.\ (\yearcite{2015ApJS..219...20L}). 
The turbulent energy of the cloud of each pixel is given approximately by
\begin{equation}
e_{\rm turb}=\frac{1}{2}m_{\rm cl}\sigma^2_{\rm 3d},
\end{equation}
where $\sigma_{\rm 3d}$ is the ${}^{13}$CO three-dimensional turbulent velocity dispersion of each pixel, which is assumed to be 
\begin{equation}
\sigma_{\rm 3d}=\sqrt{3}\sigma_{v}.
\end{equation} 
The total turbulent energy of the cloud is then given by
\begin{equation}
\mathcal{E}_{\rm turb}=\sum{e_{\rm turb}}.
\end{equation}

\begin{table}[h!]\caption{Cloud parameters of Orion A}
	\begin{tabular}{ccccccc}\hline
		\label{cloudpara}
	Subregion &    $M_{\rm cl}$	&${}^{13}\rm CO\ \sigma_{v}$	&$\mathcal{E}_{\rm turb}$\\
			&$10^3\ M_{\odot}$	&\ km s$^{-1}$				&10$^{46}$\ erg\\\hline	
	OMC 2/3	& 5.8				&0.9						&13.0		\\
	OMC 4/5 	& 9.6				&1.3						&31.2		\\
	L1641-N	& 5.0				&1.5						&21.0		\\
	NGC 1999& 4.5			&1.2						&13.6		\\
 	OMC 1	& 7.1				&1.7						&28.2		\\\hline
	Total	& 32.0				&						&107.0		\\\hline
	\end{tabular}
\end{table}

\subsubsection{Position angle of outflows}\label{PA}

\begin{figure}[h!]
 \includegraphics[keepaspectratio, width=8cm]{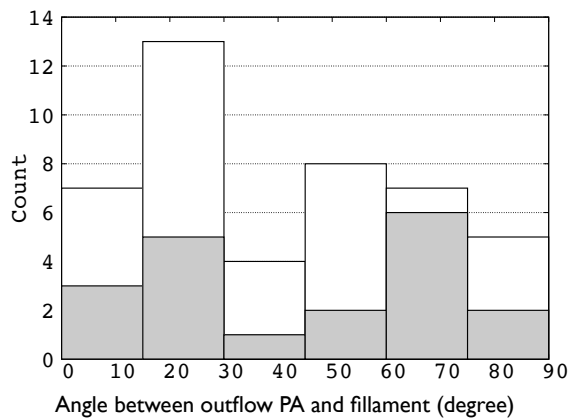}
 	\caption
	{
	Distribution of the observed angle between outflow position angle and the cloud filament (15$^\circ$ bins) for all outflows in Orion A (white filled) and for northern OMC 2/3 outflows (gray filled).
	} 
	\label{pa}
\end{figure}

Prototars formed in filamentary structures may have a characteristic outflow direction to a filament: parallel or perpendicular (Stephens et al. \yearcite{2017ApJ...846...16S}).  
According to Stephens et al. (\yearcite{2017ApJ...846...16S}), the projected angles between the outflows and filaments in the Perseus molecular cloud complex are not consistent with being either mostly parallel or perpendicular; they are consistent with being random or a mix of parallel and perpendicular. 
We also compare the position angles of the outflows to the direction of the cloud filament in Orion A. 
For simplification, we define the direction of the cloud filament on the plane of the sky as PA $=\ -20^{\circ}$ for $\delta$ $\geq -$\timeform{5D05'00".00} (J2000.0), PA $=\ 10^{\circ}$ for $-$\timeform{5D05'00".00} (J2000.0) $>$ $\delta$ $\geq -$\timeform{5D40'00".00} (J2000.0) and PA $=\ -20^{\circ}$ for $\delta$ $\leq -$\timeform{5D40'00".00} (J2000.0) (see figure \ref{out}), and we determine the relative  position angles of the outflows with respect to the cloud filament.  
The results are shown in figure \ref{pa}.
There is no clear significant difference between the north and the south of the clouds, and the results do not show any correlation between the position angles of the outflows and elongations of the cloud filaments.

\subsection{Comparison with previous outflow searches}

In this section, we compare our results of outflow search results with those of previous surveys. 

\subsubsection{OMC 2/3}
In OMC 2/3, ${}^{12}$CO\ ({\it J} = 1--0) outflow surveys were conducted by several authors: Aso et al. (\yearcite{2000ApJS..131..465A}) with the Nobeyama 45m telescope, Williams et al. (\yearcite{2003ApJ...591.1025W}) with the FCRAO 14m telescope and Berkeley-Illinois-Maryland Association (BIMA) array, and 
Shimajiri et al. (\yearcite{2008ApJ...683..255S}, \yearcite{2009PASJ...61.1055S}) with the Nobeyama Millimeter Array (NMA) in the OMC 2 FIR 3, 4, 5, and 6 regions. 
${}^{12}$CO\ ({\it J} = 2--1)  outflow surveys were conducted by Chini et al. (\yearcite{1997ApJ...474L.135C}) with the NRAO 12m telescope and Takahashi et al. (\yearcite{2008ApJ...688..344T}) with the Atacama Submillimeter Telescope Experiment (ASTE).
Takahashi et al. (\yearcite{2008ApJ...688..344T}) also observed the ${}^{12}$CO\ ({\it J} = 3--2) line.

All outflows detected by Chini et al. (\yearcite{1997ApJ...474L.135C}), Aso et al. (\yearcite{2000ApJS..131..465A}), and Williams et al. (\yearcite{2003ApJ...591.1025W}) are also detected in this study, except for an outflow associated with MMS 8.
The outflow from MMS 8 was detected by Williams et al. (\yearcite{2003ApJ...591.1025W}) as a bipolar outflow, while our map only reveals a blue lobe located around MMS 8.
We find a counter red lobe that is at the opposite side of the HOPS 75 protostar located at $\sim$\timeform{1'} south of MMS 8. 
We therefore identify these blue- and red-shifted lobes as a bipolar outflow associated with HOPS 75 (our No. 8). 
Two independent H$_2$ jets associated with MMS 11 and MMS 15 were detected by Davis et al. (\yearcite{2009A&A...496..153D}), and we identify two single-lobe outflows associated with MMS 11 (red-shifted lobe) and MMS 15 (blue-shifted lobe).
These blue- and red-shifted lobes were identified as a bipolar outflow associated with FIR 1 by Williams et al. (\yearcite{2003ApJ...591.1025W}) and Takahashi et al. (\yearcite{2008ApJ...688..344T}).
We detect a red-shifted single outflow associated with FIR 6c (No. 18). 
This outflow was clearly identified as bipolar outflows associated with FIR 6c by Takahashi et al. (\yearcite{2008ApJ...688..344T}) and Shimajiri et al. (\yearcite{2009PASJ...61.1055S}).
We detect 12 out of 14 outflows detected by Takahashi et al. (\yearcite{2008ApJ...688..344T}).
In the two outflows not detected by this study, the velocity range detected as an outflow in the (3--2) line is buried in ambient clouds in the (1--0) line.
We detect a blue-shifted lobe located at $\sim$\timeform{4'} north-west of FIR 6d (No. 17), but Shimajiri et al. (\yearcite{2009PASJ...61.1055S}) detected bipolar outflows associated with FIR 6d. 
The sizes of the blue-shifted lobes associated with FIR 6d detected in this study and by Shimajiri et al. (\yearcite{2009PASJ...61.1055S}) are different.
Outflow No. 2, which is a blue-shifted single-lobe outflow, was not detected in the previous searches.

While the outflow mass obtained in section 4.3, which is considers the correction for optical depth, is in good agreement with that found by Takahashi et al. (\yearcite{2008ApJ...688..344T}), 
the momentum of the outflows of No. 13 obtained in our study is one order of magnitude higher than that by Takahashi et al. (\yearcite{2008ApJ...688..344T}), respectively.
The total mass and momentum of outflows in OMC 2/3 in our sturdy are 107\% and 29\% of those determined by Takahashi et al. (\yearcite{2008ApJ...688..344T}).
These discrepancies are mainly from the differences in our estimation method.
Our estimations of mass and momentum of the outflows are based on the channel maps (details described in section \ref{properties}), 
while Takahashi et al. (\yearcite{2008ApJ...688..344T}) estimated the momentum of outflows by simply multiplying the total mass with the maximum velocity of outflows.
Thus, the momenta obtained by Takahashi et al. (\yearcite{2008ApJ...688..344T}) should be regarded as upper limits.

\subsubsection{OMC 4/5}
In OMC 4/5, no systematic search of outflows had been conducted previously.
Thus, all 11 outflows identified in this region are new detections.

\subsubsection{L1641-N}
Our results in the central region of L1641-N agree with the previous ${}^{12}$CO\ ({\it J} = 2--1) and ${}^{12}$CO\ ({\it J} = 1--0) observations conducted by Stanke and Williams (\yearcite{2007AJ....133.1307S}) and Nakamura et al. (\yearcite{2012ApJ...746...25N}).  
In addition, we find three new outflows in this region.
Outflow No. 33 is a blue-shifted single-lobe outflow associated with IRS 50, categorized as a PMS star with H$_2$ jets located in the central region of L1641-N.
No. 37 and No. 38 are single-lobe outflows located $\sim$\timeform{15'} south of the central region of L1641-N. 

\subsubsection{NGC 1999}
In the NGC 1999 region, we detect a well-collimated CO outflow associated with V380 Ori-NE, which had already been identified by previous searches (e.g., Davis et al. \yearcite{2000MNRAS.318..952D} (${}^{12}$CO\ ({\it J} = 3--2) and ({\it J} = 4--3)) and Choi et al. \yearcite{2017ApJS..232...24C} (SiO ($v$=0, {\it J}=1--0))).
Our results for the outflow surveys around HH 1/2 in NGC 1999 agree with the ${}^{12}$CO\ ({\it J} = 1--0) and ({\it J} = 2--1) surveys conducted by Moro-Mart{\'{\i}}n et al. (\yearcite{1999ApJ...520L.111M}).  
This study reveals that a faint blue-shifted emission located  $\sim$\timeform{2'} east of HH 1/2 is an outflow driven by IRS 121 (No. 43). 
In addition, we find a new outflow (No. 44) in this region. 

The masses of the outflows associated with V380 Ori-NE and HH 1/2 are consistent with those determined by Davis et al. (\yearcite{2000MNRAS.318..952D}) and Moro-Mart{\'{\i}}n et al. (\yearcite{1999ApJ...520L.111M}), respectively.

\section{Discussion}

\subsection{Feedback of outflows in Orion A}

With our sample of the outflows of Orion A, we estimate the dynamical effect of the outflows on the molecular clouds. 
Table \ref{total} compares the kinetic energy of the cloud turbulence to the energy ejected by the outflows in each subregion. 
It should be noted that in table \ref{total}, each outflow parameter is corrected for the inclination angle to be
57.3$^{\circ}$, which is the average value for the case of random inclination angles (see section \ref{properties1}).     
The total energy of the outflows is only $\sim$1.9\% of the turbulent energy of the cloud, excluding OMC 1.
This is similar to the Taurus outflows reported by Li et al.\ (\yearcite{2015ApJS..219...20L}).
In the Taurus molecular cloud, the energy of all outflows detected by Li et al.\ (\yearcite{2015ApJS..219...20L}) is $\sim$1.2\% of the turbulent energy of the cloud.

We also compare the momentum of the outflows with that of the cloud turbulence, $\mathcal{P}_{\rm turb}=(2M_{\rm cl}\mathcal{E}_{\rm turb})^{1/2}$, and the results are listed in table \ref{total}. 
The total momentum of the outflows is $\sim$0.4\% of the cloud turbulence.

Besides molecular outflows, molecular shells may be another driver of turbulence in parent molecular clouds (Churchwell et al. \yearcite{2006ApJ...649..759C}).
Molecular shells show expanding spherical structures of molecular gas and are driven by protostars or intermediate-mass stars (e.g., see Arce at al. \yearcite{2011ApJ...742..105A} and Offner \& Arce \yearcite{2015ApJ...811..146O}). 
In Orion A, Feddersen et al. (\yearcite{2018ApJ...862..121F}) identified molecular shells (see figure \ref{13}) and estimated that the total energy and momentum ejection rate of the shells, excluding OMC 1, were 36.7 $\times$ 10$^{33}$ erg s$^{-1}$ and 36.3 $\times$ 10$^{-3}\ M_{\odot}$ km s$^{-1}$ yr$^{-1}$, respectively.   
Those results are comparable with the energy and momentum ejection rate of the outflows.

We also investigate whether the outflows and shells have sufficient energy and momentum to maintain the turbulence in the cloud.
We estimate the turbulent dissipation rate as
\begin{equation}
\dot{\mathcal{E}}_{\rm turb}=0.5\frac{\mathcal{E}_{\rm turb}}{t_{\rm diss}},
\end{equation}
where the factor of 0.5 is derived from Mac Low (equation 8; \yearcite{1999ApJ...524..169M}), and $t_{\rm diss}$ is the turbulent dissipation time which is estimated by the following method.
With the cloud size ${R_{\rm cl}}$, the dissipation time of turbulence is  given by
\begin{equation}
t_{\rm diss}=\frac{R_{\rm cl}}{\bar{\sigma}_{v}},
\label{tdiss}
\end{equation} 
where $\bar{\sigma}_{v}$ is the ${}^{13}$CO averaged one-dimensional velocity dispersion at each subregion (McKee \& Ostriker \yearcite{2007ARA&A..45..565M}) (see fig \ref{sp}).
For $R_{\rm cl}$, the half thicknesses of the filaments are estimated to be 1.0, 1.3, 1.2, 1.4, and 1.4 pc for OMC 2/3, OMC 1, OMC 4/5, L1641-N, and NGC 1999 subregions, respectively, from the 5$\sigma$ contour in the integrated intensity map of the ${}^{12}$CO.
Accordingly, equation \ref{tdiss} gives us $t_{\rm diss}$ = 1.1, 0.7, 0.9, 0.8, and 1.1 $\times$ 10$^{6}$ yr, respectively, 
and the turbulent dissipation rates are then estimated to be ${\dot{\mathcal{E}}_{\rm turb}}$ = 1.9, 6.4, 5.5, 4.2, and 2.0 $\times$ 10$^{33}$ erg s$^{-1}$, respectively.
Table \ref{total} compares the dissipation rate of the cloud turbulence to the energy ejection rate of the outflows in each subregion. 
Excluding OMC 1, the total energy ejection rate of the outflows is approximately 235\% of the total dissipation rate of the cloud turbulence.
The momentum dissipation rate of the cloud is described as
\begin{equation}
\dot{\mathcal{P}}_{\rm turb}=0.6\frac{\mathcal{P}_{\rm turb}}{t_{\rm diss}},
\end{equation}
where the factor of 0.6 is derived from Mac Low (equation 8; \yearcite{1999ApJ...524..169M}).
According to this formula, we estimated the momentum dissipation rate of each subregion and compared them with each momentum ejection rate of the outflows. 
The results are listed in table \ref{total}.
Excluding OMC 1, the total momentum ejection rate of the outflows is approximately 36\% of the momentum dissipation rate of the cloud turbulence. 

Note that the Orion cloud complex is assumed to be located on the edge of a large-scale expanding bubble called ``the Orion-Eridanus bubble'' (e.g., see \textcolor{red}{Reynolds et al. \yearcite{1979ApJ...229..942R}, Bally et al. \yearcite{2008hsf1.book..459B},} Wilson et al. \yearcite{2005A&A...430..523W} and Pon et al. \yearcite{2014MNRAS.444.3657P}), and its contributions to the cloud kinematics, including turbulence, may also have to considered.
However, in this paper, we omit these large-scale effects, and we simply estimate $t_{\rm diss}$ from the cloud crossing time using the speed of sound.

The total energies and momentum ejection rates of the outflows and shells are respectively 5.1 and 1.6 times larger than those of the cloud turbulence. 
This means that 20\% and 60\% of the energy and momentum, respectively, must be converted into cloud turbulence to
compensate for their dissipations. 
These values seem high, but the conversion efficiencies are highly uncertain,
and whether the outflows and shells can drive the molecular cloud turbulence still remains an open question.
If the conversion efficiencies are smaller than these values, the outflows and shells in Orion A cannot maintain the cloud turbulence, 
and other agents such as s large-scale HI bubble \textcolor{red}{(its energy is estimated to be 3.7$\times$10$^{51}$ erg by Brown et al. \yearcite{1995A&A...300..903B})}, may be responsible for maintaining the turbulence  \textcolor{red}{(Pon et al. \yearcite{2014MNRAS.444.3657P})}.   
However even in such a case, the outflows may play an important role in dispersing dense cores at smaller scales. 
As seen in Figure 5, the outflows are preferentially distributed in the ridge of the molecular clouds, suggesting that the outflows including high velocity jets may have a significant impact on the turbulent gas motions in the densest but small portion of molecular clouds.

\begin{table}\caption{Energy, momentum, and ejection rate of outflows and cloud turbulence}\label{total}
	\begin{tabular}{ccccccccccccccc}\hline
		Subregion&$\mathcal{E}_{\rm turb}$&$\mathcal{E_{\rm flow}}$\footnotemark[*]&&$\mathcal{P}_{\rm turb}$&$\mathcal{P_{\rm flow}}$\footnotemark[*]&&$\dot{\mathcal{E}}_{\rm turb}$&$\dot{\mathcal{E}_{\rm flow}}$\footnotemark[$*$]&$\dot{\mathcal{E}}_{\rm shell}$\footnotemark[$\dag$]&&$\dot{\mathcal{P}}_{\rm turb}$&$\dot{\mathcal{P}_{\rm flow}}$\footnotemark[*]&$\dot{\mathcal{P}}_{\rm shell}$\footnotemark[$\dag$]\\

	&\multicolumn{2}{c}{10$^{46}$erg}&&\multicolumn{2}{c}{$10^{3}M_{\odot}\rm\ km\ s^{-1}$}&&\multicolumn{3}{c}{10$^{33}$erg s$^{-1}$}&&\multicolumn{3}{c}{$10^{-3}M_{\odot}\rm\ km\ s^{-1}\ yr^{-1}$}\\\hline
	OMC 2/3		& 13.0	&0.80				&&8.7		&0.093	&&1.9	&17.4						&4.0		&&4.7		&6.1						&6.0\\
	OMC 4/5		& 31.2	&0.19				&&17.4		&0.019	&&5.5	&4.3							&26.8	&&11.6		&1.2						&22.1\\
	L1641-N		& 21.0	&0.18				&&10.3		&0.017	&&4.2	&5.0							&3.5		&&7.7		&1.4						&4.5\\
	NGC 1999	& 13.6	&0.34				&&7.8		&0.027	&&2.0	&5.3							&2.6		&&4.3		&1.5						&3.7\\\hline
	Subtotal		& 78.8	&1.51				&&44.2		&0.156 	&&13.6	&32.0						&36.9	&&28.3		&10.2					&36.3\\\hline
 	OMC 1		& 28.2	&15\footnotemark[$\ddag$]&&14.2		&		&&6.4	&4000\footnotemark[$\ddag$]		&7.2		&&12.2		&566\footnotemark[$\ddag$]	&8.6\\\hline
	\end{tabular}
	\begin{tabnote}
	\footnotemark[*]Corrected for the inclination angle.\\
	\footnotemark[$\dagger$]Feddersen et al. (\yearcite{2018ApJ...862..121F}).\\ 
	\footnotemark[$\ddagger$]These values are from Snell et al. (\yearcite{1984ApJ...284..176S}) ${}^{12}$CO\ ({\it J} = 1--0), Kwan \& Scoville (\yearcite{1976ApJ...210L..39K}) ${}^{12}$CO\ ({\it J} = 1--0), and Zapata et al. (\yearcite{2005ApJ...630L..85Z}) ${}^{12}$CO\ ({\it J} = 2--1). 
	In OMC 1, the massive velocity outflows ($\sim100\rm\ km\ s ^{-1}$) in Orion KL account for a large fraction of their energy and momentum. 
	The outflows in OMC 1 have sufficient energy and momentum to maintain cloud turbulence in OMC 1.\\
	\end{tabnote}
\end{table}

\section{Summary}

We conducted mapping observations of the main 2 deg$^2$ regions of Orion A by ${}^{12}$CO\ ({\it J} = 1--0) and ${}^{13}$CO\ ({\it J} = 1--0) using the Nobeyama 45-m telescope to investigate the outflow feedback into the parent molecular clouds.
The main results of this study are as follows:  

\begin{enumerate}

 \item Based on a systematic procedure, we identified 44 ${}^{12}$CO outflows associated with the {\it Spitzer} YSOs in Orion A, and of these, 17 are new detections.
 
 \item The ratio of brightness temperature of the ${}^{12}$CO to ${}^{13}$CO lines suggests that the optical depth of the ${}^{12}$CO is $\sim$5 in the detected outflows. 
 
 \item We estimated the kinematic properties of the detected outflows. 
 The momentum and energy of each outflow range from 4.0 $\times$ 10$^{-2}$ $M_{\odot}$ km s$^{-1}$ to 1.7 $\times$ 10 $M_{\odot}$ km s$^{-1}$ and 1.9 $\times$ 10$^{42}$ erg to 7.8 $\times$ 10$^{44}$ erg, respectively.
  The total momentum and energy of the outflows are 1.6 $\times$ 10$^2$ $\ M_{\odot}$ km s$^{-1}$ and 1.5 $\times$ 10$^{46}$ erg, respectively.
 
 \item We compared the momenta and energy ejection rates of the outflows and shells with the momentum and energy dissipation rate of the cloud turbulence. 
The total momentum and energy ejection rate of the outflows are 36\% and 235\% of those of the cloud turbulence. 
The total momentum and energy ejection rate of the shells in Orion A are estimated to be 128\%  and 271\% of those of the cloud turbulence. 
 The total momentum and energy ejection rates of the outflows and shells in Orion A are 1.6 and 5.1 times larger than the momentum and energy dissipation rates of the cloud turbulence. 
The cloud turbulence cannot be sustained by the outflows and shells unless the efficiencies of energy and momentum conversion are as high as 20\% and 60\%, respectively.

 \end{enumerate}

\bigskip
\begin{ack}
This study was conducted as part of a large projects by the Nobeyama Radio Observatory (NRO), which is a branch of the National Astronomical Observatory of Japan, National Institute of Natural Sciences. We thank the NRO staff for both operating the 45 m instrument and helping us with the data reduction. 
This study was supported by NAOJ ALMA Scientific Research Grant Number 2017-04A.
Part of this study was supported by a Grant-in-Aid for Scientific Research of Japan (JP17H02863).
TT was supported by JSPS KAKENHI Grant Number JP17K14244.
YS received support from the ANR (project NIKA2SKY, grant agreement ANR-15-CE31-0017).  
MM was supported by JSPS KAKENHI Grant Numbers JP17H01103 and JP18H05441.
\end{ack}

\appendix　　　

\section*{Previously known outflows}

In the Appendix, we show the integrated images and P--V diagrams of the known outflows.

\begin{figure}[h!]
 \includegraphics[keepaspectratio,width=16cm]{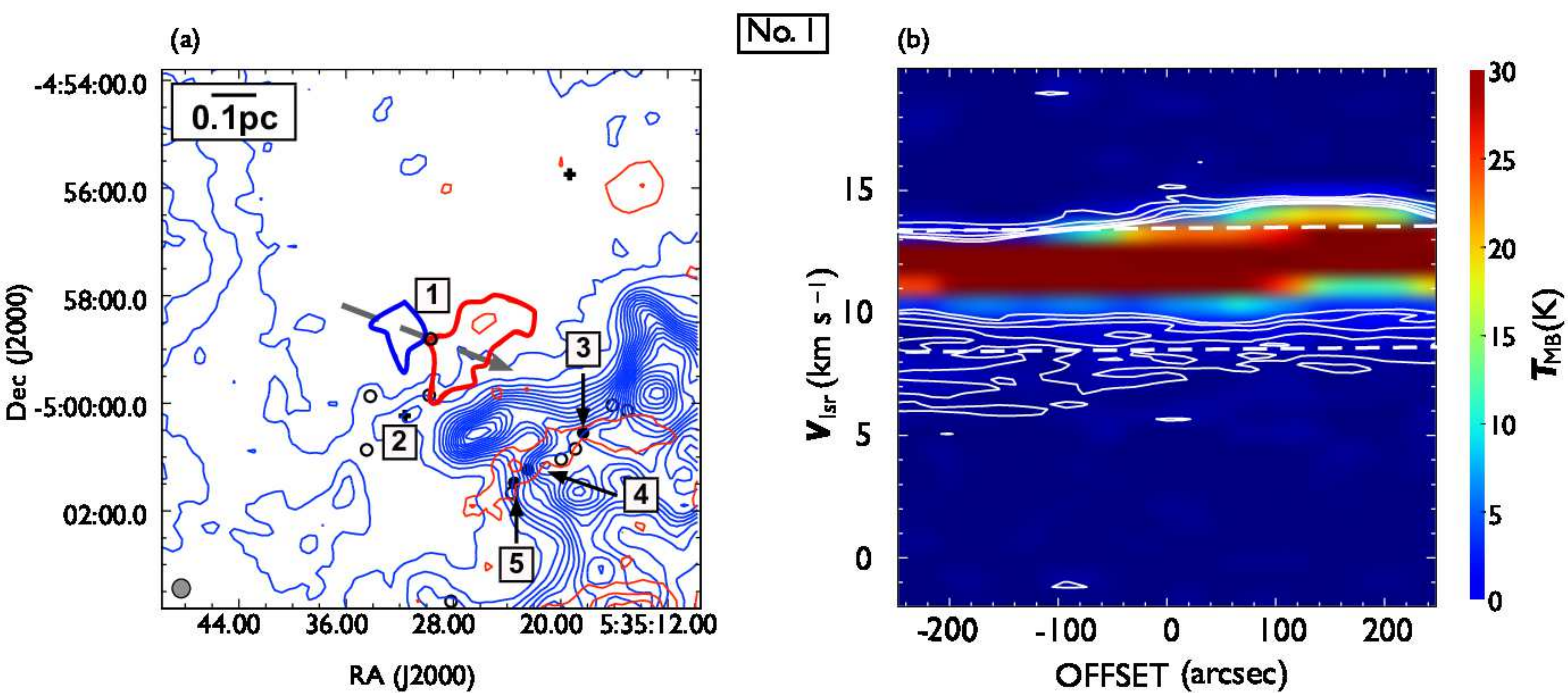}
	\caption{
	The same as in figure \ref{1} but for outflow No. 1. In panel (a), the blue- and red-shifted integrated intensity velocity ranges are 0.0 km s$^{-1}$ to 8.2 km s$^{-1}$ and 13.2 km s$^{-1}$ to 20.2 km s$^{-1}$, respectively.}
\end{figure}

\begin{figure}[h!]
 \includegraphics[keepaspectratio,width=16cm]{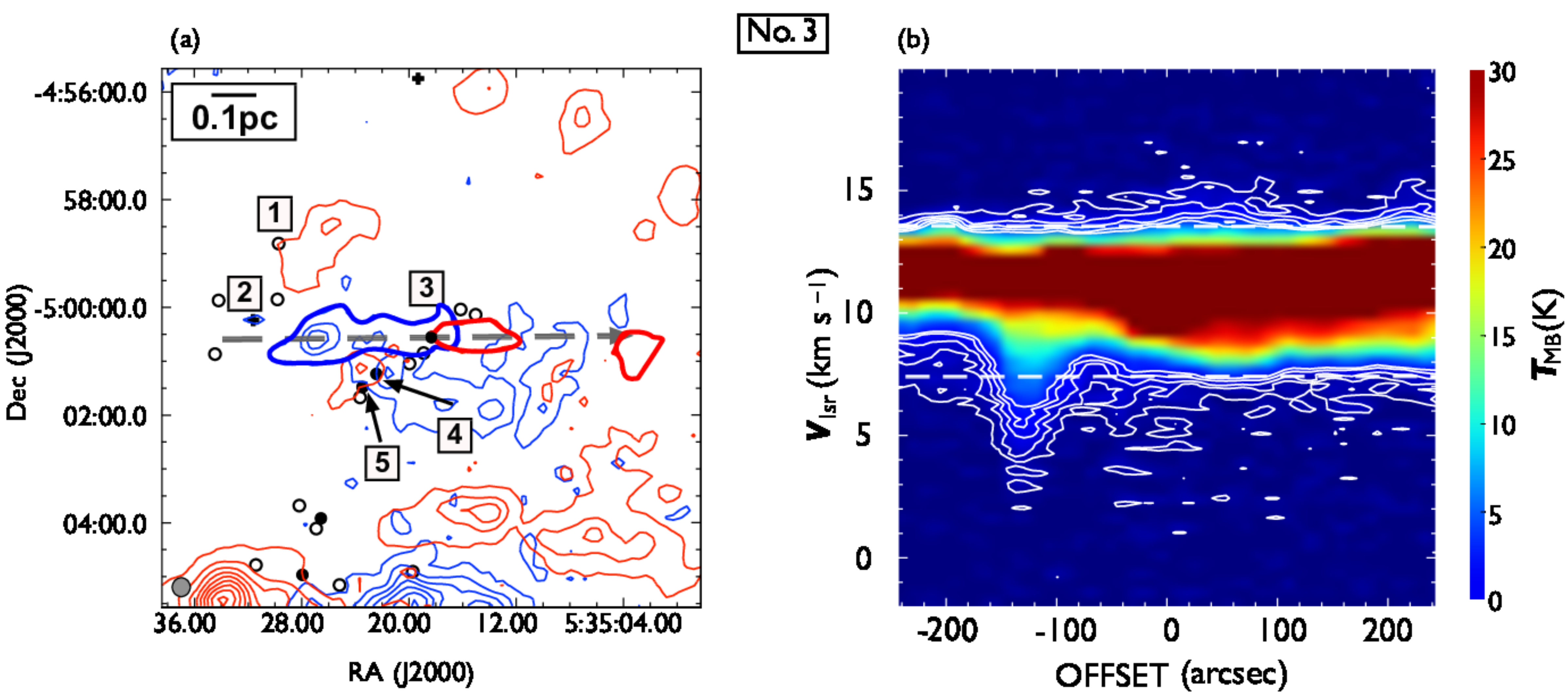}
	\caption{
	The same as in figure \ref{1} but for outflow No. 3. In panel (a), the blue- and red-shifted integrated intensity velocity ranges are 0.0 km s$^{-1}$ to 7.4 km s$^{-1}$ and 13.6 km s$^{-1}$ to 20.2 km s$^{-1}$, respectively.}
	\label{3b}
\end{figure}

\begin{figure}[h!]
 \includegraphics[keepaspectratio,width=16cm]{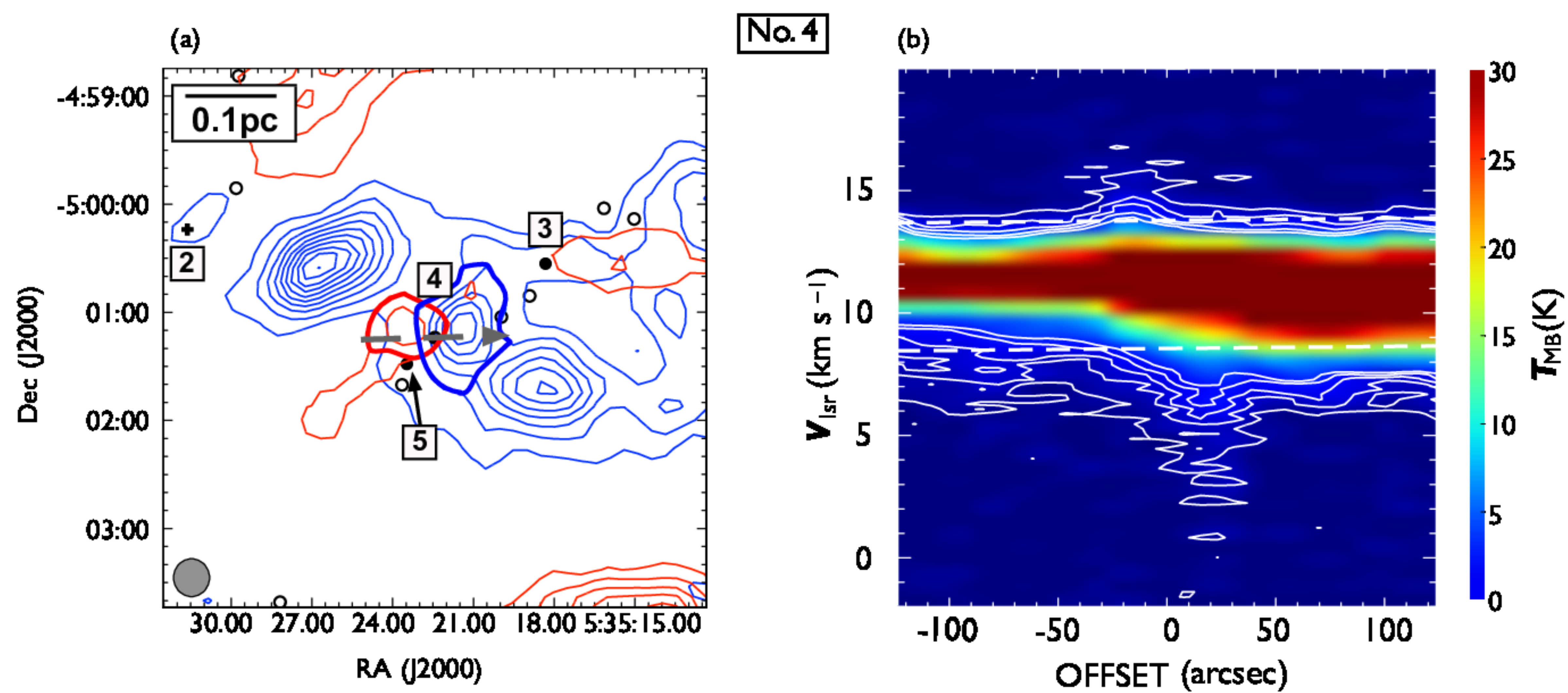}
	\caption{
	The same as in figure \ref{1} but for outflow No. 4. In panel (a), the blue- and red-shifted integrated intensity velocity ranges are 0.0 km s$^{-1}$ to 8.2 km s$^{-1}$ and 13.8 km s$^{-1}$ to 20.2 km s$^{-1}$, respectively.}
\end{figure}

\begin{figure}[h!]
 \includegraphics[keepaspectratio,width=16cm]{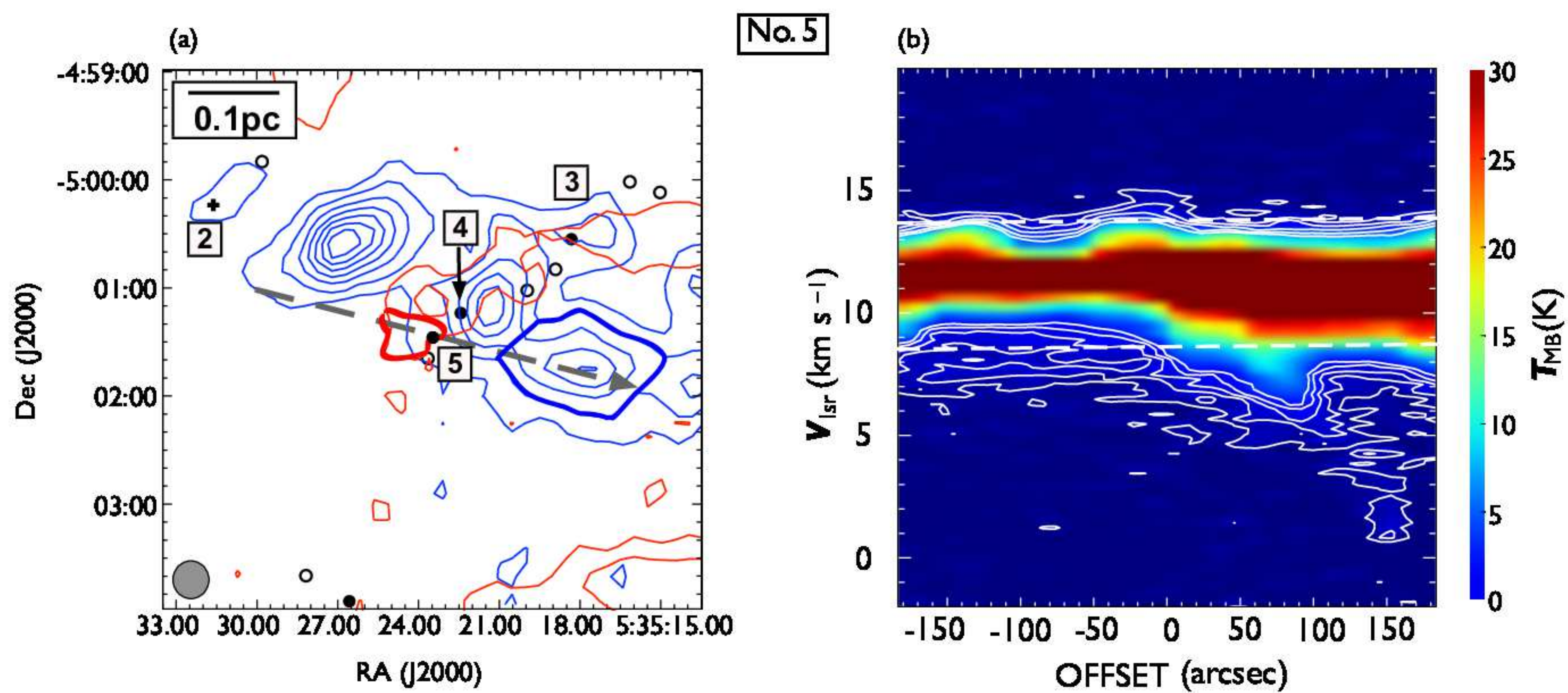}
	\caption{
	The same as in figure \ref{1} but for outflow No. 5. In panel (a), the blue- and red-shifted integrated intensity velocity ranges are 0.0 km s$^{-1}$ to 8.6 km s$^{-1}$ and 14.0 km s$^{-1}$ to 20.2 km s$^{-1}$, respectively.}
	\label{5b}
\end{figure}

\begin{figure}[h!]
 \includegraphics[keepaspectratio,width=16cm]{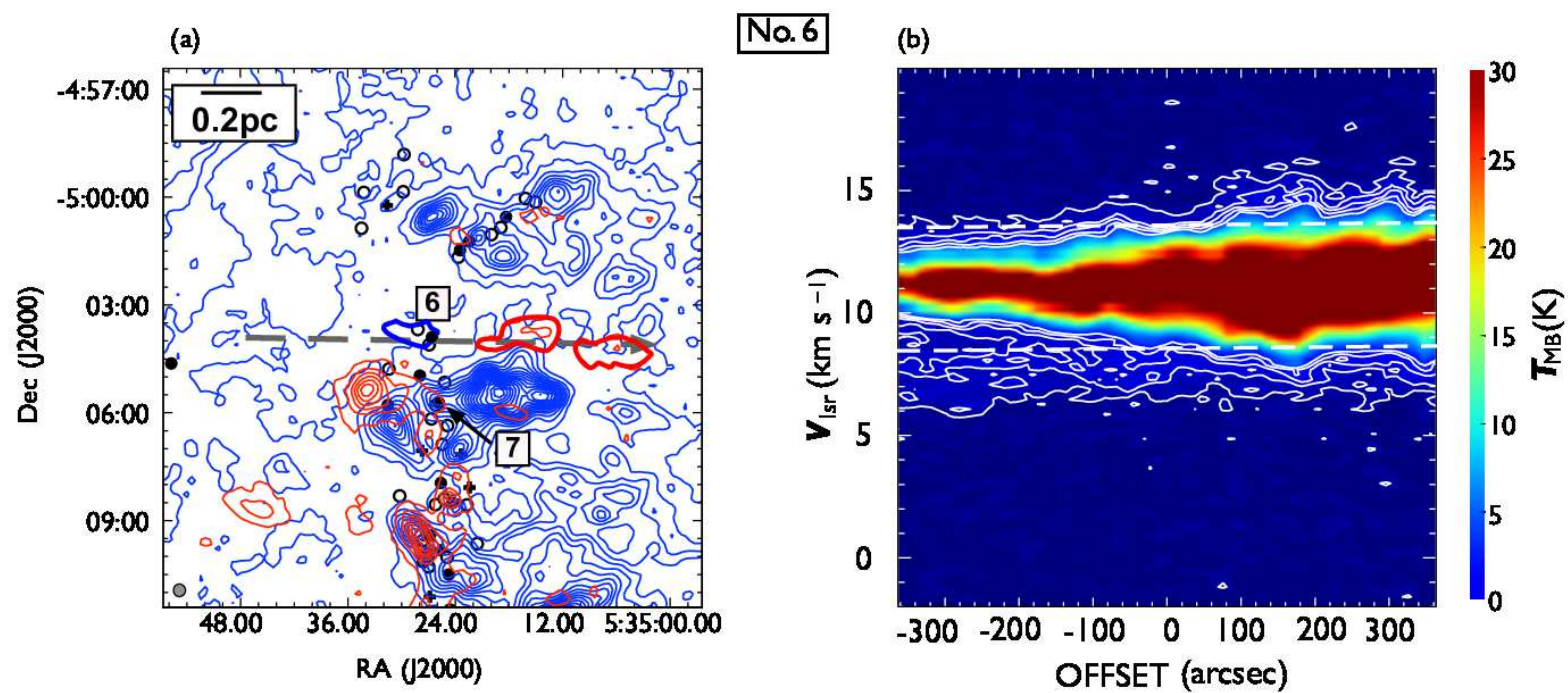}
	\caption{
	The same as in figure \ref{1} but for outflow No. 6. In panel (a), the blue- and red-shifted integrated intensity velocity ranges are 0.0 km s$^{-1}$ to 8.5 km s$^{-1}$ and 13.5 km s$^{-1}$ to 20.2 km s$^{-1}$, respectively.}
\end{figure}

\begin{figure}[h]
 \includegraphics[keepaspectratio,width=16cm]{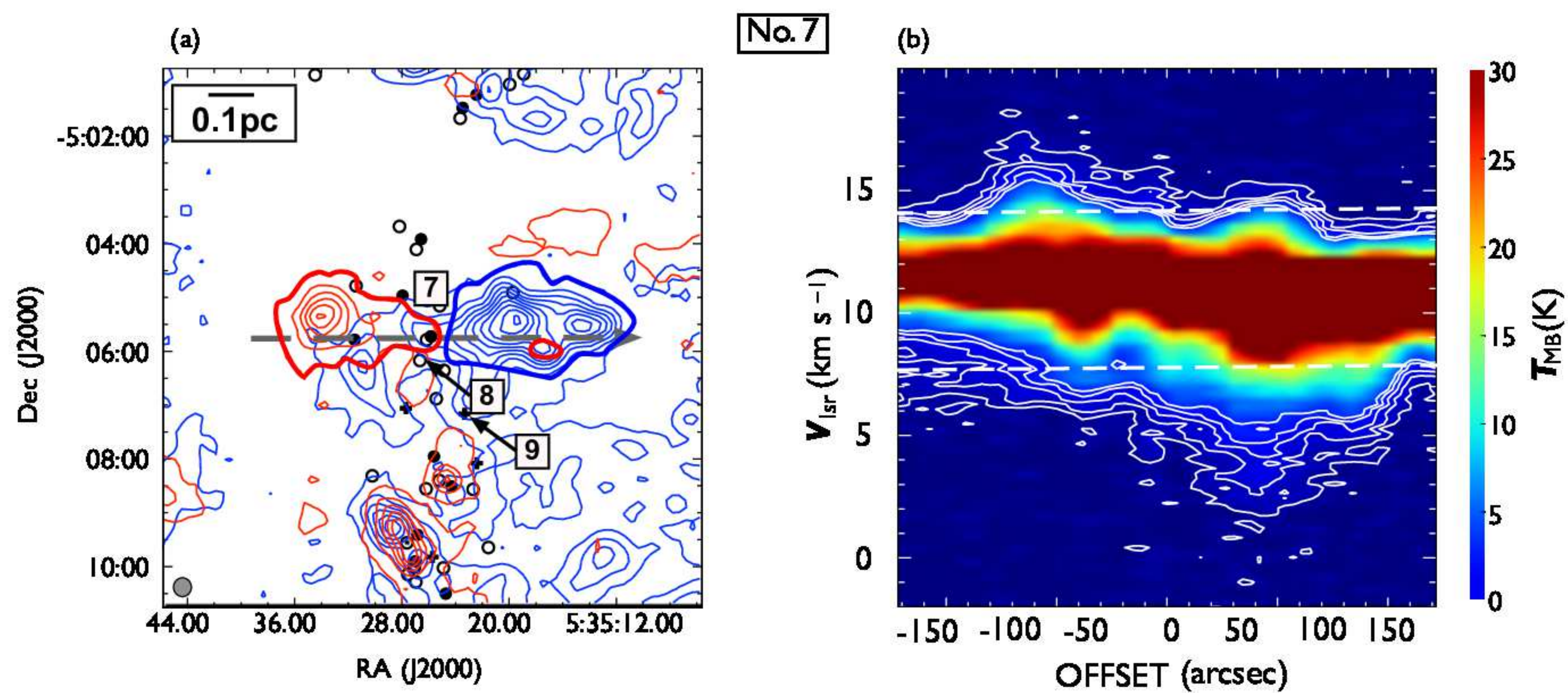}
	\caption{
	The same as in figure \ref{1} but for outflow No. 7. In panel (a), the blue- and red-shifted integrated intensity velocity ranges are 0.0 km s$^{-1}$ to 7.8 km s$^{-1}$ and 14.1 km s$^{-1}$ to 20.2 km s$^{-1}$, respectively.}
\end{figure}

\begin{figure}[h]
 \includegraphics[keepaspectratio,width=16cm]{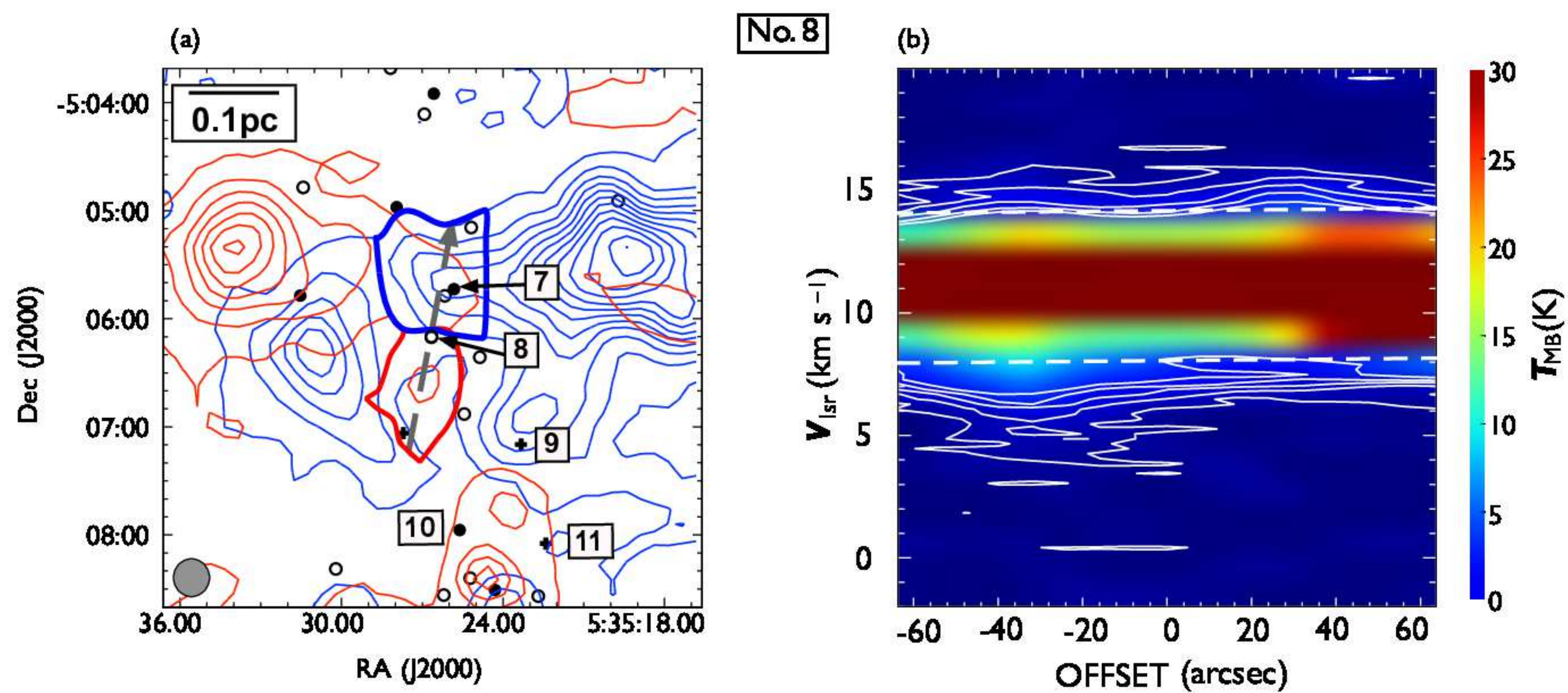}
	\caption{
	The same as in figure \ref{1} but for outflow No. 8. In panel (a), the blue- and red-shifted integrated intensity velocity ranges are 0.0 km s$^{-1}$ to 8.1 km s$^{-1}$ and 14.1 km s$^{-1}$ to 20.2 km s$^{-1}$, respectively.}
\end{figure}

\begin{figure}[h]
 \includegraphics[keepaspectratio,width=16cm]{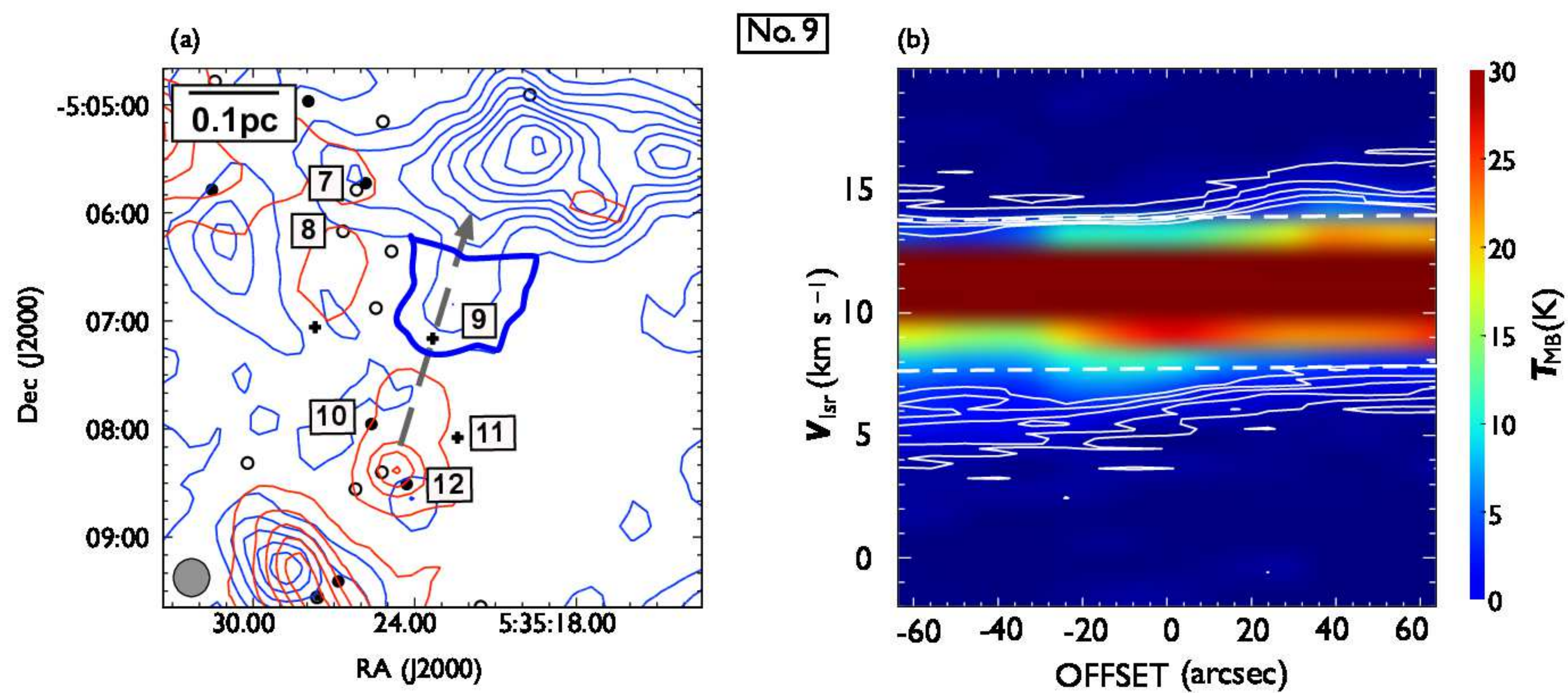}
	\caption{
	The same as in figure \ref{1} but for outflow No. 9. In panel (a), the blue- and red-shifted integrated intensity velocity ranges are 0.0 km s$^{-1}$ to 7.6 km s$^{-1}$ and 13.8 km s$^{-1}$ to 20.2 km s$^{-1}$, respectively.}
\end{figure}

\begin{figure}[h]
 \includegraphics[keepaspectratio,width=16cm]{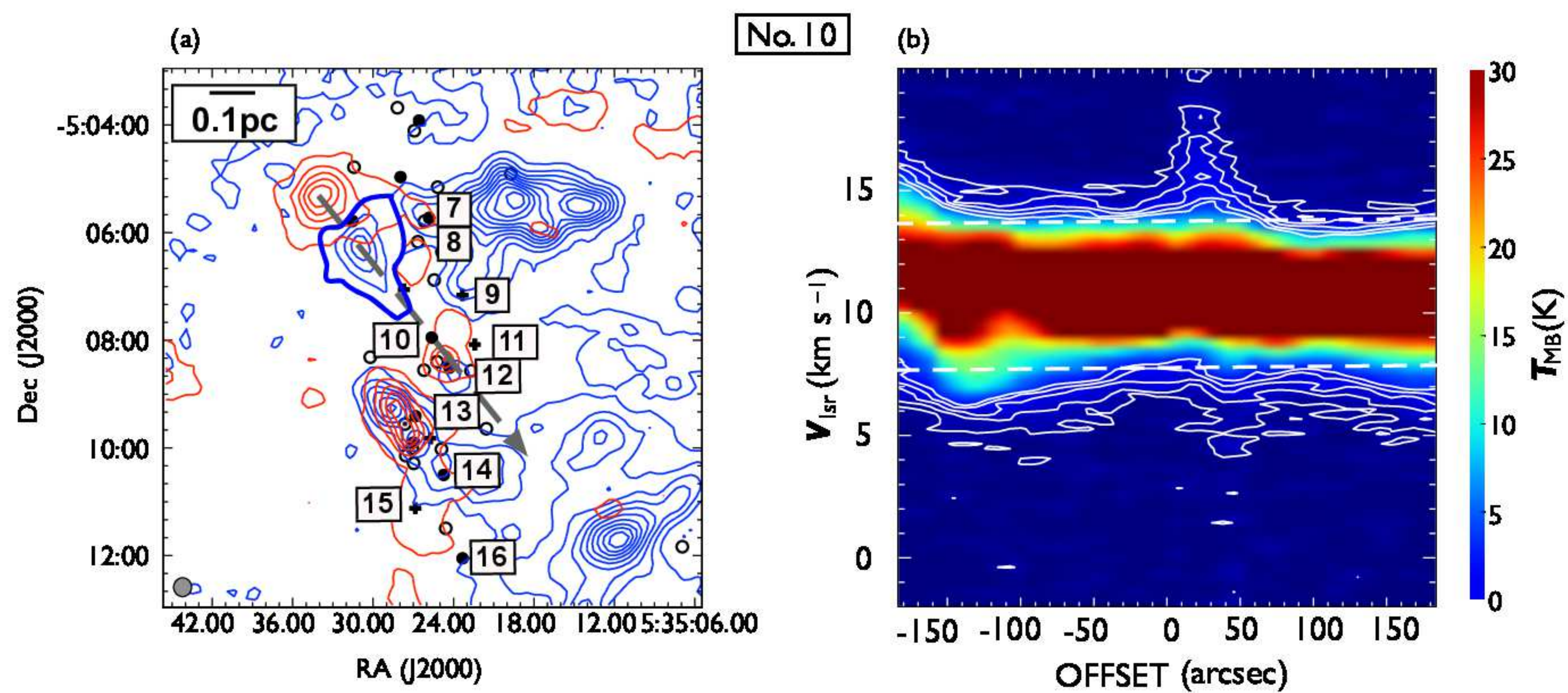}
	\caption{
	The same as in figure \ref{1} but for outflow No. 10. In panel (a), the blue- and red-shifted integrated intensity velocity ranges are 0.0 km s$^{-1}$ to 7.8 km s$^{-1}$ and 13.7 km s$^{-1}$ to 20.2 km s$^{-1}$, respectively.}
\end{figure}

\begin{figure}[h]
 \includegraphics[keepaspectratio,width=16cm]{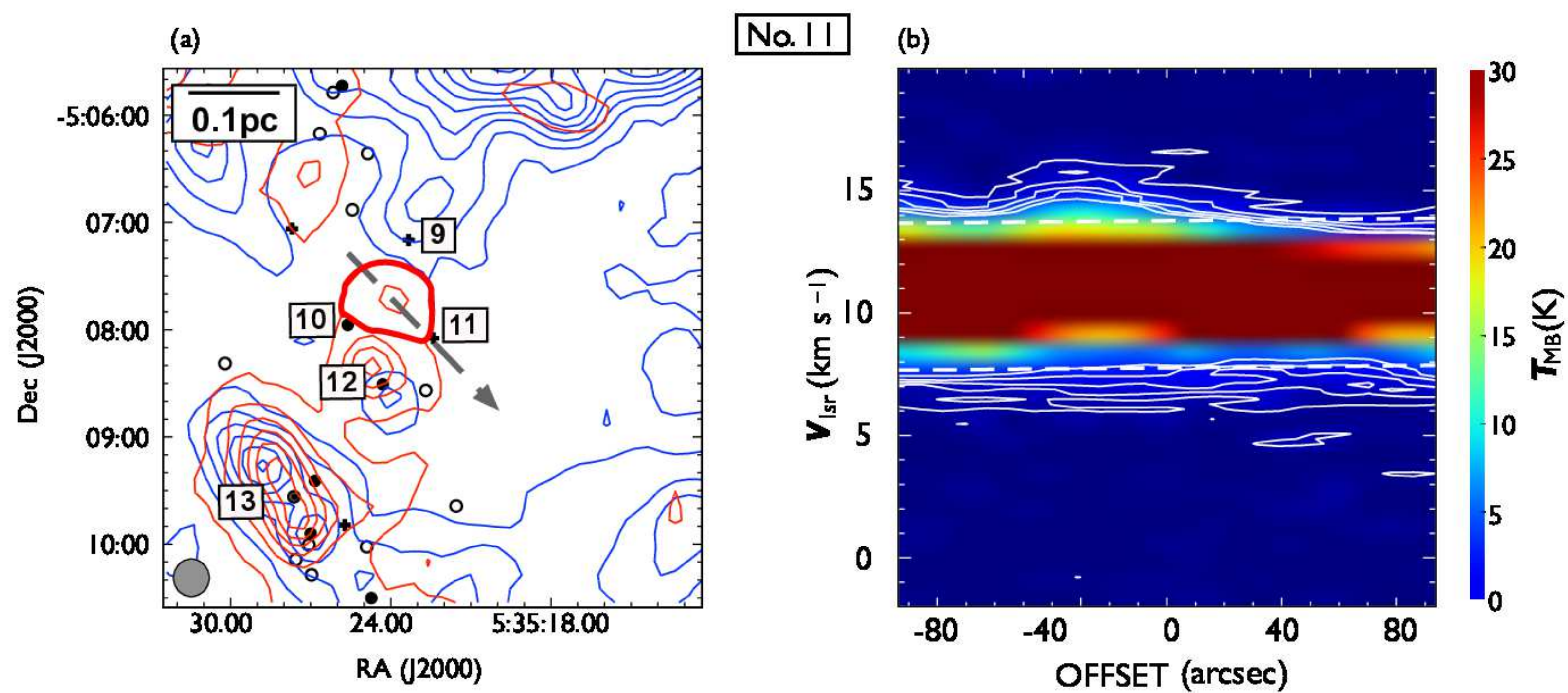}
	\caption{
	The same as in figure \ref{1} but for outflow No. 11. In panel (a), the blue- and red-shifted integrated intensity velocity ranges are 0.0 km s$^{-1}$ to 7.7 km s$^{-1}$ and 13.8 km s$^{-1}$ to 20.2 km s$^{-1}$, respectively.}
\end{figure}

\begin{figure}[h]
 \includegraphics[keepaspectratio,width=16cm]{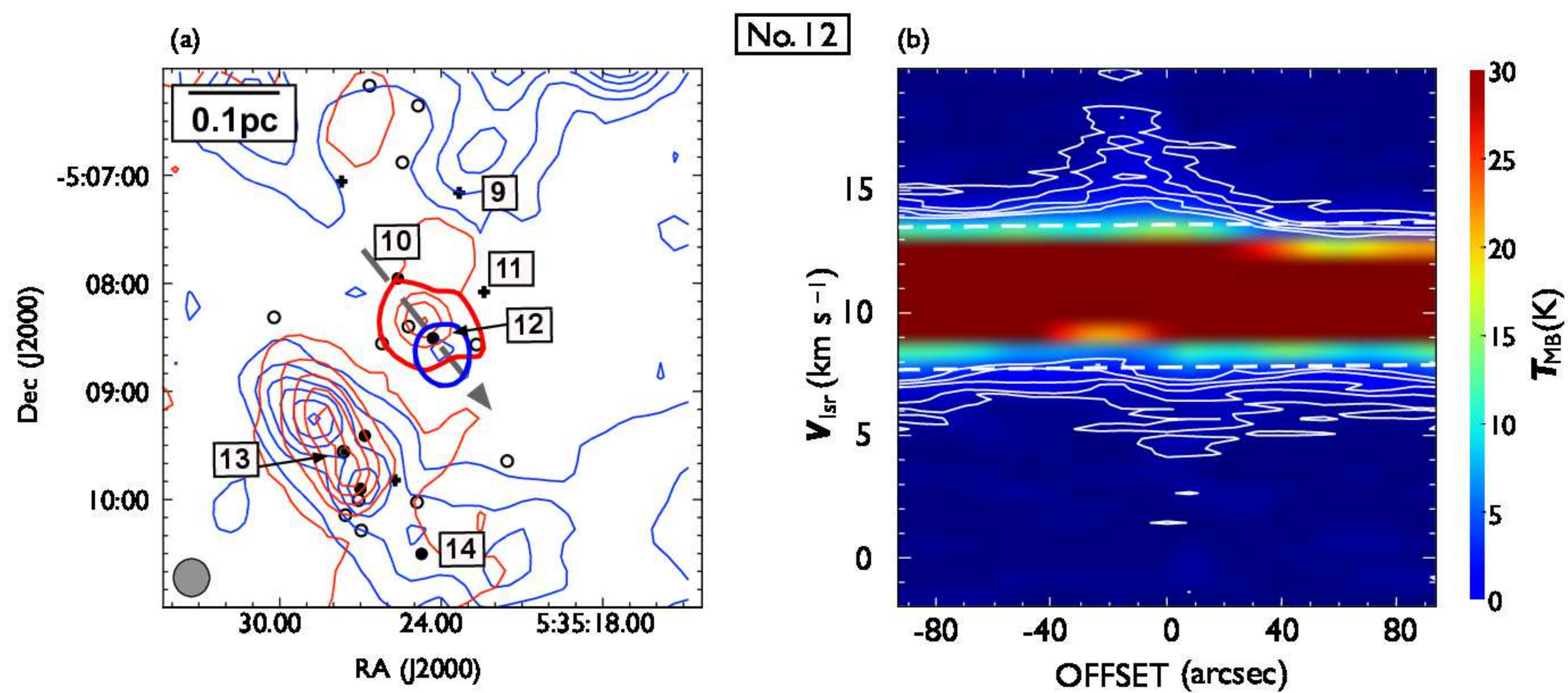}
	\caption{
	The same as in figure \ref{1} but for outflow No. 12. In panel (a), the blue- and red-shifted integrated intensity velocity ranges are 0.0 km s$^{-1}$ to 7.8 km s$^{-1}$ and 13.5 km s$^{-1}$ to 20.2 km s$^{-1}$, respectively.}
\end{figure}

\begin{figure}[h]
 \includegraphics[keepaspectratio,width=16cm]{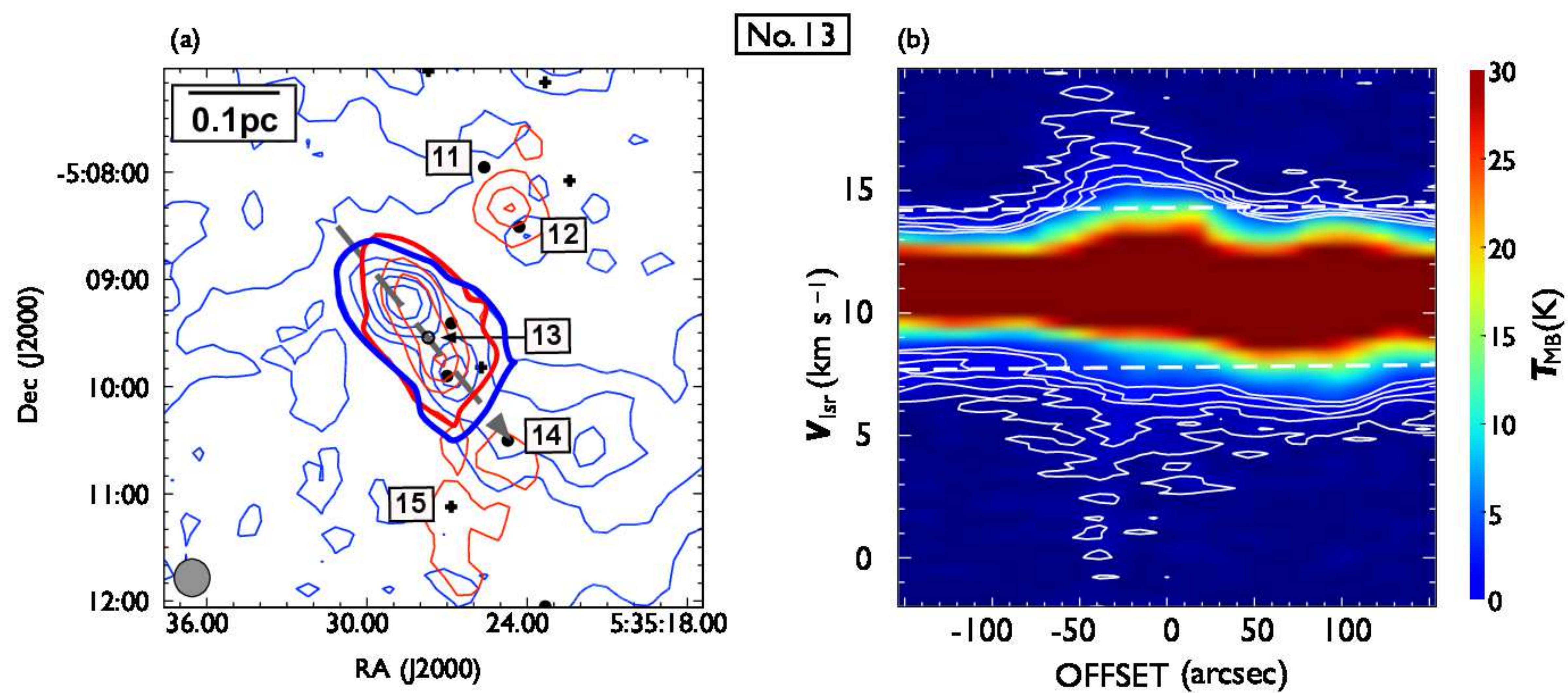}
	\caption{
	The same as figure \ref{1}, but for outflow No. 13. In panel (a), the blue- and red-shifted integrated intensity velocity ranges are 0.0 km s$^{-1}$ to 7.7 km s$^{-1}$ and 14.2 km s$^{-1}$ to 20.2 km s$^{-1}$, respectively.}
\end{figure}

\begin{figure}[h]
 \includegraphics[keepaspectratio,width=16cm]{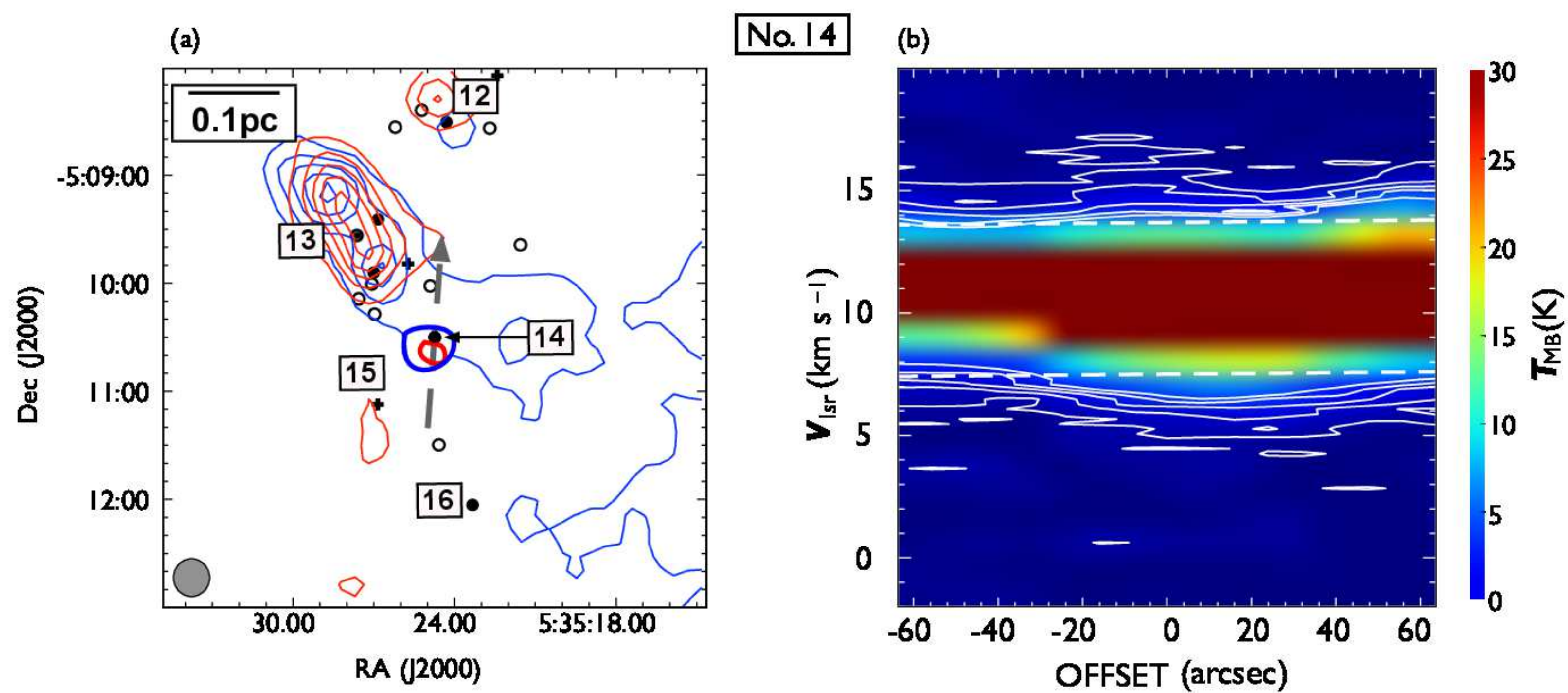}
	\caption{
	The same as in figure \ref{1} but for outflow No. 14. In panel (a), the blue- and red-shifted integrated intensity velocity ranges are 0.0 km s$^{-1}$ to 7.3 km s$^{-1}$ and 13.6 km s$^{-1}$ to 20.2 km s$^{-1}$, respectively.}
\end{figure}

\begin{figure}[h]
 \includegraphics[keepaspectratio,width=16cm]{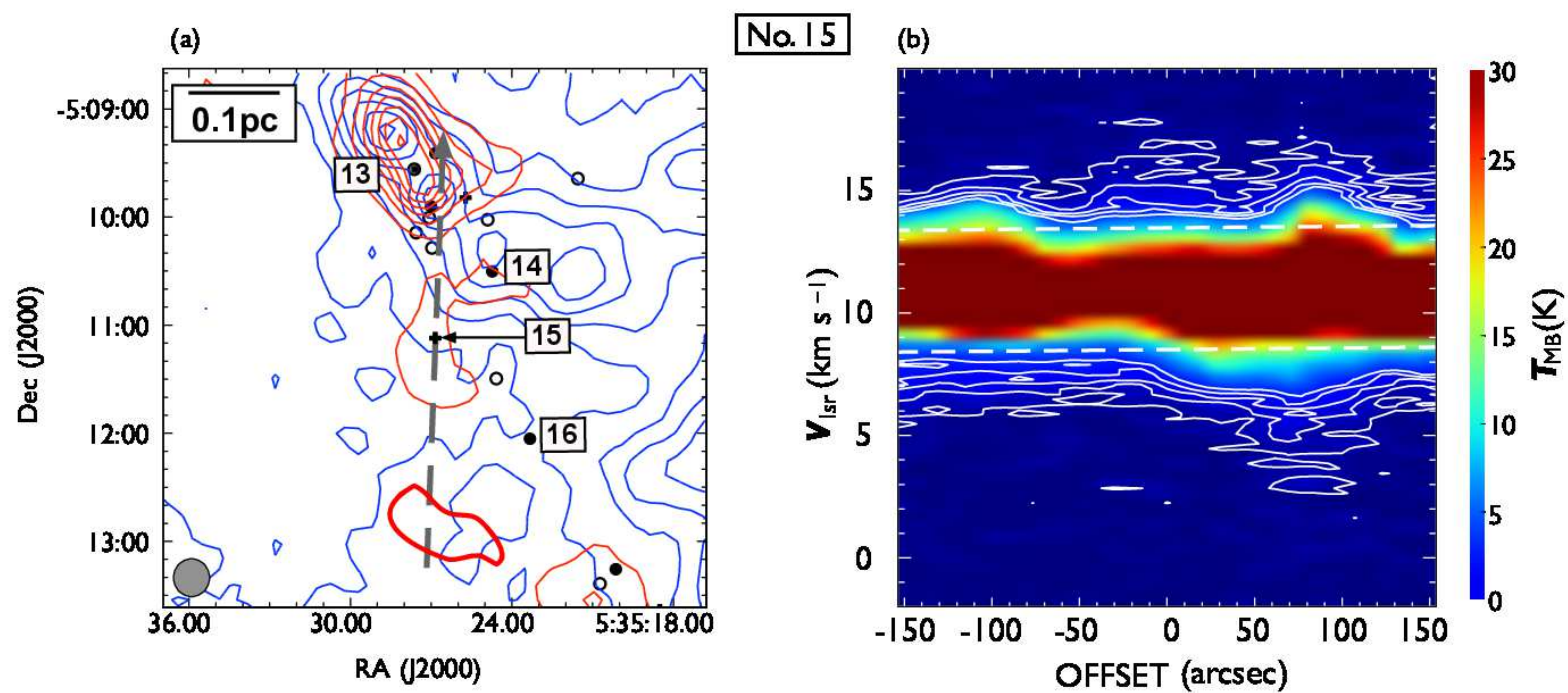}
	\caption{
	The same as in figure \ref{1} but for outflow No. 15. In panel (a), the blue- and red-shifted integrated intensity velocity ranges are 0.0 km s$^{-1}$ to 8.4 km s$^{-1}$ and 13.3 km s$^{-1}$ to 20.2 km s$^{-1}$, respectively.
	The red-shifted emission south of the central source is considered to be part of outflow No. 16.}
\end{figure}

\begin{figure}[h]
 \includegraphics[keepaspectratio,width=16cm]{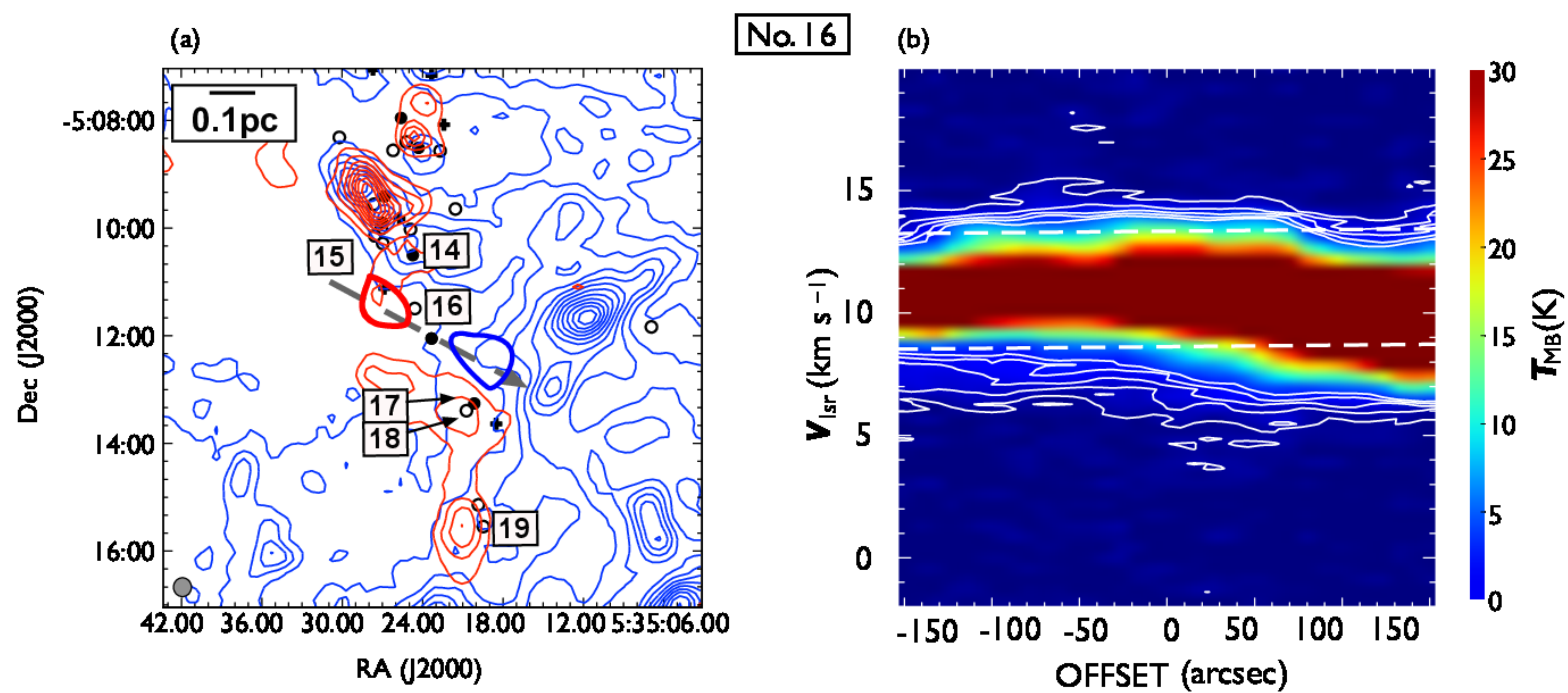}
	\caption{
	The same as in figure \ref{1} but for outflow No. 16. In panel (a), the blue- and red-shifted integrated intensity velocity ranges are 0.0 km s$^{-1}$ to 8.3 km s$^{-1}$ and 13.2 km s$^{-1}$ to 20.2 km s$^{-1}$, respectively.}
\end{figure}

\begin{figure}[h]
 \includegraphics[keepaspectratio,width=16cm]{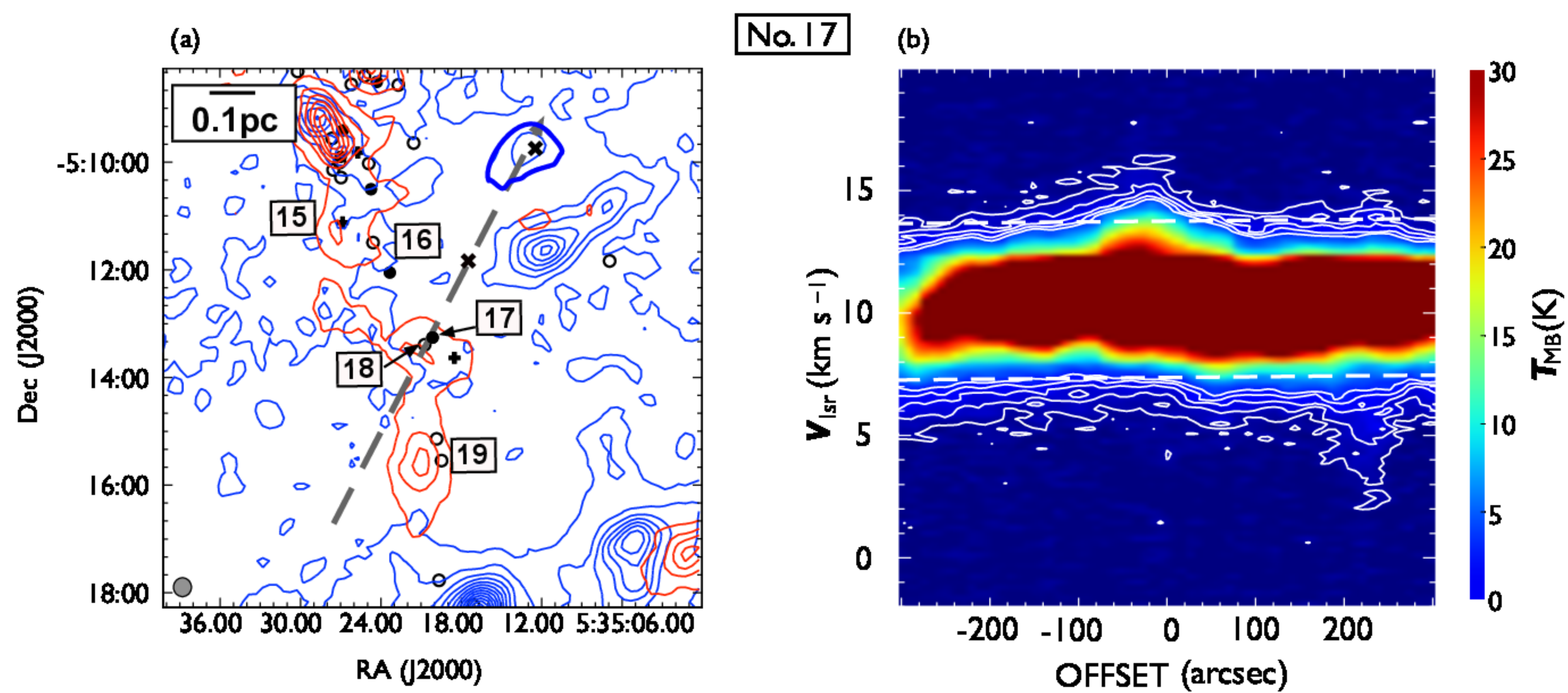}
	\caption{
	The same as in figure \ref{1} but for outflow No. 17. In panel (a), the blue- and red-shifted integrated intensity velocity ranges are 0.0 km s$^{-1}$ to 7.2 km s$^{-1}$ and 13.6 km s$^{-1}$ to 20.2 km s$^{-1}$, respectively.
	The black cross represents the location of the emission peak of H$_2$ jets detected by Davis et al. (\yearcite{2009A&A...496..153D}).
	The position angle of outflows represented by the arrow in panel (a) and that of the H$_2$ jet described by Davis et al. (\yearcite{2009A&A...496..153D}) are in agreement.
	}
\end{figure}

\begin{figure}[h]
 \includegraphics[keepaspectratio,width=16cm]{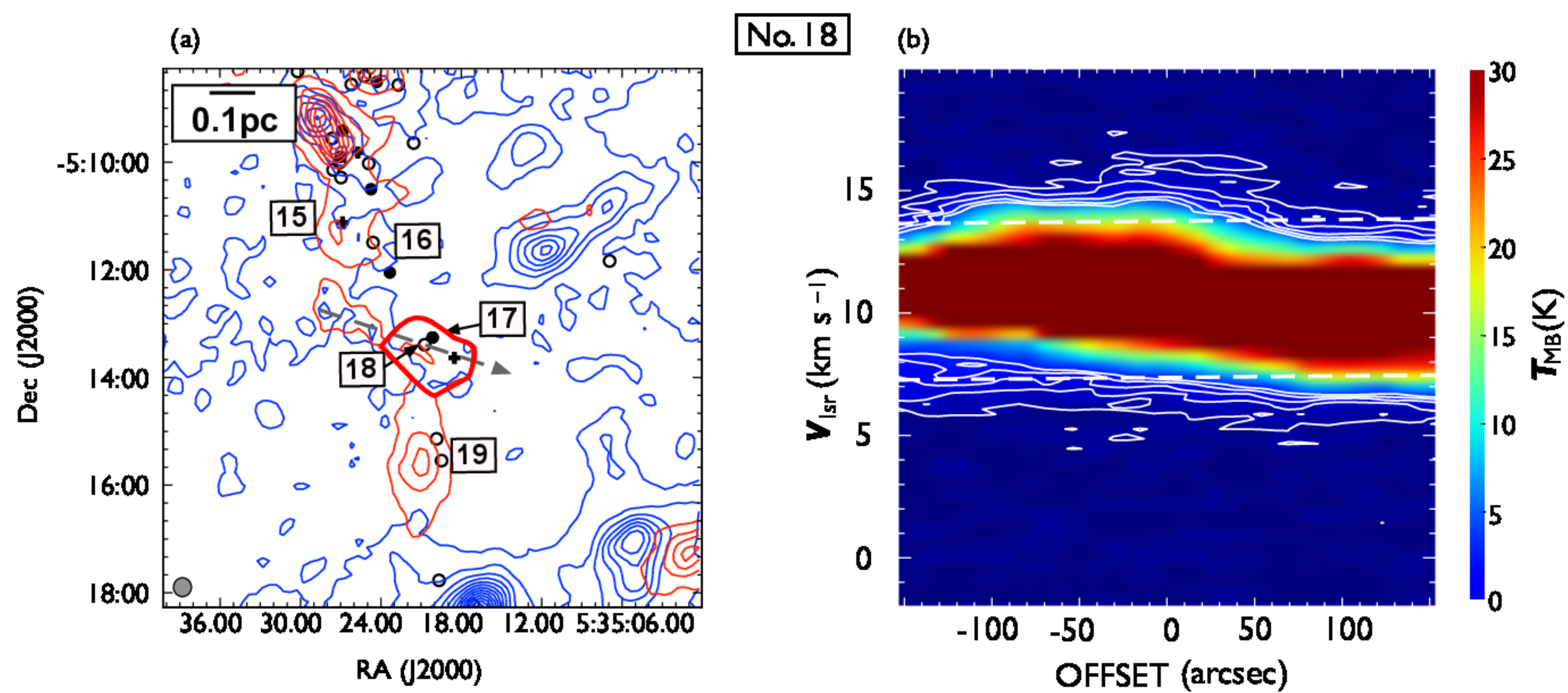}
	\caption{
	The same as in figure \ref{1} but for outflow No. 18. In panel (a), the blue- and red-shifted integrated intensity velocity ranges are 0.0 km s$^{-1}$ to 7.2 km s$^{-1}$ and 13.6 km s$^{-1}$ to 20.2 km s$^{-1}$, respectively.}
\end{figure}

\begin{figure}[h]
 \includegraphics[keepaspectratio,width=16cm]{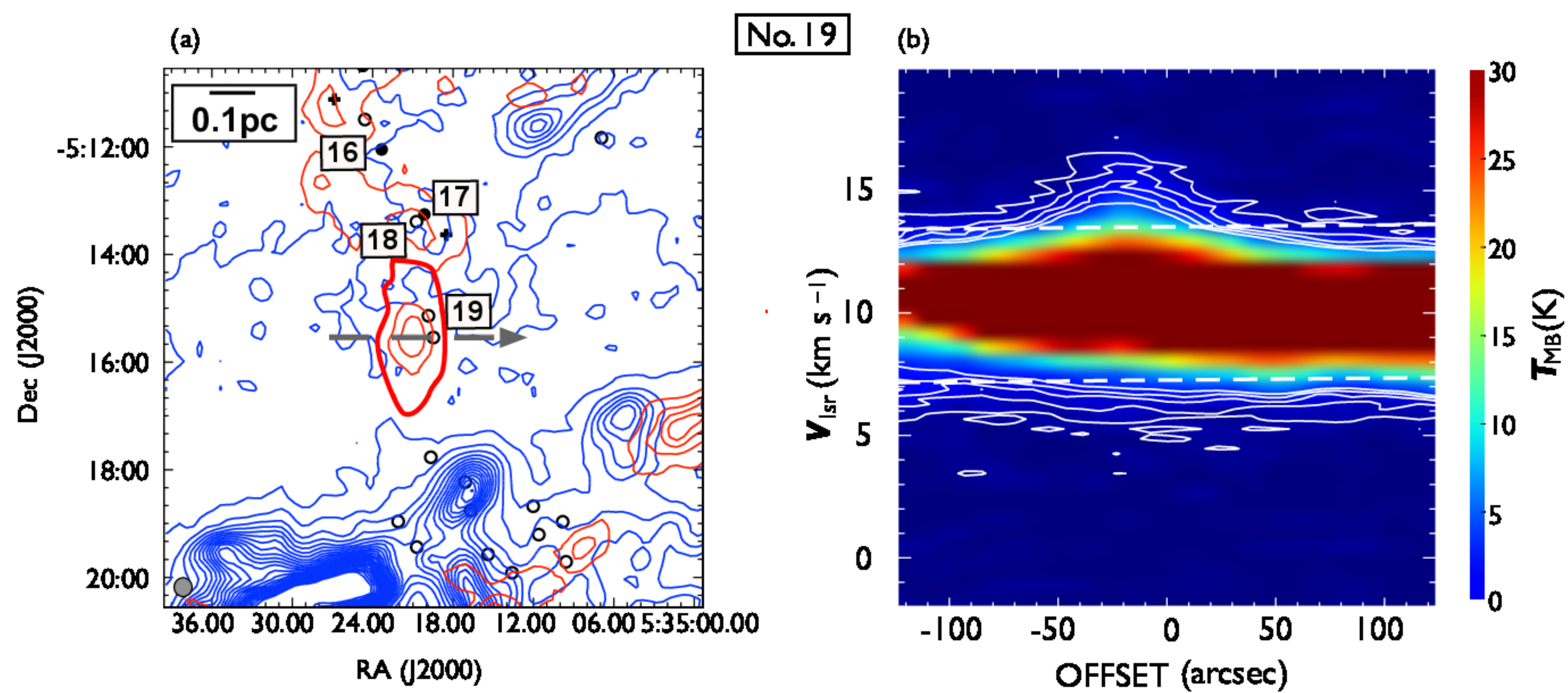}
	\caption{
	The same as in figure \ref{1} but for outflow No. 19. In panel (a), the blue- and red-shifted integrated intensity velocity ranges are 0.0 km s$^{-1}$ to 7.1 km s$^{-1}$ and 13.5 km s$^{-1}$ to 20.2 km s$^{-1}$, respectively.}
\end{figure}

\begin{figure}[h]
 \includegraphics[keepaspectratio,width=16cm]{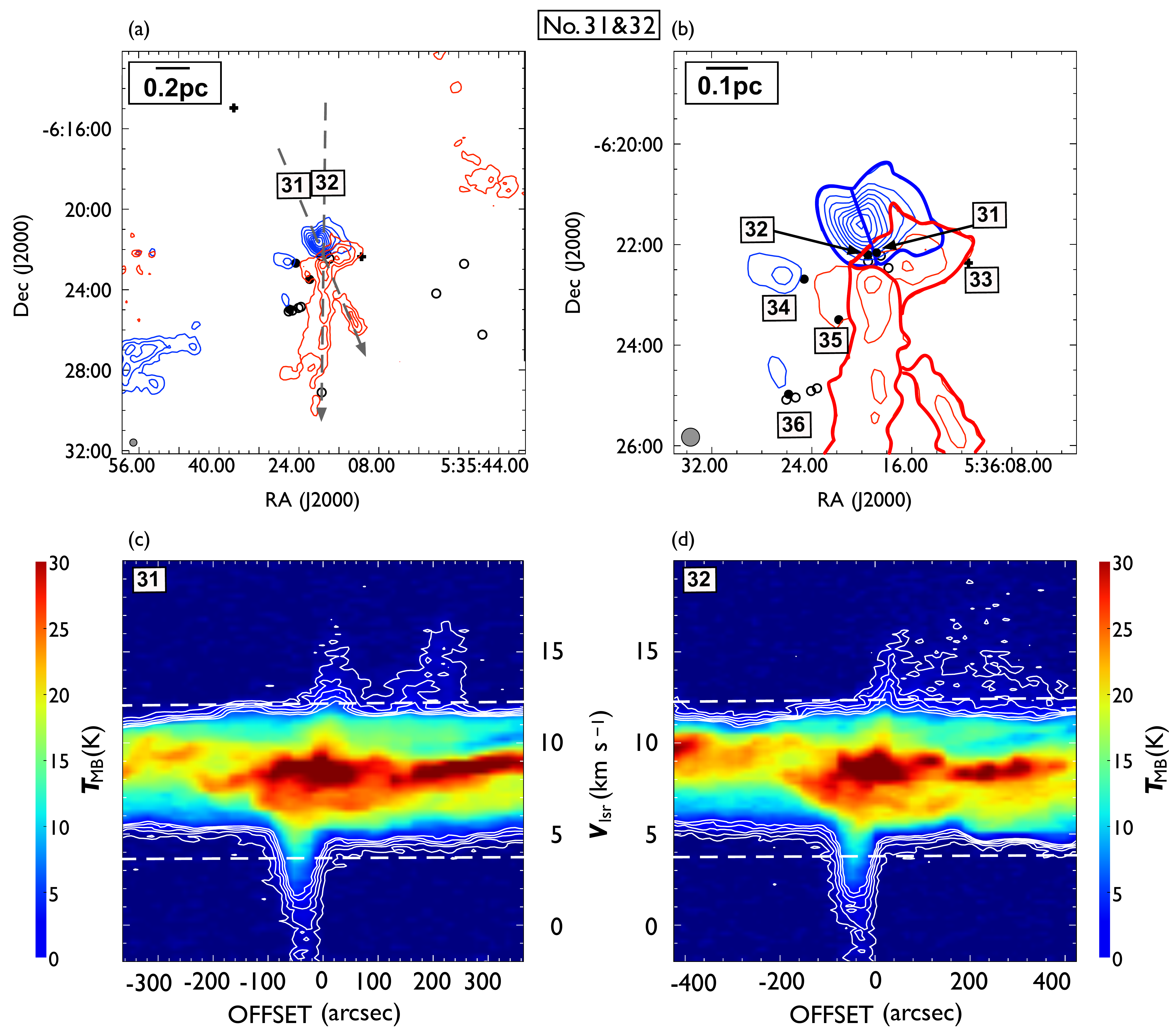}
	\caption{
	The same as in figure \ref{1} but for outflow No. 31 and 32. In panel (a), the blue- and red-shifted integrated intensity velocity ranges are -1.9 km s$^{-1}$ to 3.8 km s$^{-1}$ and 12.3 km s$^{-1}$ to 20.2 km s$^{-1}$, respectively.
	Panel (b) is a close up view of the center of panel (a).
	Panel (c) and (d) are the P--V diagrams of No. 31 and No. 32, respectively.}
\end{figure}

\begin{figure}[h]
 \includegraphics[keepaspectratio,width=16cm]{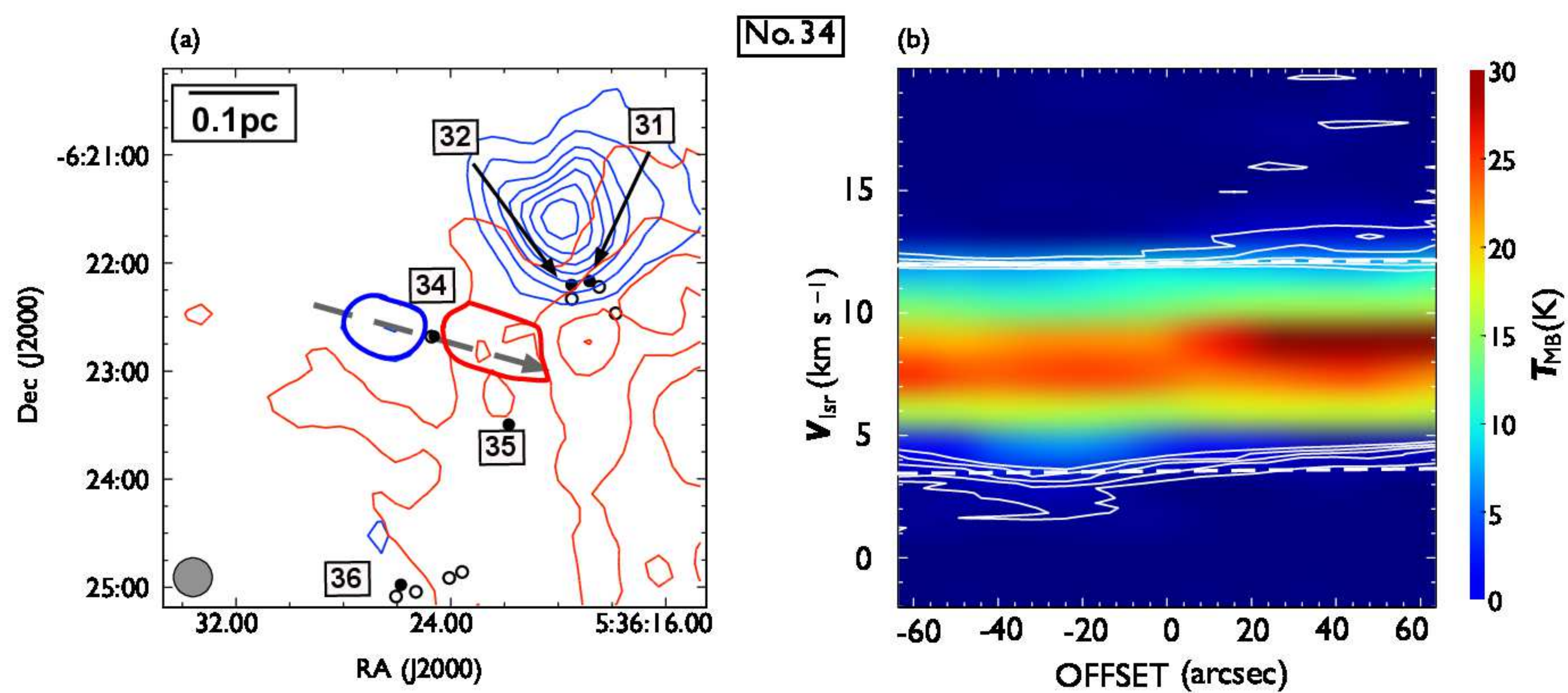}
	\caption{
	The same as in figure \ref{1} but for outflow No. 34. In panel (a), the blue- and red-shifted integrated intensity velocity ranges are -1.9 km s$^{-1}$ to 3.4 km s$^{-1}$ and 12.0 km s$^{-1}$ to 20.2 km s$^{-1}$, respectively.}
\end{figure}

\begin{figure}[h]
 \includegraphics[keepaspectratio,width=16cm]{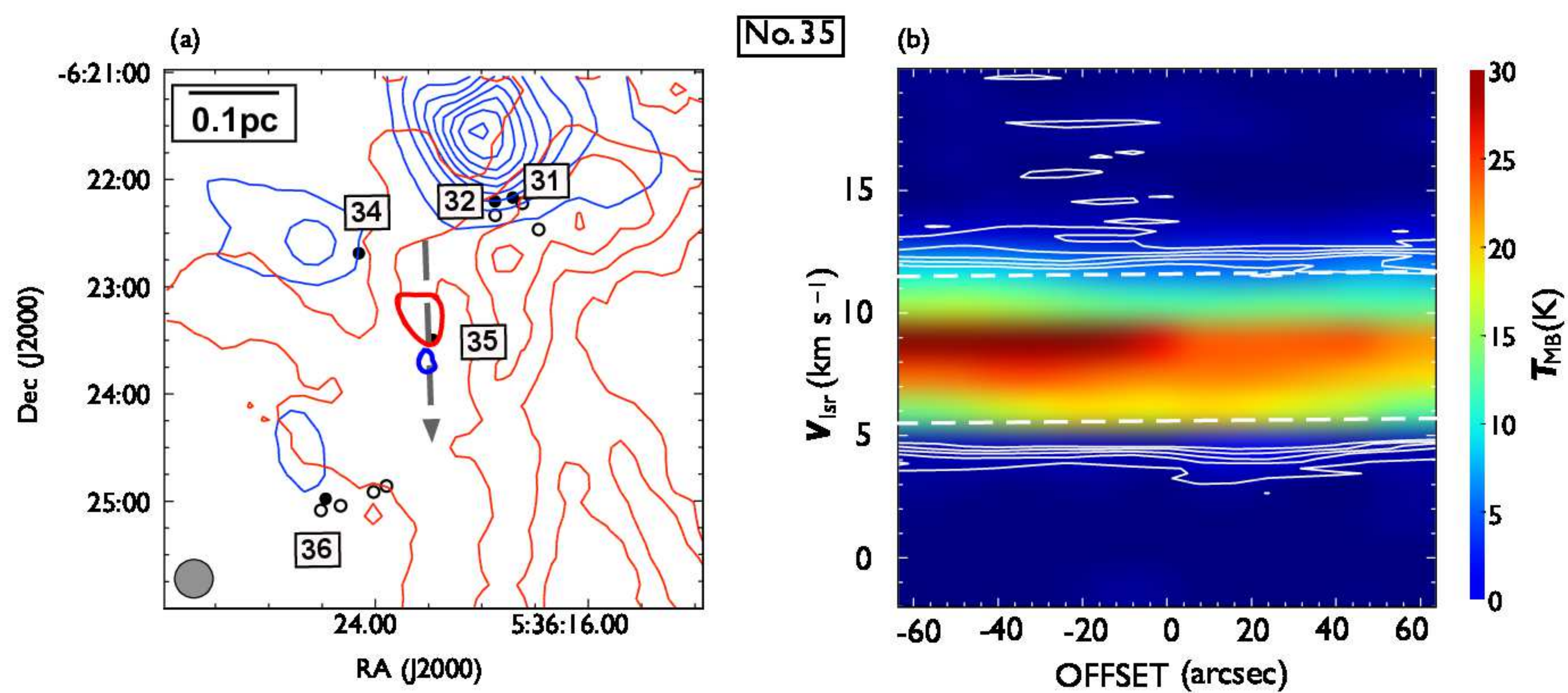}
	\caption{
	The same as in figure \ref{1} but for outflow No. 35. In panel (a), the blue- and red-shifted integrated intensity velocity ranges are -1.9 km s$^{-1}$ to 5.5 km s$^{-1}$ and 11.4 km s$^{-1}$ to 18.0 km s$^{-1}$, respectively.}
\end{figure}

\begin{figure}[h!]
 \includegraphics[keepaspectratio,width=16cm]{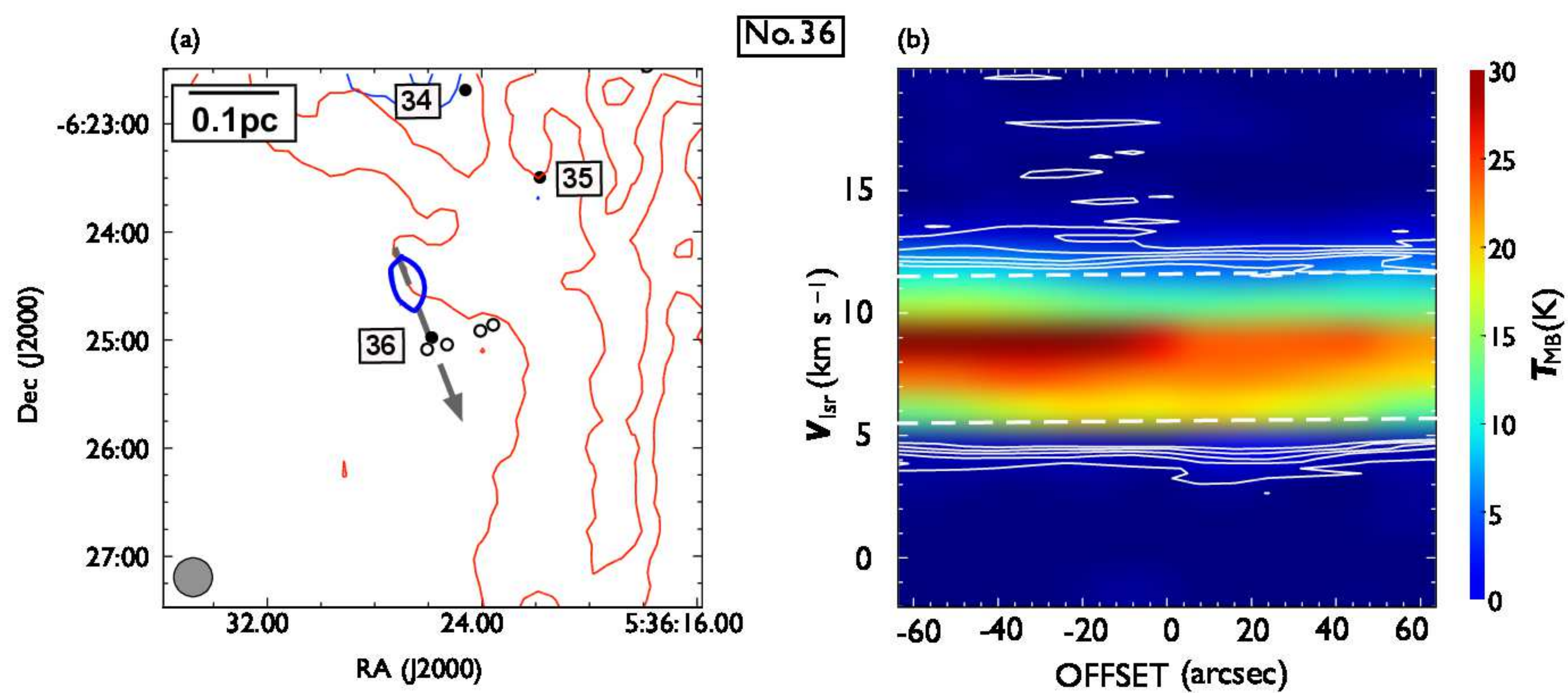}
	\caption{
	The same as in figure \ref{1} but for outflow No. 36. In panel (a), the blue- and red-shifted integrated intensity velocity ranges are -1.9 km s$^{-1}$ to 3.4 km s$^{-1}$ and 11.9 km s$^{-1}$ to 18.0 km s$^{-1}$, respectively.}
\end{figure}

\begin{figure}[h!]
 \includegraphics[keepaspectratio,width=16cm]{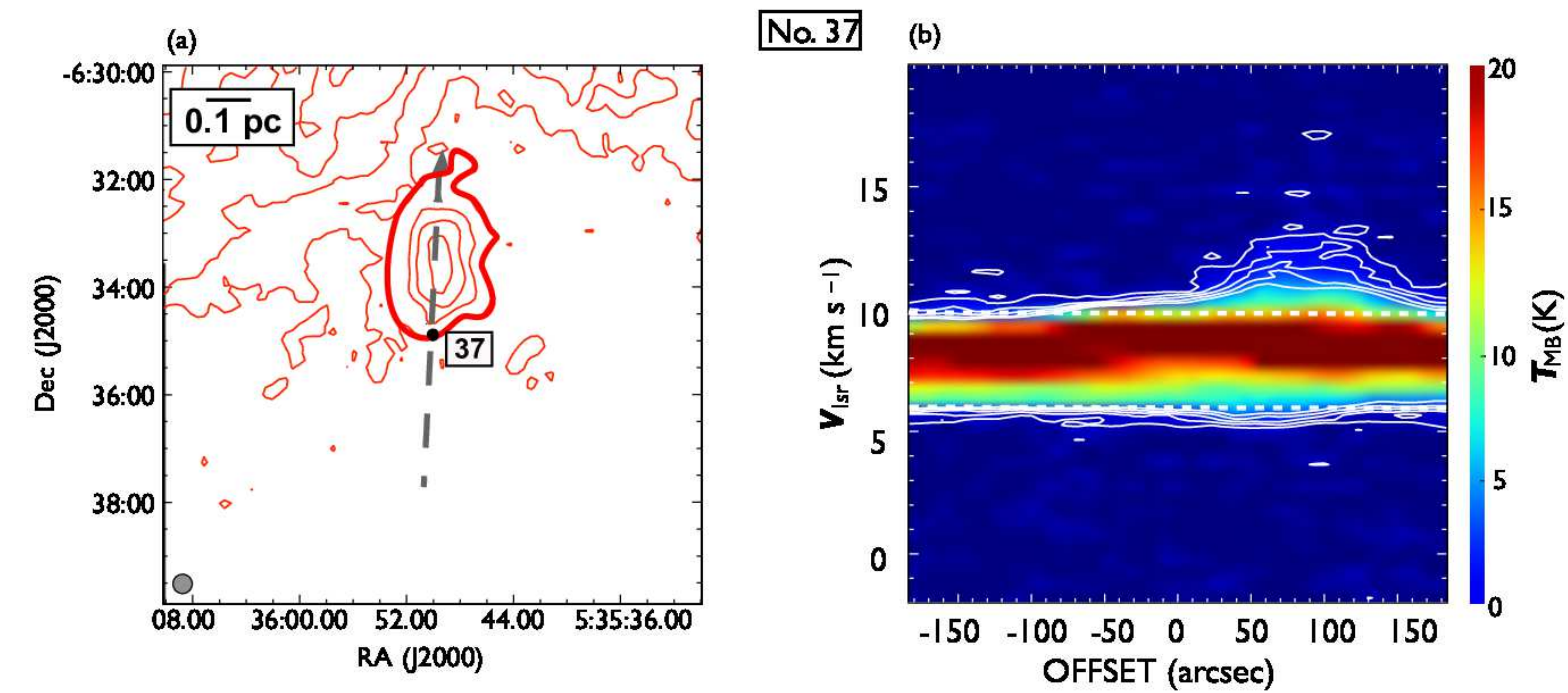}
	\caption{
	The same as in figure \ref{1} but for outflow No. 39. In panel (a), the blue- and red-shifted integrated intensity velocity ranges are -1.9 km s$^{-1}$ to 4.9 km s$^{-1}$ and 10.4 km s$^{-1}$ to 18.0 km s$^{-1}$, respectively.}
\end{figure}

\begin{figure}[h!]
 \includegraphics[keepaspectratio,width=16cm]{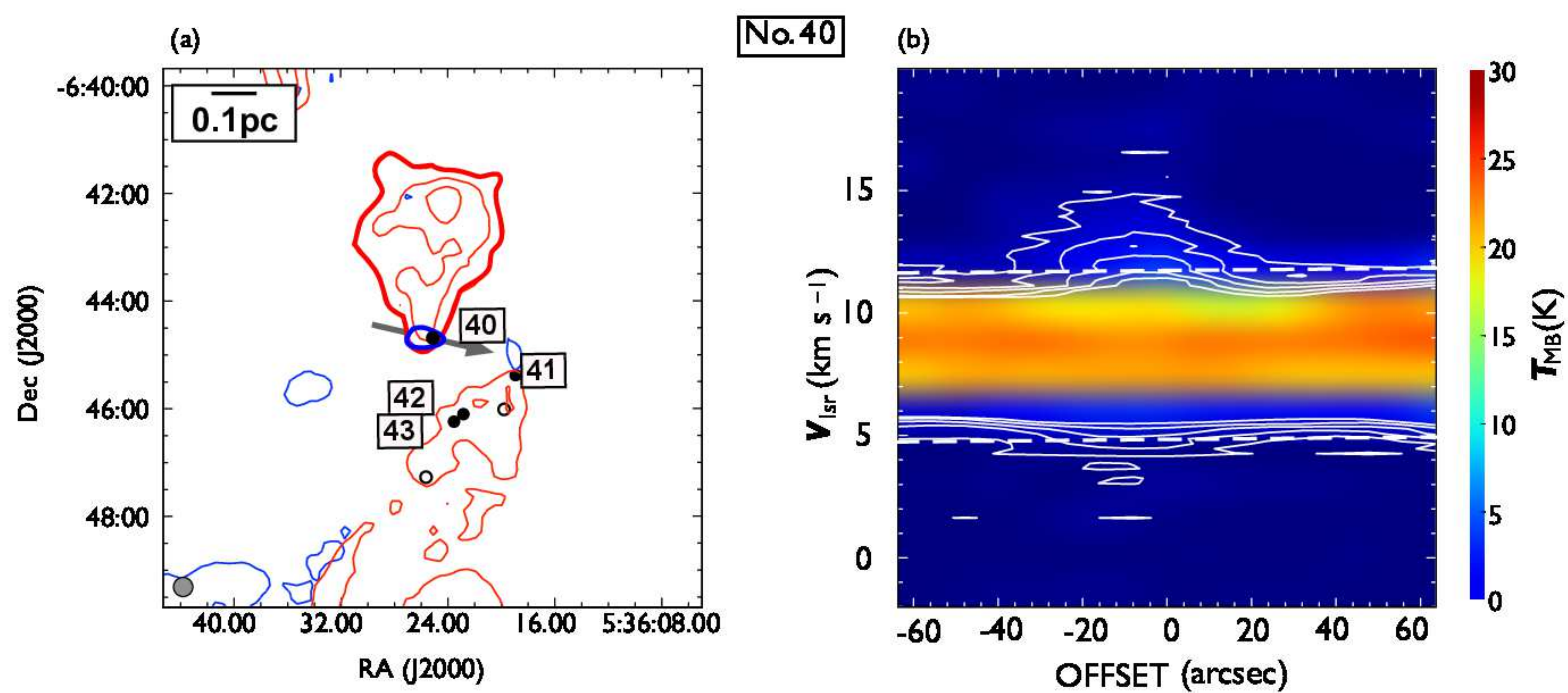}
	\caption{
	The same as in figure \ref{1} but for outflow No. 40. In panel (a), the blue- and red-shifted integrated intensity velocity ranges are -1.9 km s$^{-1}$ to 4.8 km s$^{-1}$ and 11.6 km s$^{-1}$ to 18.0 km s$^{-1}$, respectively.}
\end{figure}

\begin{figure}[h!]
 \includegraphics[keepaspectratio,width=16cm]{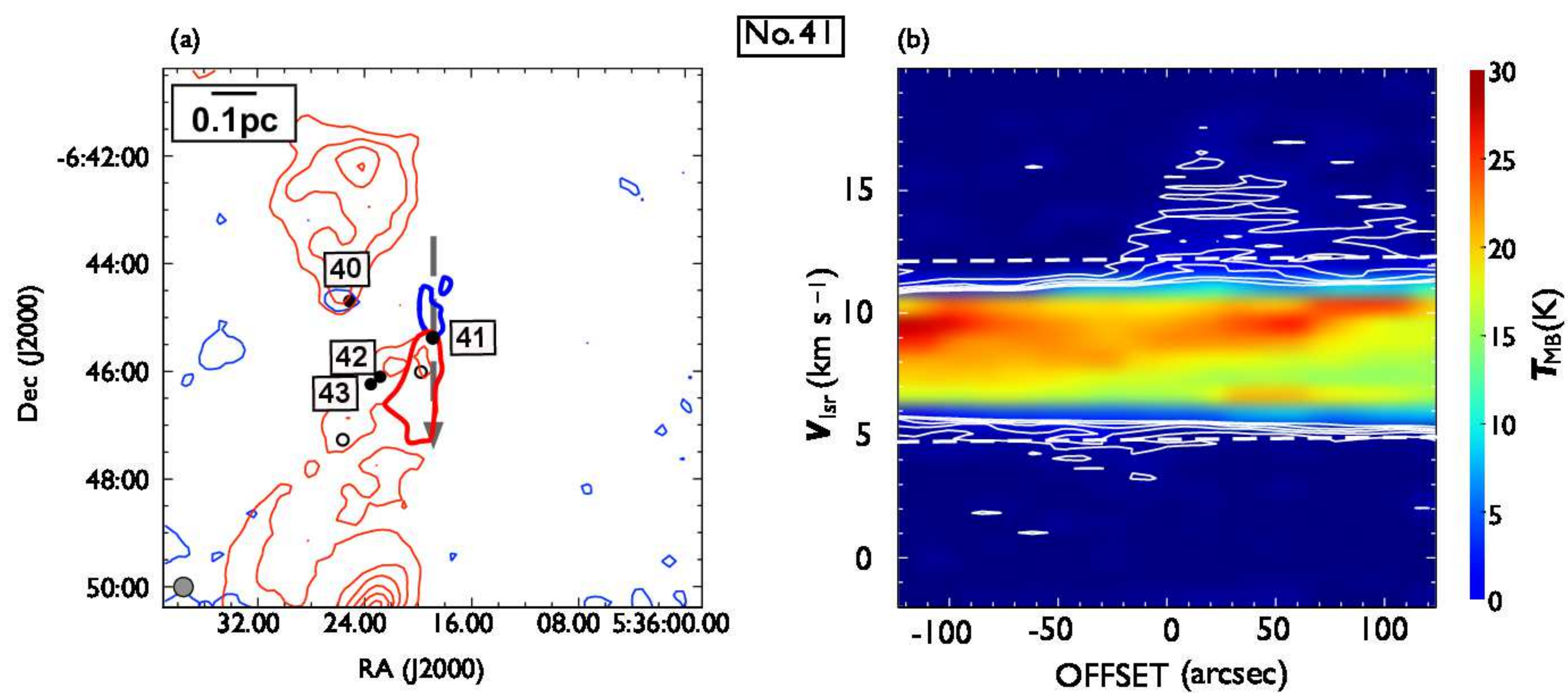}
	\caption{
	The same as in figure \ref{1} but for outflow No. 41. In panel (a), the blue- and red-shifted integrated intensity velocity ranges are -1.9 km s$^{-1}$ to 4.9 km s$^{-1}$ and 12.1 km s$^{-1}$ to 20.2 km s$^{-1}$, respectively.}
\end{figure}

\clearpage

\begin{figure}[h!]
 \includegraphics[keepaspectratio,width=16cm]{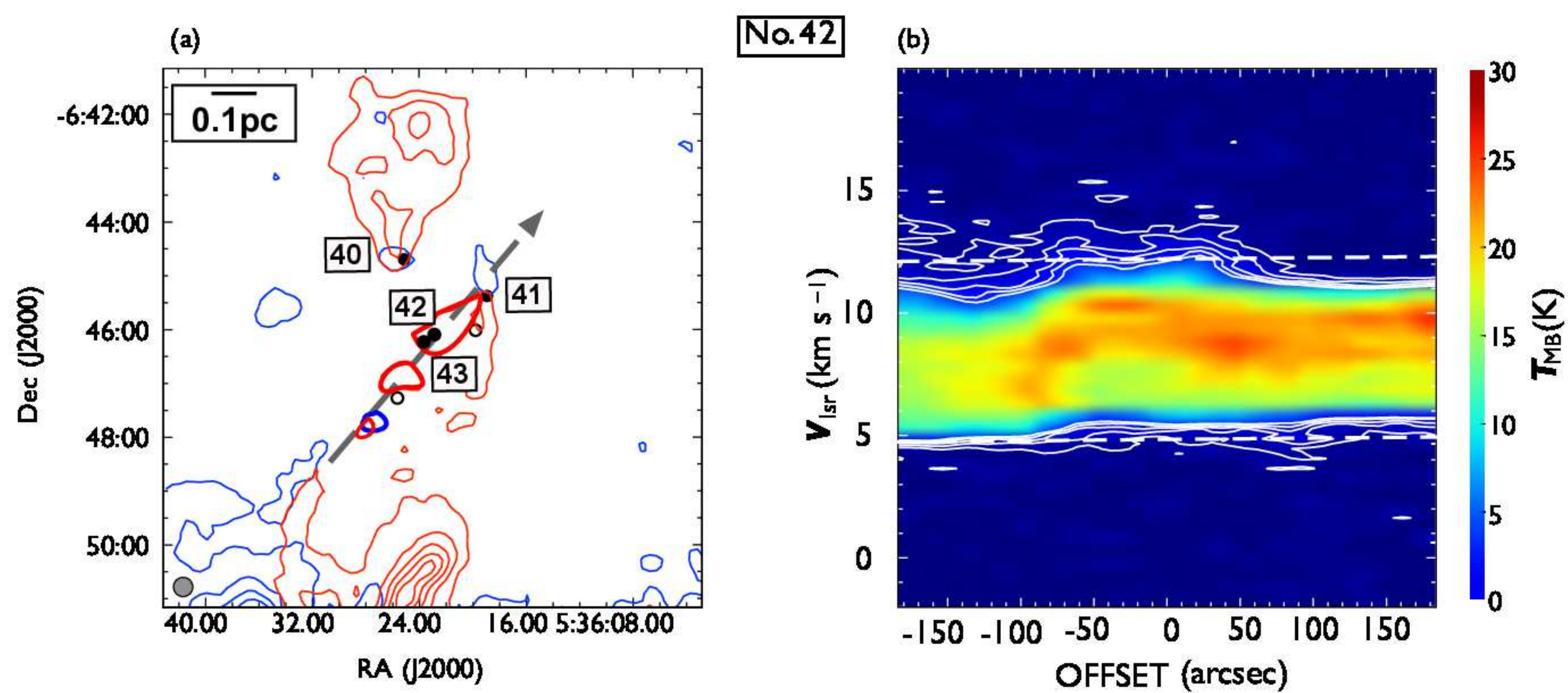}
	\caption{
	The same as in figure \ref{1} but for outflow No. 42. In panel (a), the blue- and red-shifted integrated intensity velocity ranges are -1.9 km s$^{-1}$ to 4.8 km s$^{-1}$ and 12.1 km s$^{-1}$ to 20.2 km s$^{-1}$, respectively.}
\end{figure}

\end{document}